\documentclass[showpacs,amsmath,amssymb]{revtex4}
\usepackage{graphicx,color,multirow}

\newcommand{\deltam}{\delta_{\rm m}}
\newcommand{\deltag}{\delta_{\rm g}}
\newcommand{\deltacb}{\delta_{\rm cb}}
\newcommand{\deltanu}{\delta_{\nu}}
\newcommand{\calS}{\mathcal{S}}
\newcommand{\calK}{\mathcal{K}}
\newcommand{\calF}{\mathcal{F}}
\newcommand{\fcb}{f_{\rm cb}}
\newcommand{\fnu}{f_{\nu}}
\newcommand{\Dcb}{D_{\rm cb}}

\newcommand{\Pcb}{P_{\rm cb}}
\newcommand{\bfv}{\mbox{\boldmath$v$}}
\newcommand{\bfq}{\mbox{\boldmath$q$}}
\newcommand{\bfx}{\mbox{\boldmath$x$}}
\newcommand{\bfk}{\mbox{\boldmath$k$}}
\newcommand{\bfr}{\mbox{\boldmath$r$}}
\newcommand{\bmf}[1]{\mbox{\boldmath$#1$}}
\newcommand{\be}{\begin{equation}}
\newcommand{\ee}{\end{equation}}
\newcommand{\ba}{\begin{eqnarray}}
\newcommand{\ea}{\end{eqnarray}}

\newcommand{\ltsim}{\protect\raisebox{-0.5ex}
  {$\:\stackrel{\textstyle <}{\sim}\:$}}
\newcommand{\gtsim}{\protect\raisebox{-0.5ex}
  {$\:\stackrel{\textstyle >}{\sim}\:$}}

\begin{document}


\title{%
Nonlinear power spectrum in the presence of  massive neutrinos:\\ 
perturbation theory approach, galaxy bias and parameter forecasts 
}%
\author{%
 Shun Saito$^1$, Masahiro Takada$^{2}$ and Atsushi Taruya$^{2,3}$
}%
\affiliation{%
 $^1$Department of Physics, School of Science, 
 The University of Tokyo, Tokyo 113-0033, Japan
}%
\affiliation{%
 $^2$Institute for the Physics and Mathematics of the Universe (IPMU), 
 The University of Tokyo, Chiba 277-8582, Japan
}%
\affiliation{%
 $^3$Research Center for the Early Universe, 
 School of Science, The University of Tokyo, Tokyo 113-0033, Japan
}%

\date{\today}

\begin{abstract}
Future or ongoing galaxy redshift surveys can put stringent constraints
on neutrinos masses via the high-precision measurements of galaxy power
spectrum, when combined with cosmic microwave background (CMB)
information.
In this paper we develop a method to model galaxy power spectrum in
the weakly nonlinear regime for a mixed dark matter (CDM plus 
finite-mass neutrinos) model, based on perturbation theory (PT) whose 
validity is well tested by simulations for a CDM model. In doing 
this we carefully study various aspects of the nonlinear clustering 
(nonlinear neutrino perturbations and the higher-order growth 
functions), and then arrive at a useful approximation allowing 
for a quick computation of the nonlinear power spectrum as in the CDM case. 
The nonlinear galaxy bias is also included
in a self-consistent manner within the PT framework. Thus the use of our 
PT model can give a more robust understanding of the measured galaxy 
power spectrum as well as allow for higher sensitivity to neutrino 
masses due to the gain of Fourier modes beyond the linear regime.  
Based on the Fisher matrix formalism, we find that BOSS or Stage-III type 
survey, when combined with Planck CMB information, gives a precision 
of total neutrino mass constraint, 
$\sigma(m_{\nu,{\rm tot}})\simeq 0.1$~eV, while Stage-IV type survey 
may achieve $\sigma(m_{\nu,{\rm tot}})\simeq 0.05~$eV, i.e. more than 
a $1$-$\sigma$ detection of neutrino masses. 
We also discuss possible systematic errors on dark
energy parameters caused by the neutrino mass uncertainty. The
significant correlation between neutrino mass and dark energy parameters
is found, if the information on power spectrum amplitude is included. More
importantly, for Stage-IV type survey, a best-fit dark energy model may
be biased and falsely away from the underlying true model by more than
the 1-$\sigma$ statistical errors, if neutrino mass is 
ignored in the model fitting.
\end{abstract}

\pacs{98.80.Es,14.60.Pq,98.65.Dx}

\maketitle

\section{Introduction}

The concordance $\Lambda$-dominated cold dark matter ($\Lambda$CDM)
model for structure formation in the Universe is remarkably successful
in describing various data sets such as cosmic microwave background (CMB)
anisotropies, Type-Ia supernova distance measurements, observations of
galaxy clustering and cluster counts, and weak gravitational lensing (e.g.,
\cite{Komatsu:2008aa}; \cite{Dodelson}). However, the concordance model
requires that the present-day energy budget of the Universe is dominated
by unknown two dark components. One is dark matter that is needed to
explain the nonlinear aspects of gravitational clustering in structure
formation, and the other is the cosmological constant contribution or
perhaps a more generalized form dubbed as dark energy, which is needed
to explain the cosmic accelerating expansion. Understanding the nature
of these dark components is one of the most important, profound problems
in cosmology as well as particle physics.

We now know that the Big-Bang relic neutrinos contribute to dark matter
energy density by some small fraction, because the neutrino oscillation
experiments \cite{exp1,exp2,exp3,exp4} have shown that
neutrinos have finite masses (also see \cite{McKeown:2004yq,Kayser:2005cd}
for a thorough review). However, the oscillation experiments are
sensitive only to mass square differences between different flavor
neutrinos, therefore the most fundamental constant of neutrinos,
absolute mass scale, is not yet known, although the lower bound on total
neutrino mass can be inferred as $m_{\nu,{\rm tot}}\gtsim 0.06$ or
$0.1$~eV for the normal and inverted mass hierarchies, respectively.  On
the other hand, the direct experiment has put only a weak upper bound on
electron neutrino mass such as $m_{\nu_e}\ltsim 2$~eV \cite{Bonn}.

Cosmological probes can give a complementary, albeit indirect, method
for constraining neutrino masses.  There are two kinds of the methods.
First is via the effect on cosmic expansion history. If neutrino species
are massive enough as $m_{\nu}\gtsim 0.5~$eV, the neutrinos became
non-relativistic before recombination epoch and then imprint
characteristic signatures onto structures of the 
CMB anisotropy spectra
\cite{Ichikawa:2004zi,Fukugita:2006rm}. On the other hand, low-redshift
geometrical probes such as Type-Ia supernovae (e.g. \cite{Kowalski}) and
the Baryon Acoustic Oscillation (BAO) \cite{Eisenstein:2005su} are
sensitive to the present-day energy density of non-relativistic
matter ($\Omega_{\rm m0}$) that is given by the sum of CDM, baryon and
neutrino contributions: $\Omega_{\rm m0}\equiv \Omega_{\rm
cdm0}+\Omega_{\rm b0}+\Omega_{\nu0}$. Therefore combining these
geometrical measurements can constrain neutrino mass: for example,
\cite{Komatsu:2008aa} already succeeded in obtaining the presumably
best-available constraint from this method, $m_{\nu,{\rm tot}}\ltsim
0.6~{\rm eV}$ (95\% C.L.).

Alternative cosmological method is using clustering information of
large-scale structure. Due to large velocities of the frozen Fermi-Dirac
distribution, neutrinos cannot cluster on scales below
the neutrino free-streaming scale that has a characteristic dependence
on neutrino mass and redshift as given by $k_{{\rm fs},i}\simeq
0.023h{\rm Mpc}^{-1}(m_{\nu,i}/0.1{\rm eV})
(\Omega_{m0}/0.23)[2/(1+z)]^{1/2}$, comparable with the BAO scales for
neutrino mass scales of interest.  As a result, the presence of
finite-mass neutrinos suppresses the amplitude of low-redshift power
spectrum on the small scales by at least the amount of a few percent,
compared to the model without finite-mass neutrinos, for a fixed
$\Omega_{\rm m0}$ \cite{Bond:1980}.  Thus given the CMB normalization of
primordial power spectrum, total neutrino mass can be explored by
measuring clustering strengths of low-redshift large-scale structure via
galaxy redshift survey \cite{Hu:1997mj,Elgaroy:2002aa,Tegmark:2006az},
weak gravitational lensing \cite{Ichiki:2008ye,Kitching:2008dp}, 
Lyman-$\alpha$ forest
power spectrum \cite{Seljak:2006bg} and potentially 21cm observations
\cite{Pritchard:2008wy}. The existing data sets have put more
stringent upper bounds on neutrino mass, 
 $m_{\nu,{\rm tot}}\ltsim 0.2$--$0.5$~eV,
than the direct experiment limit, although some residual systematics are
under discussion. 

There are a number of ongoing and planned galaxy redshift surveys such as 
WiggleZ \footnote{http://wigglez.swin.edu.au/}, FMOS \cite{Sumiyoshi:2009bf},
BOSS \footnote{http://cosmology.lbl.gov/BOSS/}, Subaru redshift survey
known as the former project WFMOS
\footnote{http://arxiv.org/ftp/astro-ph/papers/0510/0510272.pdf}, HETDEX
\footnote{http://www.as.utexas.edu/hetdex/}, EUCLID
\footnote{http://sci.esa.int/science-e/www/object/index.cfm?fobjectid=42266},
and JDEM \footnote{http://www.science.doe.gov/hep/hepap/feb2007/
hepap\_bennett\_feb07.pdf}.  The primary scientific target of these
surveys is exploring the nature of dark energy via the BAO
experiment. At the same time these surveys promise to achieve the
high-precision measurements of galaxy power spectrum amplitudes to a
percent level precision at each wavenumber bins, and therefore offer 
a possibility to dramatically improve cosmological constraints including
neutrino masses
\cite{Takada:2005si,Hannestad:2007cp,Abdalla:2007ut}.

Thus large-scale structure probes are very promising, however, the main
obstacle is nonlinear effects such as nonlinear gravitational
clustering, galaxy bias and redshift distortion.  Recent theoretical
studies have shown that, even at scales as large as $\sim 150 h^{-1}$Mpc
relevant for both the BAO and neutrino free-streaming scale, the
standard linear theory, which gives remarkably successful agreement with
CMB measurements, ceases to be accurate. The nonlinear effects are
found to be significant compared to the precision of future surveys, using
$N$-body simulations
\cite{Jeong:2006xd,Angulo:2007fw,Takahashi:2008wn,Sanchez:2008iw,
Seo:2008yx,Jeong:2008rj,Nishimichi:2008ry} and analytical studies
inspired from perturbation theory
\cite{McDonald:2006hf,Valageas:2006bi,
Matarrese:2007wc,Nishimichi:2007xt,Crocce:2007dt,Matsubara:2007wj,
Taruya:2007xy,Pietroni:2008jx,Nomura:2008tm,Carlson:2009it,Taruya:2009ir}.
However in most of these studies the
contribution of finite-mass neutrinos were ignored. The nonlinear effect of
finite-mass neutrinos on the power spectrum needs to be understood in
order to attain the full potential of future surveys, which is also important
to minimize the possible systematic error on BAO experiments caused by 
the incorrect assumption that neutrinos are massless. 

Therefore the aim of this paper is developing a formulation to model
nonlinear galaxy power spectrum in a mixed dark matter (CDM plus
finite-mass neutrinos) model, based on 
 standard perturbation theory (SPT)  
(see \cite{Bernardeau:2001qr} for a thorough review of perturbation
theory for a CDM model). Here we mean by ``standard'' that the
next-order corrections to the power spectrum, i.e. the one-loop
corrections, are included. 
In doing this we carefully study various
aspects of the nonlinear clustering: estimate the nonlinear neutrino
perturbations by solving the collision-less Boltzmann equation hierarchies and
study the higher-order growth functions of CDM plus baryon perturbations
that have complicated scale- and redshift-dependences similarly to the
linear-order growth rate. Then, given the detailed
assessment of various effects, we will arrive at a useful
approximation to compute the nonlinear matter power spectrum 
whose results were highlighted in
\cite{Saito:2008bp}.
We then include a modeling of nonlinear galaxy bias self-consistently
within the SPT framework 
following the method developed in
\cite{McDonald:2006mx}. Thus, while the nonlinear redshift distortion
effect is not yet included, our model of the galaxy power spectrum can
be compared to the actual measurement such as that in
\cite{Tegmark:2006az}, where the redshift distortion effect is removed
using the Finger-of-God compression algorithm \cite{Tegmark:2004}. For
preparation of such a study we will demonstrate parameter forecasts for
neutrino mass constraints expected from 
some of galaxy surveys listed above, paying a particular 
attention on the correlation between neutrino mass and dark energy
parameters in the weakly nonlinear regime. We also discuss a possible
systematic error in the constraints on the dark energy parameter caused 
by the neutrino mass uncertainty.

The structure of this paper is as follows. In \S~\ref{sec:Perturbation
theory} we develop the formulation of SPT method for computing the
nonlinear matter power spectrum. In \S~\ref{sec:neutrino _suppression}
we then study the effect of finite-mass neutrinos on the matter power
spectrum by varying the neutrino masses within the range inferred from
the constraints. After including a model of nonlinear galaxy bias based
on perturbation theory in \S~\ref{subsec:Nonlinear galaxy biasing}, we
study parameter forecasts of neutrino masses and dark energy parameters
using the Fisher matrix formalism in
\S~\ref{sec:Forecast}. \S~\ref{sec:Conclusion} is devoted to summary and
discussion. Unless explicitly stated, throughout this paper we assume
the concordance $\Lambda$CDM-like cosmology with finite-mass neutrino
contribution, which is consistent with the WMAP results
\cite{Komatsu:2008aa}. The fiducial model is: the density parameters are 
$\Omega_{\rm m0}=0.24,\ \Omega_{\rm m0}h^{2}=0.1277,$ and 
$ \Omega_{\rm b0}h^{2}=0.0223$.  
The neutrino effect is studied by varying the neutrino mass scale. 
For simplicity the number of neutrino species is assumed to be  
$N_{\nu}=3$ because the matter power spectrum is sensitive to 
the sum of neutrino masses, $m_{\nu,{\rm tot}}=N_{\nu}m_{\nu}$. 
We assume a flat universe and consider $w_0=-1$ 
for dark energy equation of state. 
For the primordial fluctuation parameters, 
the amplitude, the tilt, and the running, are set to 
$\Delta^2_{\mathcal R}=2.35\times 10^{-9}$, $n_{\rm S}=1.0$, 
and $\alpha_{\rm S}=0$, respectively.

\section{Perturbation theory for nonlinear matter power spectrum 
in a MDM model} 
\label{sec:Perturbation theory}

\subsection{Preliminaries} 
\label{subsec:Preliminaries}

First we write down basic equations to describe structure formation 
in a MDM model. Throughout this paper, we focus on the evolution 
of matter fluctuations consisting of MDM (CDM plus massive neutrinos) 
and baryon: 
\be
\deltam \equiv
\frac{\delta\rho_{\rm c}+\delta\rho_{\rm b}+\delta\rho_{\nu}}
{\rho_{\rm m}} 
=\fcb \deltacb + \fnu \deltanu, 
\label{eq:deltam}
\ee 
where the subscript `m', `c', `b', `$\nu$' and `cb' stand for total matter, 
CDM, baryon, massive neutrinos, and CDM plus baryon, respectively, and 
$\delta_{\rm cb}$ and $\delta_\nu$
denote their density perturbations. 
The coefficients, $\fcb$ and $\fnu$, are the fractional contributions 
of each component to
the present-day total matter density: 
\be
\fcb=\frac{\Omega_{\rm c0}+\Omega_{\rm b0}}{\Omega_{\rm m0}}, 
\quad
\fnu=\frac{\Omega_{\nu0}}{\Omega_{\rm m0}}
    =1-f_{\rm cb}
    \simeq\frac{m_{\nu,{\rm tot}}}{94.1\Omega_{\rm m0}h^{2}}, 
\ee
with the density parameter, $\Omega_{i0}$, being defined as
$\Omega_{i,0}\equiv 8\pi G\rho_i(t_0)/(3H_{0}^{2})\ (i={\rm m,c,b,\nu})$ 
where the parameter, $h$, is dimensionless Hubble constant defined as 
$H_{0}=100h~{\rm kms^{-1}Mpc^{-1}}$. 
In the limit of $\fcb\to 1$, the results shown below 
 recover a CDM model which does not contain 
massive neutrinos. 
The evolution of homogeneous and isotropic universe is controlled by 
CDM, baryon, massive neutrinos and dark energy whose equation of 
state is  
simply assumed to be constant in time: 
$p_{\rm DE}=w_{0}\rho_{\rm DE}$
where $w_0$ is referred to as the equation of state parameter. 
Then, the background Friedman equations become 
\ba
H^2 &=& \frac{8\pi G}{3} \left(\rho_{\rm m}+\rho_{\rm DE} \right), \\
\dot{H}&=&-\frac{3}{2}\,H^2(1+w_{0}), 
\ea
where the dot $\dot{\ }$ denotes the derivative  with respect to 
cosmic time, $t$: 
$\dot{a}=da/dt$, and the Hubble parameter, $H$, is defined as 
$H\equiv \dot{a}/a$. \par

We are specifically concerned with the nonlinear growth of matter 
perturbations, $\delta_{\rm m}$. Let us first consider the contribution
of CDM plus baryon perturbations, $\delta_{\rm cb}$, to the total matter
perturbations. 
In order to evaluate the nonlinear evolution of 
$\delta_{\rm cb}$, we treat the CDM plus baryon 
components as a single pressure-less fluid. 
The continuity equation and
the Euler equation for the CDM plus baryon fluctuations in Fourier space 
are given in
\cite{Bernardeau:2001qr,Taruya:1999vq} as
\ba
&&H^{-1}\frac{\partial\delta_{\rm cb}(\bfk;t)}{\partial t} + 
\theta_{\rm cb}(\bfk;t) =\,-\int\frac{d^3\bfk'}{(2\pi)^3}\,
\alpha(\bfk',\bfk-\bfk')\,\delta_{\rm cb}(\bfk-\bfk';t)\,
\theta_{\rm cb}(\bfk';t), 
\label{eq:perturb1}
\\
&&H^{-1}\frac{\partial\theta_{\rm cb}(\bfk;t)}{\partial t} + 
\frac{1}{2}(1-3w_{0}\Omega_{\rm DE})\theta_{\rm cb}(\bfk;t) 
+\frac{1}{2}(1-\Omega_{\rm DE})\delta_{\rm m}(\bfk;t) 
\nonumber
\\
&&\quad\quad\quad\quad\quad\quad\quad\quad\quad~~~
=\,-\frac{1}{2}\,\int\frac{d^3\bfk'}{(2\pi)^3}\,
\beta(\bfk',\bfk-\bfk')\,\theta_{\rm cb}(\bfk-\bfk';t)\,
\theta_{\rm cb}(\bfk';t), 
\label{eq:perturb2}
\ea
where the velocity divergence $\theta_{\rm cb}$ is defined as 
$\theta_{\rm cb}=\nabla\cdot \bfv_{\rm cb}/(aH)$ in real space. 
Note that we assume an irrotational flow, 
i.e. the vorticity is neglected \cite{Pueblas:2008uv}. 
The Fourier kernels to describe the nonlinear mode coupling, $\alpha$ 
and $\beta$, are defined as
\ba
\alpha(\bfk_1,\bfk_2)\equiv 1+\frac{\bfk_1\cdot\bfk_2}{|\bfk_1|^2},
\quad\quad
\beta(\bfk_1,\bfk_2)\equiv
\frac{(\bfk_1\cdot\bfk_2)|\bfk_1+\bfk_2|^{2}}{|\bfk_1|^2|\bfk_2|^2}. 
\ea
Taking the time-derivative of Eq.(\ref{eq:perturb1}) 
and also using the Euler equation (\ref{eq:perturb2})
yield the second-order differential equation for $\delta_{\rm cb}$:
\ba
&&\ddot{\deltacb} + 2H\dot{\deltacb}
-\frac{3}{2}H^2(1-\Omega_{\rm DE})\,\deltam
\nonumber\\
&&\quad\quad\quad
=-\int\frac{d^3\bfk'}{(2\pi)^3}\,\Bigl[\,
\alpha(\bfk',\bfk-\bfk') 
\Bigl\{\, \left[H\,\deltacb(\bfk-\bfk')\theta_{\rm cb}(\bfk')\right]^{.}
+2H^2\,\deltacb(\bfk-\bfk')\theta_{\rm cb}(\bfk')\,\Bigr\}\Bigr.
\nonumber\\
&&\quad\quad\quad\quad\quad\quad\quad\quad\quad
\left.+\frac{1}{2}H^2\,\beta(\bfk',\bfk-\bfk')\,
\theta_{\rm cb}(\bfk-\bfk')\theta_{\rm cb}(\bfk')
\right]. 
\label{eq:2nd_df}
\ea
Thus 
Eq.~(\ref{eq:2nd_df}) contains $\delta_{\rm m}=\fcb \deltacb + \fnu \deltanu$ 
and cannot be solved unless 
the neutrino fluctuation field, $\delta_{\nu}$, 
is specified. 

So let us move on to discussion on the neutrino perturbations. 
Unlike CDM and baryon, 
the finite-mass  neutrinos have 
a large velocity dispersion following the frozen Fermi-Dirac distribution, 
and cannot be treated as fluids. 
Therefore, exactly speaking, it is necessary to 
solve Eq.~(\ref{eq:2nd_df}) coupled with 
the collision-less Boltzmann equations for neutrino perturbations 
that include the nonlinear terms. 
This is still computationally expense, especially for solving the
nonlinear Boltzmann equations. 
Here we rather consider the approximated method for solving the
nonlinear perturbations as described in the next subsection, and will
also assess an accuracy of the approximation. 

In our method we focus on the linearized collision-less Boltzmann 
equations for neutrino perturbations \cite{Ma:1995ey}:
\ba
 && \Psi'_{0}=-\frac{qk}{a\epsilon}\Psi_{1}
                   +H\phi\frac{d\ln f_{0}}{d\ln q}, 
                   \label{eq:Vlasov0}\\
 && \Psi'_{1}=\frac{qk}{3a\epsilon}(\Psi_{0}-2\Psi_{2})
                   -\frac{\epsilon k}{3aq}\phi\frac{d\ln f_{0}}{d\ln q},
                   \label{eq:Vlasov1}\\
 && \Psi'_{\ell} = \frac{qk}{(2\ell+1)a\epsilon}
                       [\ell\Psi_{\ell-1}-(\ell+1)\Psi_{\ell+1}]
                       \ \ (\ell\ge 2), \label{eq:Vlasov2}
\ea
where the variables, $q$ and $\epsilon$, are comoving 3-momentum and 
proper energy defined as 
$\epsilon\equiv(q^{2}+a^{2}m_{\nu,i}^{2})^{1/2}$, respectively,
and the function $\phi$ is the gravitational potential perturbation 
under the conformal Newtonian gauge (see below). 
The superscript ' denotes the derivative with respect to 
conformal time. 
The function $f_0$ is the zeroth-order (isotropic) Fermi-Dirac
distribution, given as $f_{0}=2/(e^{\epsilon/aT}+1)$, and the function
$\Psi$ is the linear-order perturbed distribution. The full phase-space
distribution function of neutrinos is given in the
linear regime as

\ba
f(x^{i},q_{j}/a,t)=f_{0}(q)[1+\Psi(x^{i},q,\hat{n}_{j},t)],
\ea
where the momentum vector is rewritten as $q_j=q\hat{n}_j$ with 
$\hat{n}_j \hat{n}^j=1$.
The variables, $\Psi_{\ell}$, appearing in the Boltzmann equations above
are the $\ell$-th moments in the Legendre expansion of $\Psi$:
\ba 
\Psi(\bfk,\hat{\bf n},q,t)=\sum^{\infty}_{\ell=0}(-i)^{\ell}
    (2\ell+1)\Psi_{\ell}(k,q,t)P_{\ell}(\hat{\bfk}\cdot\hat{\bf n}),
\ea
where $P_\ell$ is the $l$-th order Legendre polynomial. 
The neutrino density perturbation is given by integrating the monopole
contribution of neutrino perturbations over momentum: 
\be
 \delta_{\nu}(k,t)=\frac{4\pi}{a^{4}f_{\nu}\rho_{\rm m}}\int q^{2}dq\, 
              \epsilon f_{0}(q)\Psi_{0}(k,q,t). \label{eq:neutrino_delta} 
\ee

The system of momentum hierarchies, Eqs.~(\ref{eq:Vlasov0}), 
(\ref{eq:Vlasov1}), and (\ref{eq:Vlasov2}), can be solved once the
gravitational potential $\phi$ is given. 
One of the Einstein equations, the Poisson equation, relates 
the potential $\phi$ to the total matter density perturbation 
$\delta_{\rm m}$ on subhorizon scales: 
\be
 -k^{2}\phi(k,t)=4\pi G a^{2}\rho_{\rm m}\delta_{\rm m}. 
\label{eqn:grav_phi}
\ee
On scales smaller than the neutrinos' free-streaming scale, $k\gtsim
k_{\rm fs}$, the neutrino perturbation would be absent, and the Poisson
equation roughly becomes 
$ -k^{2}\phi(k,t)\approx 4\pi G a^{2}\rho_{\rm m}f_{\rm cb}\delta_{\rm cb}$. 
Thus on these small scales
the dynamics of neutrino perturbations are governed 
by the CDM plus baryon perturbations. 
We have so far written down all the basic equations that govern the
dynamics of density perturbations for each components, 
$\deltacb$ and $\deltanu$. 
A quantity that is more relevant for actual large-scale structure probes
such as galaxy clustering is the power
spectrum of total matter including nonlinear corrections:
\be
\langle \deltam(\bfk;t)\deltam(\bfk';t)\rangle
=(2\pi)^{3}\delta_{D}(\bfk+\bfk')P_{\rm m}(k;t). 
\ee
The power spectrum, $P_{\rm m}$, 
is defined in terms of the density perturbations of CDM, baryon and 
neutrino perturbations as
\be
P_{\rm m}(k;t) = 
\fcb^2\,P_{\rm cb}(k;t) + 2\fcb\fnu\,P_{\rm cb\nu}(k;t) 
+ \fnu^2\,P_{\nu}(k;t),
\label{eq:total_matter_pk}
\ee
where $P_{\rm cb\nu}(k)$ is the cross spectrum between 
$\deltacb$ and $\deltanu$.

\subsection{On the treatment of neutrino perturbation } 
\label{subsec:assumption}
Strictly speaking, in order to compute the total matter perturbation,
$\delta_{\rm m}$, in the nonlinear regime, 
we need to solve Eq.~(\ref{eq:2nd_df}) coupled with 
nonlinear collision-less Boltzmann equations for massive neutrinos, 
which seems computationally expensive. 
In order to avoid this obstacle, in this paper we employ a simple 
approximation that allows to analytically compute the
nonlinear power spectrum in a MDM model based on the standard 
perturbation theory (SPT) (see \cite{Saito:2008bp} 
for the similar discussion).

Let us begin with recalling characteristic properties of neutrino 
clustering on scales up to $k\sim 0.1~h{\rm Mpc}^{-1}$.
Firstly, the neutrino perturbations contribute to nonlinear gravitational
clustering via its contribution to the gravitational potential, where, 
implied in Eq.~(\ref{eqn:grav_phi}), the perturbation of {\em physical}
neutrino density, $\delta\rho_\nu=\bar{\rho}_\nu\delta_\nu$, affects the
gravitational potential. Thus the contribution is
suppressed by a small factor $f_\nu$, currently limited as $f_\nu\ltsim
0.05$ \cite{Komatsu:2008aa}, even if the density perturbations of CDM
and neutrinos are in similar amplitudes as predicted by the adiabatic
structure formation scenario.  Secondly, the neutrino perturbations would
tend to stay in the linear regime due to the large velocity dispersion,
at least on scales relevant for the BAO scales.
These facts suggest that the nonlinear power spectrum arises mainly from
the nonlinear perturbations of CDM plus baryon. Thus we model the
nonlinear matter spectrum based on SPT
(see below), but including only the linear-order 
perturbations of neutrinos: 
\be
P^{\rm NL}_{\rm m}(k;t) = 
\fcb^2\,P^{\rm NL}_{\rm cb}(k;t) + 2\fcb\fnu\,P^{\rm L}_{\rm cb\nu}(k;t) 
+ \fnu^2\,P^{\rm L}_{\nu}(k;t), \label{eq:total_Pk}
\ee
where the spectra with superscript ``NL'' denote the nonlinear spectra
described below, and the spectra with ``L'' are the linear-order
spectra. With this assumption, the neutrino perturbations can be
precisely computed for given initial conditions by using the publicly
available codes such as CMBFAST \cite{Seljak:1996} and
CAMB \cite{Lewis:1999bs}.
The validity of our assumption is studied in detail in
Appendix~\ref{sec:validity_neutrino_linear}. Here we briefly summarize the
result. As explained around Eqs.~(\ref{eq:Vlasov2}) and
(\ref{eqn:grav_phi}), 
nonlinear clustering of neutrino perturbations is driven by nonlinear
gravitational potential supported by CDM plus baryon perturbations, in a
CDM dominated structure formation model. Therefore, 
the nonlinear correction to neutrino perturbations can be qualitatively
estimated by solving the linearized Boltzmann equations
(\ref{eq:Vlasov0})-(\ref{eq:Vlasov2}), where the nonlinear gravitational
potential due to the total matter density perturbations given by
Eq.~(\ref{eq:total_Pk}) is inserted into the gravitational force term
(ignoring the nonlinear neutrino perturbations).
The results are shown in Fig.~\ref{fig.A}. Nonlinear clustering indeed
causes a nonlinear evolution of neutrino perturbations, deviating from
the linear theory prediction.  The nonlinear effect causes greater
amplitudes of the neutrino perturbations on larger $k$ and at lower
redshifts; e.g. the fractional difference between the linear and
non-linear density perturbations $\delta_\nu^{\rm NL}/\delta^{\rm
L}_\nu$ reaches to $\sim 10\%$ on $k\ltsim 1~h$Mpc$^{-1}$ at $z=0$ for
$f_\nu=0.05$. However, the nonlinear effect on the total matter power
spectrum is suppressed by additional small factor $f_\nu$ as implied in
Eq.~(\ref{eq:total_Pk}). In conclusion the nonlinear correction to the
total matter power spectrum is smaller than one percent level in the
amplitude for a range of neutrino masses, $f_\nu\ltsim 0.05$.
For these reasons, throughout this paper, 
we employ the approximation (\ref{eq:total_Pk}),
where the neutrinos affect nonlinear power spectrum of total matter via
the effect on the growth rates of CDM plus baryon perturbations as
described in the next section.\par

\subsection{Perturbation Theory Approach} 
\label{subsec:Perturbative approach}
In this subsection we develop a method to compute nonlinear power
spectrum of CDM plus baryon perturbations, 
$P_{\rm cb}(k;t)$ in a MDM model based on perturbation theory. 
First, in order to solve Eq.~(\ref{eq:2nd_df}), 
we expand the density and velocity
perturbations in a perturbative manner: 
\be
\delta_{\rm cb} = \delta_{\rm cb}^{(1)} + 
\delta_{\rm cb}^{(2)} + 
\delta_{\rm cb}^{(3)} +\cdots, 
\quad\quad
\theta_{\rm cb} = \theta_{\rm cb}^{(1)} + 
\theta_{\rm cb}^{(2)} + 
\theta_{\rm cb}^{(3)} +\cdots, 
\label{eq:pert_sol}
\ee
where the superscript `($i$)' denotes the $i$-th order perturbation. 
Here, we include the next-to-leading order corrections for 
$P_{\rm cb}(k;t)$, which are expressed as 
\be
 P_{{\rm cb}}(k;t)=P^{{\rm L}}_{{\rm cb}}(k;t)
 +P^{(13)}_{{\rm cb}}(k;t)
 +P^{(22)}_{{\rm cb}}(k;t). 
\ee
The first term $P^{{\rm L}}_{{\rm cb}}$ denotes the linear power spectrum of 
CDM plus baryon. The last two terms describe the nonlinear 
corrections, the so-called one-loop corrections, and the superscript 
`(13)' and `(22)' denote the multiplied order of perturbations, 
$\langle\deltacb^{(1)}\deltacb^{(3)}\rangle$ and 
$\langle\deltacb^{(2)}\deltacb^{(2)}\rangle$. 
We thus include contributions up to the third-order perturbations. \par

Inserting the formal solutions (\ref{eq:pert_sol}) into Eq.~(\ref{eq:2nd_df})
gives, at the lowest order of perturbations,
the differential equation for $\deltacb^{{(1)}}$:
\be
\ddot{\deltacb}^{(1)} + 2H\dot{\deltacb}^{(1)}
-\frac{3}{2}H^2(1-\Omega_{\rm DE})\,\deltam^{(1)}=0. 
\ee
This equation can be straightforwardly solved, 
together with the linearized Boltzmann equation for 
neutrino perturbations (e.g., \cite{Lesgourgues:2006nd}).
In this paper we use CAMB \cite{Lewis:1999bs} to obtain the 
accurate solution of $\deltacb^{{(1)}}$.
Before moving on to the higher-order perturbations of $\delta_{\rm cb}$, 
for convenience of our discussion, 
we formally write down the linear-order solutions of density and 
velocity perturbations expressed as \cite{Eisenstein:1997jh,Hu:1997vi}:
\be
\deltacb^{(1)}(\bfk;t) = \Dcb(k;t) \,\hat{\Delta}(\bfk), 
\quad\quad
\theta_{\rm cb}^{(1)}(\bfk;t) = 
-\frac{d\Dcb(k;t)}{d\ln a} \,\hat{\Delta}(\bfk), 
\label{eq:1st_sol} 
\ee 
where the quantity $\hat{\Delta}(\bfk)$ represents the initial 
perturbation variables at
an early epoch  $t_{\rm ini}$, sufficiently in the linear regime, 
e.g. the Compton-drag epoch. The ensemble average
gives the initial power spectrum: 
\be
\langle\hat{\Delta}(\bfk)\hat{\Delta}(\bfk')\rangle = (2\pi)^{3}
\delta_D(\bfk+\bfk')\,\,P^{\rm L}_{\rm cb}(k;t_{\rm ini}). 
\ee
The effect of massive neutrinos can thus be described as the
scale-dependent growth function, $D_{\rm cb}(k;t)$. 
At wavenumbers smaller than the neutrino 
free-streaming scale, $k_{\rm fs}$, the neutrinos can cluster 
together with CDM and baryon. On the other hand, at $k>k_{\rm fs}$, 
the growth of CDM plus baryon perturbations is suppressed due to 
the weaker gravitational force caused by the lack of neutrino 
perturbations. 
Thus the growth function has asymptotic behaviors given 
in \cite{Lesgourgues:2006nd} as 
\be
 \Dcb(k;t) \propto \left\{
 \begin{array}{cc}
 D_{1}(t) & (k\ll k_{\rm fs})\\
 D_{1}(t)^{1-p} & (k\gg k_{\rm fs})
 \end{array}\right. ,
\ee
where $D_{1}(z)$ is the growth rate for a CDM model without massive
neutrinos, but with the same matter density $\Omega_{\rm m0}$ to that of
the MDM model (in this case the growth rate has no scale dependence), 
and the parameter $p$
is defined as $p\equiv (5-\sqrt{25-24f_{\nu}})/4$. 

We now consider the second- and third-order perturbations. 
Substituting the linear solutions into the r.h.s of 
Eq.~(\ref{eq:2nd_df}) yields  
the differential equation for the 
second-order perturbation $\deltacb^{(2)}$: 
\ba
&&\ddot{\deltacb}^{(2)} + 2H\dot{\deltacb}^{(2)}
-\frac{3}{2}H^2(1-\Omega_{\rm DE})\,f_{\rm cb}\deltacb^{(2)}
=\int\frac{d^3\bfk_1d^3\bfk_2}{(2\pi)^3}\,\,\delta_D(\bfk-\bfk_1-\bfk_2) 
\hat{\Delta}(\bfk_1)\hat{\Delta}(\bfk_2)
\nonumber\\
&&\quad\quad\quad\quad\times \,\,\Bigl[\,
\alpha(\bfk_1,\bfk_2) 
\Bigl\{\, \left[H\,\frac{d\Dcb(k_1)}{d\ln a}\Dcb(k_2)\right]^{.}+
2H^2\,\frac{d\Dcb(k_1)}{d\ln a}\Dcb(k_2)\,\Bigr\}
\nonumber\\
&&\quad\quad\quad\quad\quad\quad\quad\quad\quad\quad\quad\quad
\quad\quad\quad\quad\quad
+\,\beta(\bfk_1,\bfk_2)\,\,\frac{1}{2}H^2\,
\frac{d\Dcb(k_1)}{d\ln a}\frac{d\Dcb(k_2)}{d\ln a}\,
\Bigr],  
\label{eq:2nd_order_eq}
\ea
where, as described in the preceding section, we have 
ignored the second-order contribution of neutrino 
perturbations, i.e. set $\delta_{\nu}^{(2)}=0$, and therefore used
$\delta_{\rm m}^{(2)}=f_{\rm cb}\deltacb^{(2)}$ in deriving the equation
above. 
The formal solutions of Eq.~(\ref{eq:2nd_order_eq}) can be written as
\ba
\deltacb^{(2)}(\bfk;t) &=& 
\int\frac{d^3\bfk_1 d^3\bfk_2}{(2\pi)^3}\,\,\delta_D(\bfk-\bfk_1-\bfk_2) 
\hat{\Delta}(\bfk_1)\hat{\Delta}(\bfk_2)
\nonumber\\
&&\quad\quad\quad\quad\quad\quad\times\,\Bigl[\,
\alpha(\bfk_1,\bfk_2) A^{(2)}_{\delta}(k_1,k_2;t) +\,\beta(\bfk_1,\bfk_2) 
B^{(2)}_{\delta}(k_1,k_2;t) \Bigr], 
\label{eq:2nd_order_sol1}
\ea
where $A^{(2)}_{\delta}$ and $B^{(2)}_{\delta}$ are 
the second-order growth functions given in detail in 
Appendix~
\ref{sec:Higher-order growth}

There are notable differences between the second-order perturbations 
in models with and without massive neutrinos. Firstly, the second-order
growth functions $A^{(2)}_\delta$ and $B^{(2)}_\delta$ 
are scale-dependent originating from the scale-dependence
of the linear growth rate. Thus additional nonlinear mode coupling
arises via the scale-dependent growth rate, in addition to via the shape of
the input linear power spectrum.  
Secondly, the gravitational force is weaker in a MDM model because 
we ignored the second-order neutrino perturbations in our method,
i.e., $\delta_{\rm m}^{(2)}=f_{\rm cb}\deltacb^{(2)}$ with $f_{\rm cb}< 1$.  
These imply that the second-order density perturbations are suppressed
compare to those of CDM model with same $\Omega_{\rm m0}$. 
We will in detail show the results below. 

Similarly, a formal solution of the third-order perturbation can be
expressed as 
\ba
\deltacb^{(3)}(\bfk;t) &=& 
\int\frac{d^3\bfk_1 d^3\bfk_2 d^{3}\bfk_3}{(2\pi)^6}\,
\delta_D(\bfk-\bfk_1-\bfk_2-\bfk_3) 
\hat{\Delta}(\bfk_1)\hat{\Delta}(\bfk_2)\hat{\Delta}(\bfk_3)
\nonumber\\
&&\times\,\left[\,\alpha_{1,23}
\{\alpha_{2,3}A^{(3)}_{\delta1,2,3}(t)
  +\beta_{2,3}B^{(3)}_{\delta1,2,3}(t)\} -\alpha_{23,1}
\{\alpha_{2,3}C^{(3)}_{\delta1,2,3}(t)
  +\beta_{2,3}D^{(3)}_{\delta1,2,3}(t)\}
\right.\nonumber\\
&&\quad\quad\quad -\beta_{1,23}
\{\alpha_{2,3}E^{(3)}_{\delta1,2,3}(t)
  +\beta_{2,3}F^{(3)}_{\delta1,2,3}(t)\}, 
\label{eq:3rd_order_sol1}
\ea
where the Fourier kernels $\alpha(\bfk_1,\bfk_2)$, $\beta(\bfk_1,\bfk_2)$
and $\alpha(\bfk_1,\bfk_{2}+\bfk_{3})$ are abbreviated as
$\alpha_{1,2}$, $\beta_{1,2}$ and $\alpha_{1,23}$ respectively, and the
third-order growth functions $\mathcal{I}^{(3)}_{\delta}\,
(\mathcal{I}=A,B,C,D,E,F)$, abbreviated as
$\mathcal{I}^{(3)}_{\delta}(k_{1},k_{2},k_{3})=
\mathcal{I}^{(3)}_{\delta1,2,3}$ and so on, are given in
Appendix~\ref{sec:Higher-order growth}.

\subsection{One-loop Corrections to $P_{\rm cb}$}
\label{subsec:One-loop corrections}
We now study the higher-order density perturbations 
for a given MDM model. 
Using the formal solutions, we can derive the explicit 
expressions for the one-loop corrections to $P_{\rm cb}$.
First let us consider $P^{(22)}_{\rm cb}$. Using
Eq.~(\ref{eq:2nd_order_sol1}), 
the ensemble average $\langle 
\deltacb^{(2)}(\bfk;t)\deltacb^{(2)}(\bfk';t)\rangle
$ yields the power spectrum $P_{\rm cb}^{(22)}$: 
\be
 P^{(22)}_{\rm cb}(k;t) = 
2\int\left.\frac{d^{3}\bfk_1}{(2\pi)^{3}}
[\alpha_{1,2}A^{(2)}_{\delta 1,2}(t)
 +\beta_{1,2}B^{(2)}_{\delta 1,2}(t)]^{2}
P^{\rm L}_{\rm cb}(k_{1},t_{\rm ini})
P^{\rm L}_{\rm cb}(k_{2},t_{\rm ini})\right|_{\bfk=\bfk_1 +\bfk_2}, 
\ee
where we have used the abbreviated expressions such as 
$A^{(2)}_{\delta}(k_{1},k_{2},t)=A^{(2)}_{\delta, 1,2}(t)$, and the
$\bfk_1$-integration has to be done under the condition $\bfk=\bfk_1+\bfk_2$. 
The prefactor 2 arises from 
the Wick's theorem in evaluating the ensemble average, 
$\langle\Delta(\bfk_1)\Delta(\bfk_2)\Delta(\bfk_3)\Delta(\bfk_4)\rangle$. 
Changing the integration variables to 
$\bfr\equiv \bfk_1/k$ and $\mu\equiv \bfk\cdot\bfk_1 /(k\,k_1)$, 
the expression of $P^{(22)}_{\rm cb}$ is rewritten as 
\ba
P_{\rm cb,\,MDM}^{(22)}(k;t) &=& \frac{k^3}{2\pi^2}
\int^{\infty}_{0}r^{2}dr P^{\rm L}_{\rm cb}(kr;t)
\int^{1}_{-1}d\mu\, P^{\rm L}_{\rm cb}(k\sqrt{1+r^{2}-2r\mu};t)
\mathcal{K}^{(2)}_{\delta}(k,r,\mu;t), 
\label{eq:Pk22_1}
\ea
where $P^{\rm L}(k; t)$ is the linear spectrum at time $t$, given in
terms of the initial spectrum as 
$P^{\rm L}(k; t)\equiv D_1(t)^2P^{\rm L}(k; t_i)$, and $\mathcal{K}$ is
the function containing the growth functions, which is defined as
\ba
\mathcal{K}^{(2)}_{\delta}(k,r,\mu;t)&=&
\left[\frac{1}{2}\left(\frac{\mu}{r}
  \frac{A^{(2)}_{\delta 1,2}(t)}{D_{\rm cb}(kr;t)
        D_{\rm cb}(k\sqrt{1+r^{2}-2r\mu};t)}\right.\right.
\nonumber\\
&&\quad\quad  +\left.\frac{1-r\mu}{1+r^{2}-2r\mu}
\frac{A^{(2)}_{\delta 2,1}(t)}
{D_{\rm cb}(kr;t)D_{\rm cb}(k\sqrt{1+r^{2}-2r\mu};t)}\right)
\nonumber\\
&&\quad \left. +\frac{\mu-r}{r(1+r^{2}-2r\mu)}
\frac{B^{(2)}_{\delta 1,2}(t)}
{D_{\rm cb}(kr;t)D_{\rm cb}(k\sqrt{1+r^{2}-2r\mu};t)}\right]^{2}.
\label{eq:2ndgrowth_def}
\ea
For the CDM model case (the case without massive neutrinos), 
i.e. the limit $f_{\rm cb}\to 1$, 
the higher-order growth functions become scale-independent as a result
of the scale-independence of the linear growth rate. In this case, 
the growth functions are well approximated as
\ba
&&\frac{A^{(2)}_{\delta 1,2}(t)}{D_{\rm cb}(kr;t)
        D_{\rm cb}(k\sqrt{1+r^{2}-2r\mu};t)}
  \to \frac{5}{7},\nonumber\\
&&\frac{B^{(2)}_{\delta 1,2}(t)}{D_{\rm cb}(kr;t)
        D_{\rm cb}(k\sqrt{1+r^{2}-2r\mu};t)}
  \to \frac{1}{7}.
\label{eq:2ndNLgrowth}
\ea
Note that the asymptotic behaviors above are exact
only in an Einstein de-Sitter model ($\Omega_{\rm m0}=1$), 
but hold an excellent approximation for a CDM model at relevant redshifts
\cite{Bernardeau:1993qu,Takahashi:2008yk}. 
Hence, for the case $f_{\rm cb}=1$, Eq.~(\ref{eq:Pk22_1})
recovers a well-known expression of the one-loop power spectrum 
$P^{(22)}_{\rm CDM}(k,t)$
for the CDM model case \cite{Makino:1992a,Jain:1994a}: 
\ba
 &&P^{(22)}_{\rm cb}~\to~P^{(22)}_{\rm cb}(k;t
 )=\frac{k^{3}}{98(2\pi)^{2}}\int^{\infty}_{0}dr\,
   P^{\rm L}_{\rm cb}(kr;t) \nonumber\\
 &&\quad\quad\quad\quad\quad\quad
   \times \int^{1}_{-1}d\mu\,P^{L}_{\rm cb}(k\sqrt{1+r^{2}-2r\mu};t)
   \frac{(3r+7\mu-10r\mu^{2})^{2}}{(1+r^{2}-2r\mu)^{2}}.
 \label{eq:P22approx}
\ea
Thus, for a MDM model, the scale-dependent growth function
(Eq.~[\ref{eq:2ndgrowth_def}]) has to be solved before obtaining the 
the power spectrum $P^{(22)}_{\rm cb}$. 
Because of this, 
the exact computation of $P_{\rm cb}^{(22)}$ is computationally expensive. 

Similarly, another one-loop power spectrum,  $P^{(13)}_{\rm cb}$,
is formally expressed as 
\ba
 P^{(13)}_{\rm cb,\,MDM}(k;t)&=&\frac{2k^{3}P^{\rm L}_{\rm cb}(k;t)}
 {(2\pi)^{2}}
 \int^{\infty}_{0} dr\; r^{2}\Pcb^{\rm L}(kr;t) 
 \mathcal{K}^{(3)}_{\delta}(k,r;t), 
\label{eq:Pk13_1}
\ea
where the growth function $\mathcal{K}^{(3)}_\delta$ is defined as
\ba
 \mathcal{K}^{(3)}_{\delta}(k,r;t)&=& \frac{1}{\Dcb(k;t)\Dcb(kr;t)^{2}}
\left[-\frac{2}{3}\left\{A^{(3)}_{\delta}(kr,k,kr;t)+\frac{1}{r^2}
                         A^{(3)}_{\delta}(kr,kr,k;t)\right\}
\right.\nonumber\\
 && \quad\quad
      -\frac{2(1+r^2)}{3r^2}\left\{B^{(3)}_{\delta}(kr,k,kr;t)+
                                   B^{(3)}_{\delta}(kr,kr,k;t)\right\}
\nonumber\\
 && \quad\quad
      +\left\{\frac{-3+r^2}{2}-\frac{(1-r^2)^2}{4r}
 \ln\left|\frac{1+r}{1-r}\right|\right\}C^{(3)}_{\delta}(kr,k,kr;t)
\nonumber\\
 && \quad\quad
      +\left\{\frac{-1-r^2}{2r^{2}}+\frac{(1-r^2)^2}{4r^3}
 \ln\left|\frac{1+r}{1-r}\right|\right\}C^{(3)}_{\delta}(kr,kr,k;t)
\nonumber\\
 && \quad\quad
      -\frac{2}{3}\left\{D^{(3)}_{\delta}(kr,k,kr;t)+
                                   D^{(3)}_{\delta}(kr,kr,k;t)\right\}
\nonumber\\
 && \quad\quad
      +\left\{\frac{1+r^2}{2r^{2}}-\frac{(1-r^2)^2}{4r^3}
 \ln\left|\frac{1+r}{1-r}\right|\right\}E^{(3)}_{\delta}(kr,k,kr;t)
\nonumber\\
 && \quad\quad
      +\left\{\frac{-1+3r^2}{2r^{4}}+\frac{(1-r^2)^2}{4r^5}
 \ln\left|\frac{1+r}{1-r}\right|\right\}E^{(3)}_{\delta}(kr,kr,k;t)
\nonumber\\
 && \quad\quad
\left.+\frac{2}{3r^2}\left\{F^{(3)}_{\delta}(kr,k,kr;t)+
                                   F^{(3)}_{\delta}(kr,kr,k;t)\right\}
\right]. 
\ea
Note that $P^{(13)}_{\rm cb}$ is evaluated by 
one-dimensional integration once the linear power spectrum and the
growth function are given.
For a model without massive neutrinos,
the growth functions are approximated as
\be
 \frac{1}{\Dcb(k;t)\Dcb(kr;t)^{2}}
\left\{A_{\delta}^{(3)},B_{\delta}^{(3)},C_{\delta}^{(3)},
 D_{\delta}^{(3)},E_{\delta}^{(3)},F_{\delta}^{(3)}\right\}
\to \left\{
 \frac{5}{18},\frac{1}{18},-\frac{1}{6},-\frac{1}{9},
 -\frac{1}{21},-\frac{2}{63}
\right\}. 
\label{eq:3rdNLgrowth}
\ee
Therefore, Eq.~(\ref{eq:Pk13_1}) recovers the expression of $P^{{(13)}}$
for the CDM model case: 
\ba
  &&P^{(13)}_{\rm cb}~\to~P^{(13)}_{\rm cb}(k;t)
 =\frac{k^{3}}{252(2\pi)^{2}}P^{\rm L}_{\rm cb}(k;t)
 \int^{\infty}_{0}dr P^{\rm L}_{\rm cb}(k;t)
 \nonumber\\
 &&\quad\quad\quad\quad\quad\quad
   \times 
   \left[\frac{12}{r^{2}}-158+100r^{2}-42r^{4}
   +\frac{3}{r^{3}}(r^{2}-1)^{3}(7r^{2}+2)
   \ln\left|\frac{1+r}{1-r}\right|\right].
 \label{eq:P13approx}
\ea
Again, an exact computation of  $P^{(13)}_{\rm cb}$  
is computationally expensive. \par

Thus evaluations of the one-loop correction spectra, $P^{(12)}_{\rm cb}$
and $P^{(13)}_{\rm cb}$, at each $k$ and each time $t$ require
high-dimension integrations, which are numerically time-consuming.
Rather, we find that Eqs.~(\ref{eq:P22approx}) and
(\ref{eq:P13approx}) serve as a good approximations to obtain the spectra
for a MDM model, if the scale-dependent linear growth rate entering into
the linear power spectrum $P^{\rm L}_{\rm cb}$ is properly taken into
account. We below give the justification. \par

\begin{figure}[t]
\begin{center}
\includegraphics[width=0.325\textwidth]{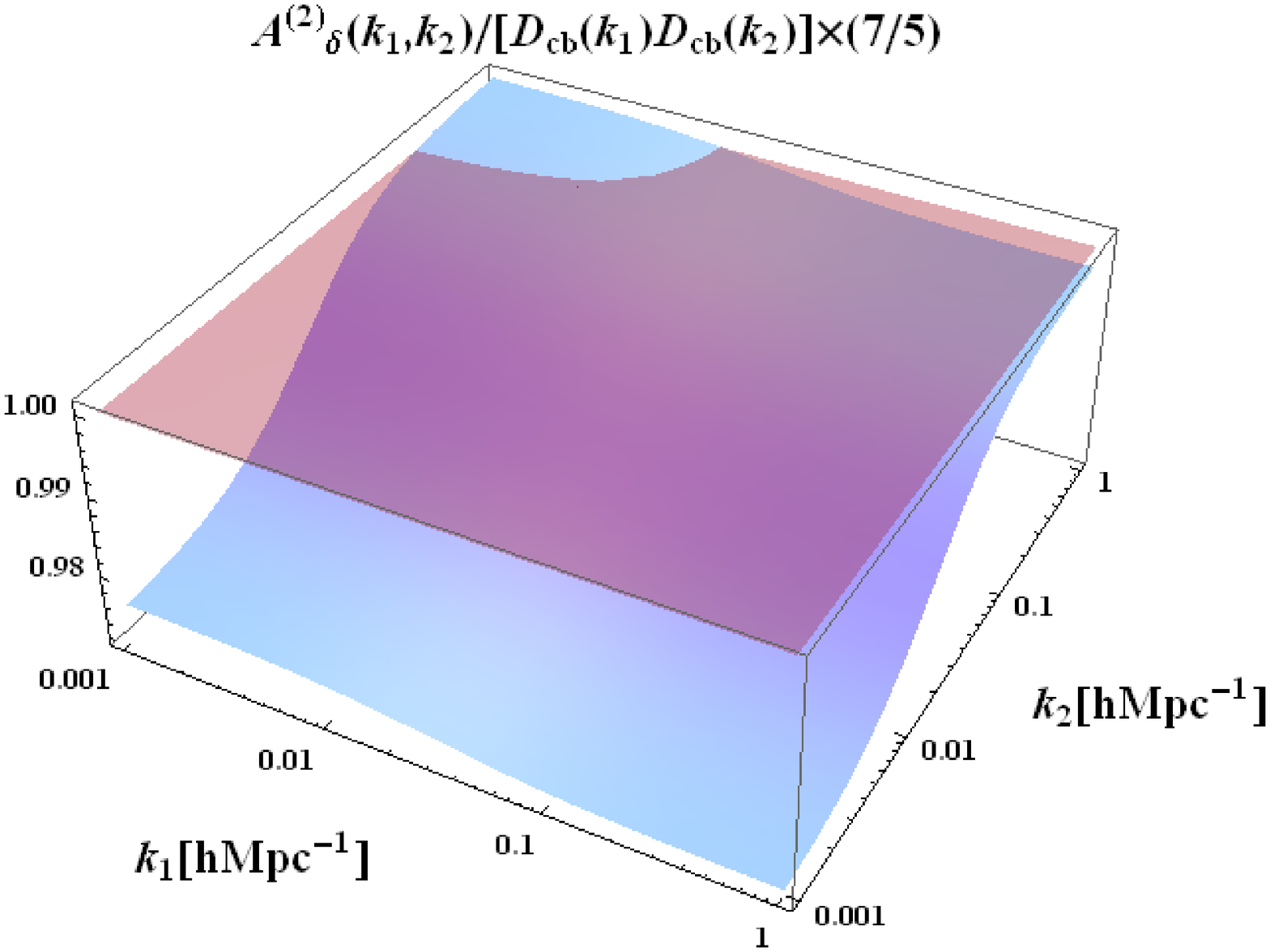}
\includegraphics[width=0.325\textwidth]{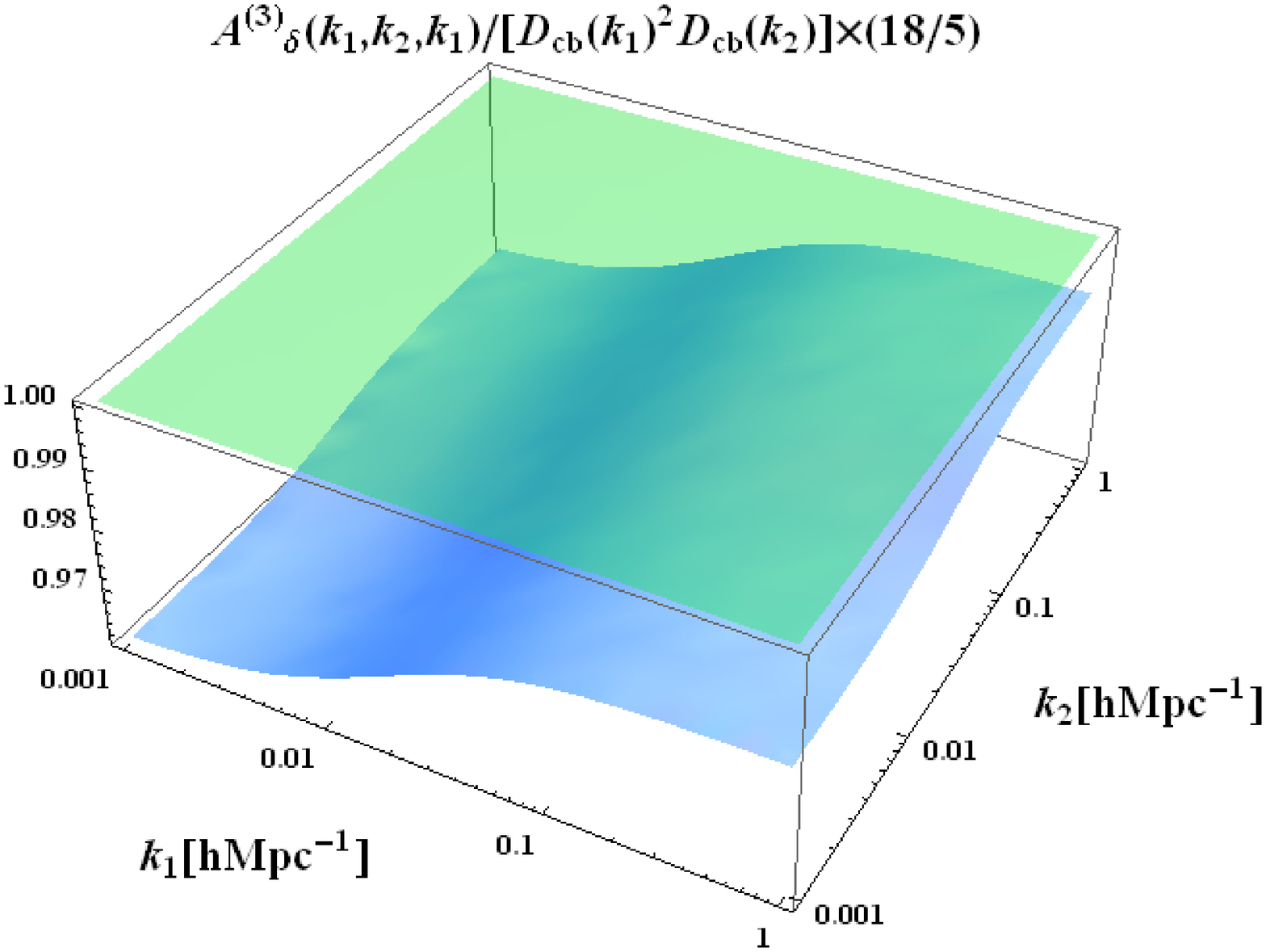}
\includegraphics[width=0.325\textwidth]{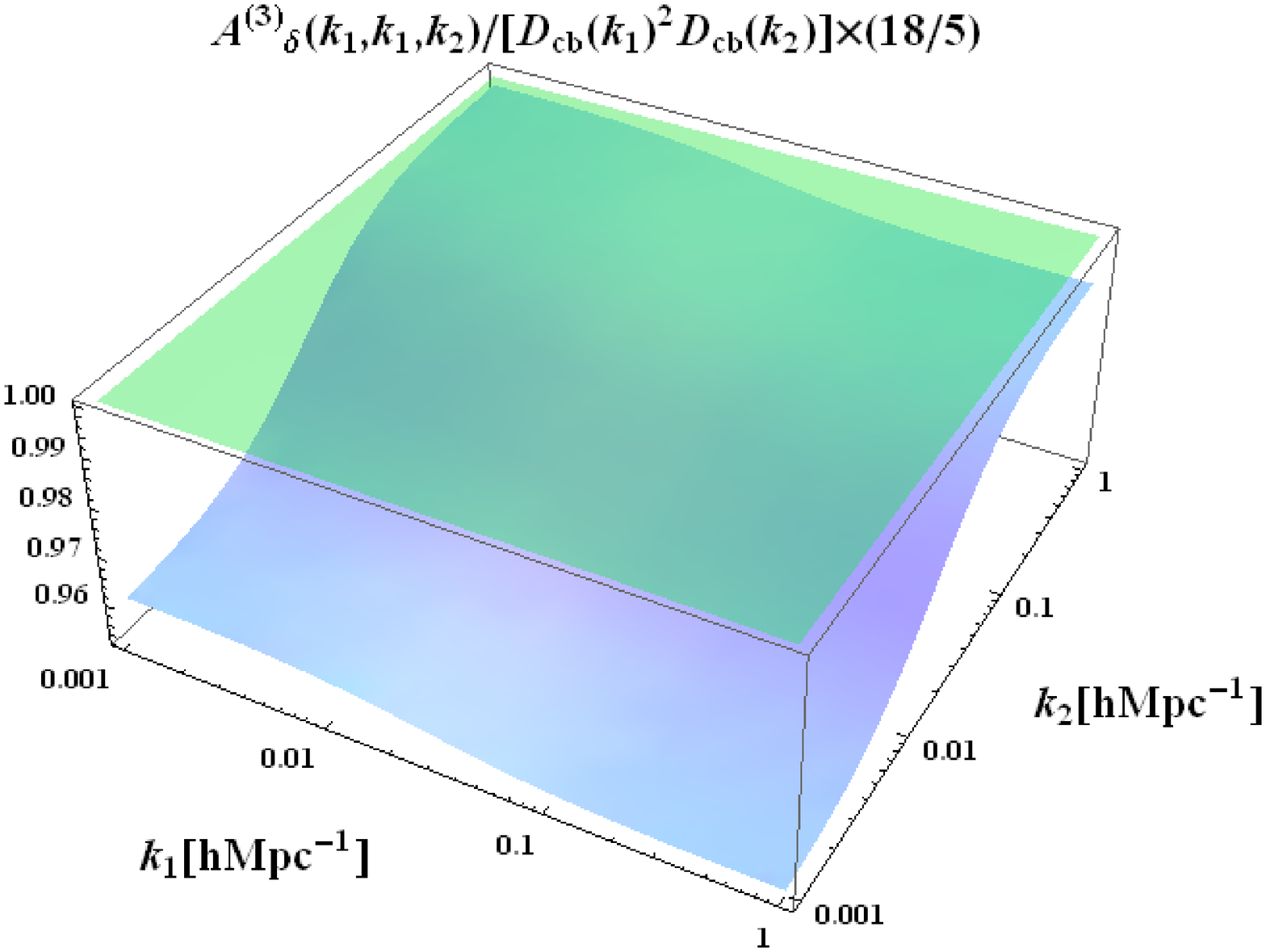}
\end{center}
\vspace*{-2em}
\caption{
 The 2nd- and 3rd-order growth functions
 at redshift $z=0$ are plotted as a function of two wavenumbers $k_1$ and
 $k_2$, for a MDM model with $f_\nu=0.05$ ($m_{\nu,{\rm tot}}\simeq 0.6~$eV).
As representative examples,
 shown here is the growth functions divided by some powers of the linear
 growth rate: 
 $A^{(2)}_\delta(k_1,k_2)/[D_{\rm cb}(k_1)D_{\rm cb}(k_2)(5/7)]$ 
 ({\em left panel}), 
 $A^{(3)}_{\delta}(k_1,k_2,k_1)/[D_{\rm cb}(k_1)^2D_{\rm cb}(k_2)(5/18)]$ 
 ({\em middle}) and 
 $A^{(3)}_{\delta}(k_1,k_1,k_2)/[D_{\rm cb}(k_1)^2D_{\rm cb}(k_2)(5/18)]$ 
 ({\em right}), respectively.  Note that specific combinations of 
 $k_i$-arguments in $A^{(3)}_\delta $ are chosen 
 because the one-loop power spectrum $P_{\rm cb}^{(13)}$ 
 (see Eq.~[\ref{eq:Pk13_1}]) depends on the growth functions
 of specific configurations. 
 The quantities shown become unity for 
 the limit $f_{\nu}=0$, i.e. a model without massive neutrinos 
 (see Eqs.~[\ref{eq:2ndNLgrowth}] and [\ref{eq:3rdNLgrowth}]), 
 which is shown by the plane in each plot.
 Therefore the deviations from unity reflect additional scale
 dependences arising from the mode-coupling. It is clear that
 scale-dependences of the higher-order growth functions are
 well-captured by combinations of the linear growth rate, and the
 approximations (\ref{eq:2ndNLgrowth}) and (\ref{eq:3rdNLgrowth}) hold
 valid with accuracy better than $\sim 5\%$ over a range of
 wavenumbers we have considered. 
} \label{fig:higher-order growth funtion}
\end{figure}

First, we study validity of the approximations~(\ref{eq:2ndNLgrowth})
and (\ref{eq:3rdNLgrowth}) for the nonlinear growth functions in a MDM
model.  Fig.~\ref{fig:higher-order growth funtion} compares the approximations
~(\ref{eq:2ndNLgrowth}) and (\ref{eq:3rdNLgrowth}) with the
results obtained by numerically solving the differential equations
(\ref{eq:eq_for_A_delta}), (\ref{eq:eq_for_B_delta}) and
(\ref{eq:eq_for_calI}) that govern time-evolution of the nonlinear
growth functions 
\footnote{To compute the results shown in
  Fig.~\ref{fig:higher-order growth funtion}, we used the fitting formula 
  of the linear growth function, $\Dcb(k;t)$, 
  developed in Eq.(12) of \cite{Hu:1997vi}, for
  computational simplicity. }.  
Here we consider the growth functions
$A_\delta^{(2)}(k_1,k_2)$ and $A_\delta^{(3)}(k_1,k_2,k_3)$ as
representative examples. 
Also note that we considered
$f_{\nu}=0.05$, corresponding to the current upper bound, 
and redshift $z=0$, where nonlinear clustering is strongly evolved. 
This figure clearly shows that the
fractional errors of the approximations are less than $\sim 5\% $ over
a wide range of wavenumbers we have considered. This agreement implies
that the scale-dependence of higher-order growth functions are
well-captured by the $k$-dependence of linear growth rate, 
$D_{\rm cb}(k; t)$. 
This level agreement was also found for other growth
functions, $B_\delta^{(3)}$,
$C_\delta^{(3)}$, $D_\delta^{(3)}$, $E_\delta^{(3)}$, 
and $F_\delta^{(3)}$.
\par

\begin{figure}[t]
\begin{center}
\includegraphics[width=0.45\textwidth]{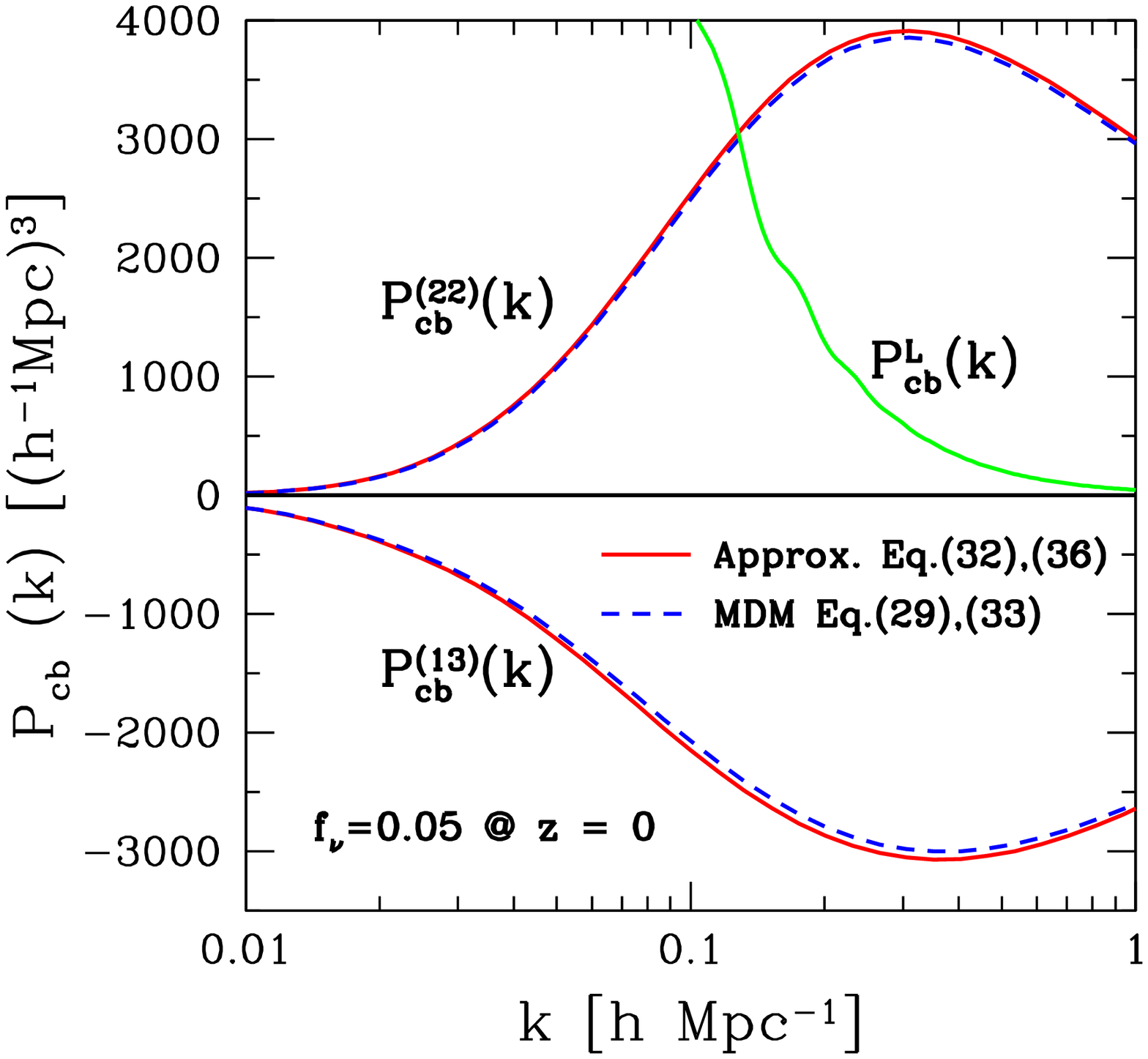}
\includegraphics[width=0.45\textwidth]{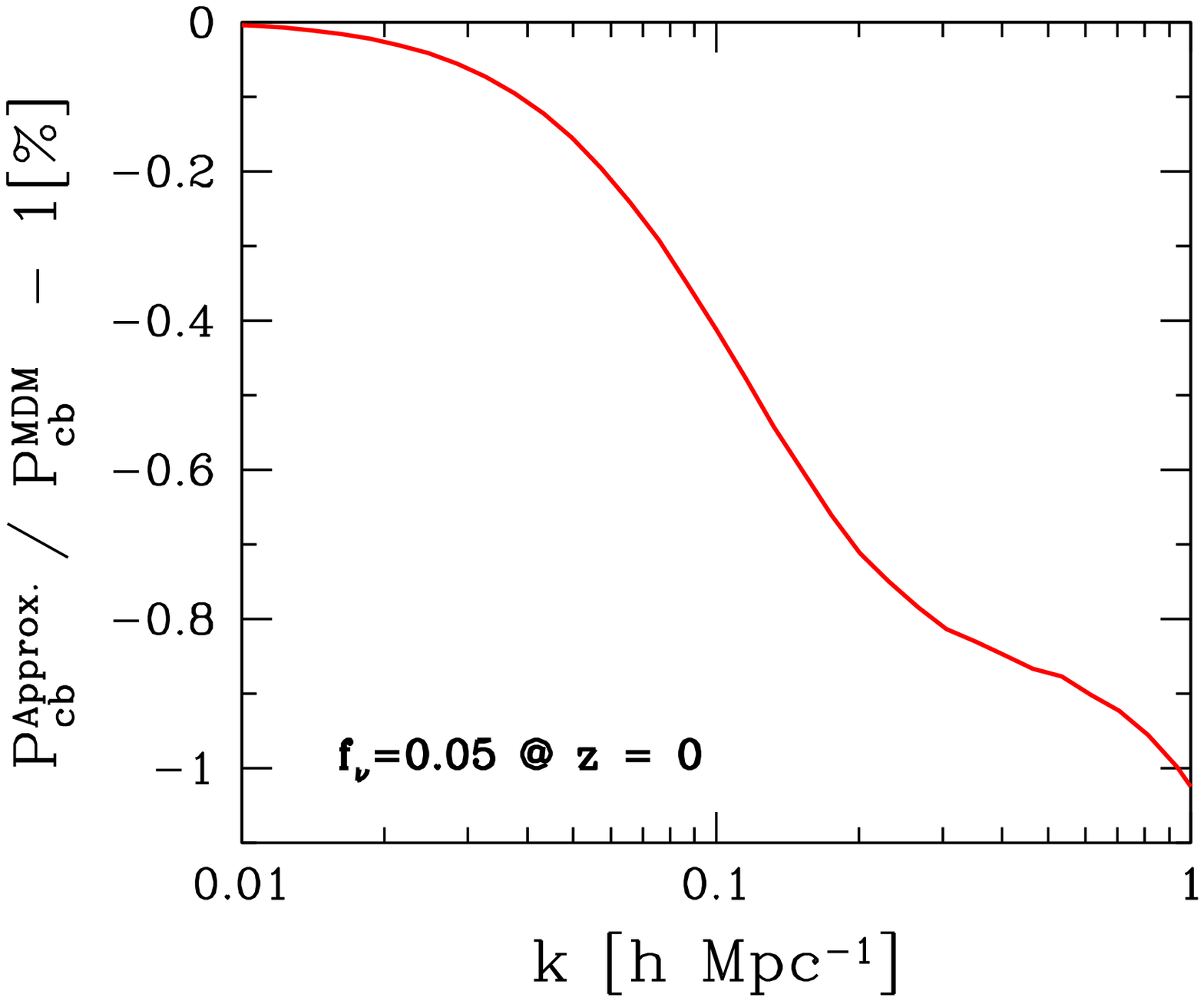}
\end{center}
\vspace*{-2em}
\caption{
 {\em Left panel}: The dashed curves show the one-loop power
 spectra of CDM plus baryon perturbations,
 $P^{(22)}_{\rm cb}$ and $P^{(13)}_{\rm cb}$, which are
 obtained by numerical integrations of Eqs.~(\ref{eq:Pk22_1}) and
 (\ref{eq:Pk13_1}), respectively, while the solid curves show the
 spectra computed using the approximations (\ref{eq:P22approx}) and
 (\ref{eq:P13approx}).
 Note that the $y$-axis is plotted in the linear scale,
 and we consider $f_\nu=0.05$ and $z=0$.  For comparison, the rightmost
 solid curve labelled as ``$P_{\rm cb}^{\rm L}$'' shows the linear power
 spectrum.
 {\em Right panel}: The fractional difference of the total matter power
 spectrum including up to the one-loop corrections is shown in the left panel:
 $P^{\rm NL}_{\rm cb}(k)=P^{\rm L}_{\rm cb}(k)+P^{(13)}_{\rm cb}(k)
 +P^{(22)}_{\rm cb}(k)$. The approximation is found to
 be accurate to better than 
 $ 1\%$ on scales 
 $k\ltsim 1h$Mpc$^{-1}$. 
}
\label{fig:diff between CDM & MDM}
\end{figure}

Fig.~\ref{fig:diff between CDM & MDM} compares 
the approximation with the full evaluation 
of one-loop power spectra, $P^{(22)}_{\rm cb}(k)$ and 
$P^{(13)}_{\rm cb}(k)$. In the left panel, 
the dashed curves show the result obtained by performing the
numerical integrations in Eqs.~(\ref{eq:Pk22_1}) and (\ref{eq:Pk13_1})
where the high-order growth functions are inserted into the
calculations of the Fourier kernels
$\calK^{(2)}_{\delta}$ and $\calK^{(3)}_{\delta}$, 
while  the solid curves are the results obtained by the 
approximations~(\ref{eq:P22approx}) and (\ref{eq:P13approx}). 
Note that we assumed $f_\nu=0.05$ as 
in Fig.~\ref{fig:higher-order growth funtion}. 
It is apparent that the absolute values of
$P^{(22)}_{\rm cb}(k)$ and $P^{(13)}_{\rm cb}(k)$ 
are slightly overestimated by the approximations, because the
higher-order growth functions are overestimated by the approximations
as implied in Fig.~\ref{fig:higher-order growth funtion}. 
The right panel shows the resulting
total power spectra of baryon plus CDM perturbations
that include up to the one-loop corrections: 
$P_{\rm cb}=P_{\rm cb}^{\rm L}+P_{\rm cb}^{(22)}+P_{\rm cb}^{(13)}$. 
It is found that the fractional error of the approximation 
is smaller than $\sim 1\%$ on scales up to $k=1h$Mpc$^{-1}$, 
for the case of $f_{\nu}=0.05$.
Thus we can conclude that the approximations to compute the one-loop
power spectra are sufficiently accurate for our purpose, over ranges of
wavenumbers, redshifts and neutrino mass scales we are interested in. 

Given the results shown in Figs.~\ref{fig:higher-order growth funtion}
and \ref{fig:diff between CDM & MDM}, we will hereafter employ the
approximations (\ref{eq:P22approx}) and (\ref{eq:P13approx})
 for computing the nonlinear
power spectrum of total matter at an arbitrary time $t$.
A brief summary is: we first compute the linear power spectra of
each components at time $t$, $P^{\rm L}_{\rm cb}(k;t)$, 
$P_{\rm \nu}^{\rm L}(k;t)$, and $P^{\rm L}_{\rm cb\nu}(k; t)$, 
for a desired MDM model, by using the publicly available 
code CAMB \cite{Lewis:1999bs}.
We then compute one-loop power spectra, $P^{(22)}_{\rm cb}(k; t)$ 
and $P^{(13)}_{\rm cb}(k; t)$, using 
Eqs.~(\ref{eq:P22approx}) and (\ref{eq:P13approx}). Then, all the
spectra are summed up to obtain the nonlinear spectrum of total matter: 
\be
 P^{\rm NL}_{\rm m}(k;z)=
f_{\rm cb}^{2}[P^{L}_{\rm cb}(k;z)+P^{(22)}_{\rm cb}(k;z)
               +P^{(13)}_{\rm cb}(k;z)]
+2f_{\rm cb}f_{\nu}P^{L}_{\rm cb\nu}(k;z)
+f_{\nu}^{2}P^{L}_{\nu}(k;z). \label{eq:Nonlinear_matter_Pk}
\ee

Before closing this sub-section, we comment on the work of
\cite{Wong:2008ws}, where a similar method for computing the one-loop
corrected power spectra for a MDM model was developed based on
perturbation theory ignoring the  nonlinear neutrino
perturbations. Although our method is qualitatively equivalent to their
method, there are several technical differences that may be worth
stressing. \cite{Wong:2008ws} first employed the analytical fitting
formula for scale-dependent linear growth function (also for the
transfer function) developed in \cite{Eisenstein:1997jh}, which is given
as a function of neutrino masses and cosmological parameters. Then,
analytical expressions for the higher-order growth functions and the
one-loop power spectra were derived. There are several inaccuracies
involved in the fitting formula. The formula becomes less accurate for
small neutrino masses as explicitly pointed out in
\cite{Kiakotou:2007pz}. 
More precisely for a case that neutrino(s) is massive enough such that the
neutrinos become non-relativistic in the radiation dominated era,
corresponding to $m_\nu\gtsim 0.6~{\rm eV}$ ($f_\nu\gtsim 0.05$) for a
$\Lambda$CDM model, the fitting formula becomes inaccurate
because it assumes a continuous
suppression in the matter growth since neutrinos became the
non-relativistic, although the suppression occurs only in the matter
dominated regime. In addition the fitting formula does not include 
BAO features in the transfer function. Therefore,
the use of the fitting formula may under-estimate the ability of future
galaxy surveys for constraining cosmological parameters, especially dark
energy parameters to which 
the observed scales of BAO peaks are
sensitive. In contrast, in our method, the nonlinear power spectrum is
obtained by inserting the linear power spectrum outputs of the numerical
Boltzmann solver such as CAMB, which takes into account the
scale-dependent growth rate as well as BAO features at high precision.
Albeit these small differences, \cite{Wong:2008ws} also verified that 
Eqs.~(\ref{eq:P22approx}) and (\ref{eq:P13approx}) are good
approximation. 

\section{Nonlinear power spectrum in a MDM model} 
\label{sec:neutrino _suppression}

In this section, 
based on the treatment developed in the previous section, 
we study effects of finite mass neutrinos on the nonlinear power 
spectrum of total matter. 

\begin{figure}[t]
\begin{center}
\includegraphics[width=0.5\textwidth]{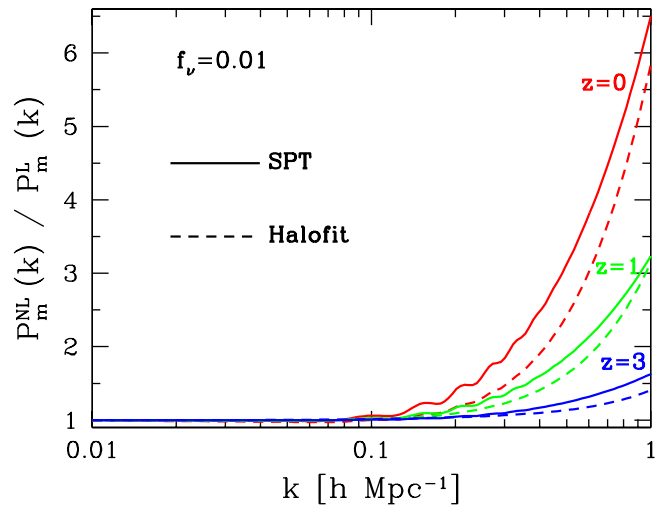}
\end{center}
\vspace*{-2em}
\caption{
 The nonlinear power spectra for a MDM model with $f_\nu=0.01$. 
 The solid curves show the SPT predictions 
divided by the linear spectra for three redshifts $z=0, 1$ and
 $3$, while the dashed curves denote the halofit results.  
}
\label{fig:SPT vs HALOFIT}
\end{figure}

In Fig.~\ref{fig:SPT vs HALOFIT} we show the SPT predictions for nonlinear
power spectra divided by the linear power spectra at three different
redshifts $z=0$, 1 and $3$, respectively, for a MDM model with 
$f_\nu=0.01$ ($m_{\nu,{\rm tot}}\simeq 0.12~$eV). 
Nonlinear gravitational clustering causes
amplitudes of the nonlinear total matter power spectrum to be enhanced,
resulting in more significant deviations from the linear
theory predictions on smaller scales and at lower redshifts.
In other words our PT model tells the range of wavenumbers and redshifts
where the linear theory is valid or equivalently the linear theory
starts to break down on wavenumbers beyond
the applicable range.
Our model predictions are also compared with the result of an empirical
method, which is the halo model approach (hereafter we call
`halofit'). In this model the nonlinear power spectrum is obtained by
mapping the input linear power spectrum based on the fitting formula
that is calibrated by numerical simulations for CDM models 
\cite{Smith:2002dz}. Recent
studies \cite{Ichiki:2008ye} employed the halofit method
to compare the model predictions to the weak lensing measurements for a
MDM model, and then derived an upper limit on total neutrino mass 
as $m_{\nu,{ \rm tot}}\ltsim 0.54~{\rm eV}$ (95\%C.L.).
The halofit power spectra are smaller in amplitudes 
than SPT by up to $\sim 10\%$
over a range of scales we consider. 
Furthermore,  SPT more washes out oscillatory BAO
features than halofit, as pointed out in the previous study
\cite{Jeong:2006xd}, where the SPT results were
shown to better
reproduce the simulations results than the halofit results. 

It should also be noted that SPT eventually ceases to be accurate at
smaller scales, and the validity needs to be carefully studied by 
using numerical simulations (e.g., \cite{Nishimichi:2008ry} for such a study
for a CDM model). Our method may be further improved by
including the higher-order perturbation contributions or using a refined
method such as the renormalized perturbation theory 
(e.g. \cite{Crocce:2007dt}) or the closure 
theory method \cite{Taruya:2007xy}. 
These are in progress and will be presented elsewhere. 

\begin{figure}[t]
\begin{center}
\includegraphics[width=0.45\textwidth]{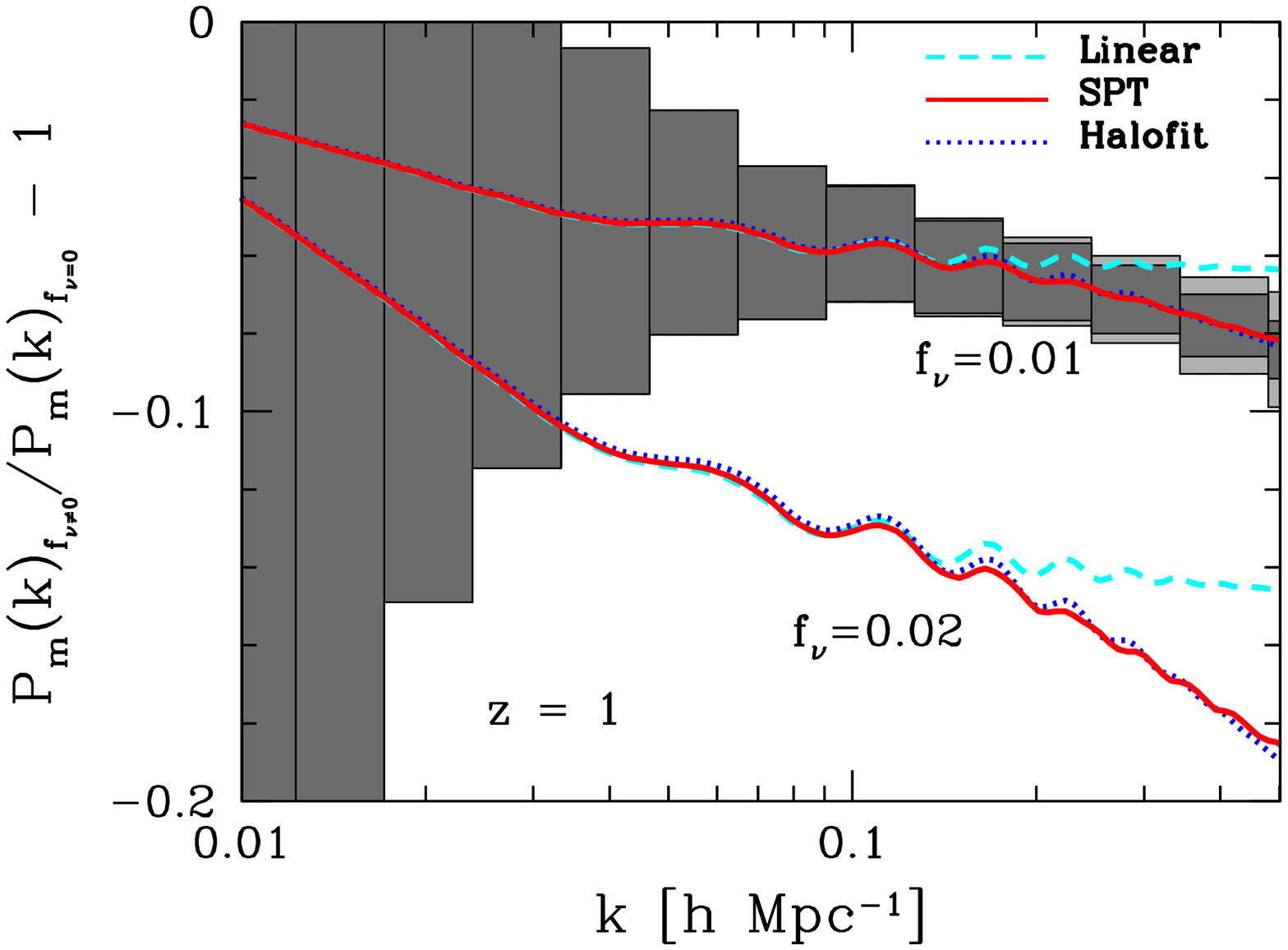}
\includegraphics[width=0.45\textwidth]{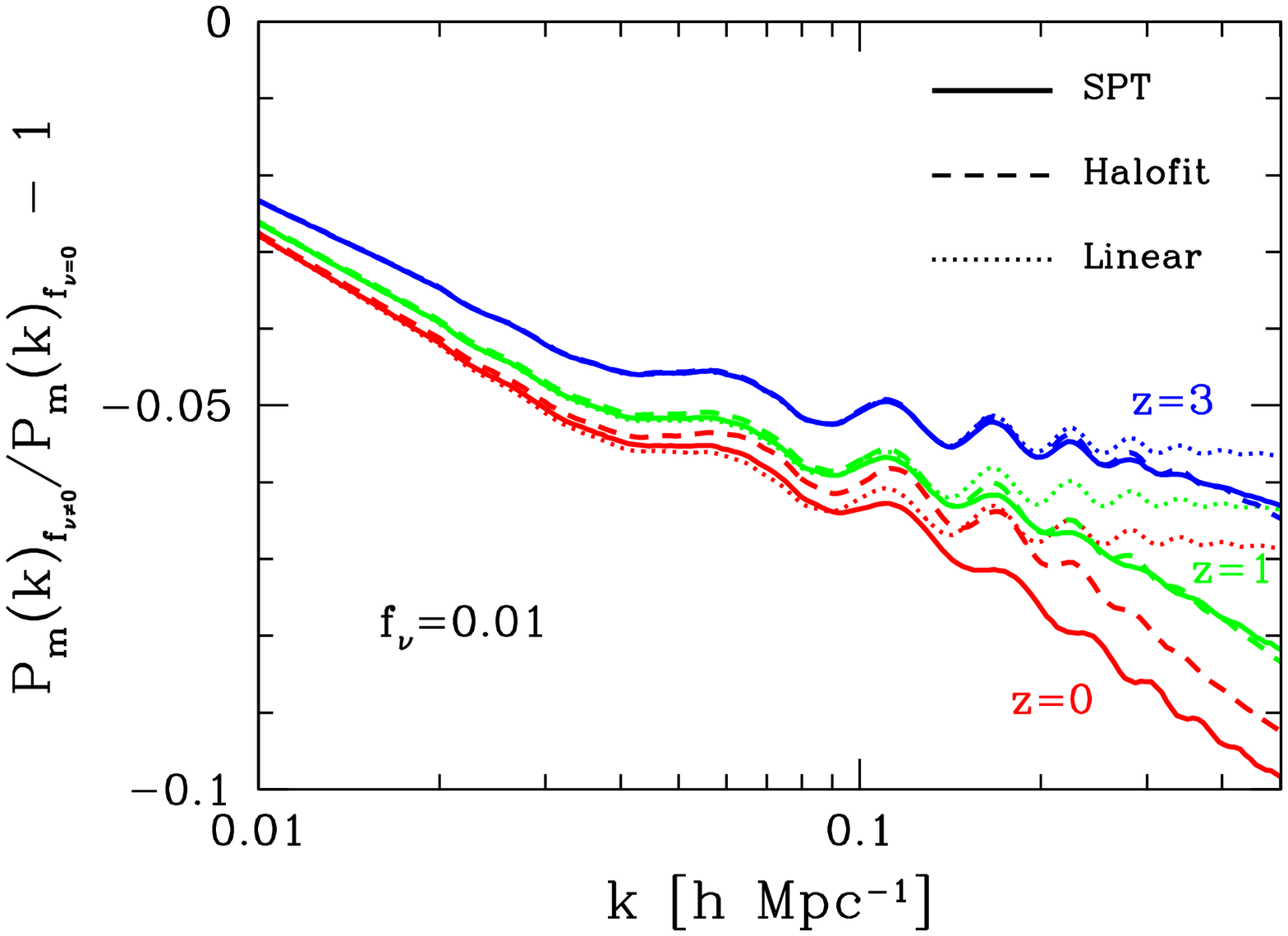}
\end{center}
\vspace*{-2em}
\caption{ {\it Left panel}: The fractional difference between mass
 power spectra with and without massive neutrino contribution.  The
 shaded boxes show the expected $1$-$\sigma$ errors on the power
 spectrum measurement for a Stage-III type survey of $z\sim 1$ slice
 that is characterized by the mean number density of galaxies and survey
 volume, $\bar{n}_{\rm g}=5\times 10^{-4}~h^3{\rm Mpc}^{-3}$ and $V_{\rm
 survey}=1.5~h^{-3}{\rm Gpc}^{3}$ (also see
 Table~\ref{table:survey}). The two models of neutrino mass,
 $f_\nu=0.01$ and 0.03 ($m_{\nu, {\rm tot}}\simeq 0.12$ and $0.36~{\rm
 eV}$, respectively) are assumed, where other cosmological
 parameters are kept fixed.
  {\it Right panel}: It is shown how the neutrino suppression
 feature in the power spectrum amplitude varies with redshifts,
 comparing the results for the SPT, linear theory and halofit.
} \label{fig:matter suppression}
\end{figure}

In the left panel of Fig.~\ref{fig:matter suppression} we compare 
the two power spectra with and without massive neutrinos, 
$P_{f_\nu\neq 0}/P_{f_\nu=0}$, 
for a fixed $\Omega_{\rm m0}$.
Note that
we show the results for $f_{\nu}=0.01~ (m_{\nu,{\rm tot}}=0.12{\rm eV})$
and $f_{\nu}=0.02~ (m_{\nu,{\rm tot}}=0.24{\rm eV})$ to study
dependences of the neutrino effect on total neutrino mass, and consider
redshift $z=1$, the central target redshift of WFMOS-like survey.  
As can be clearly seen, the massive neutrinos imprint characteristic,
scale-dependent suppression features onto the power spectrum
shape. Comparing the linear theory and SPT results manifests that the
nonlinear power spectrum has increasing suppression on scales $k\gtsim
0.1h$Mpc$^{-1}$, where the linear theory predicts a constant suppression
roughly given as $P_{f_{\nu}\neq 0}/P_{f_{\nu}= 0} \sim -8f_{\nu}$
\cite{Hu:1997mj}.
This enhanced suppression effect can be understood as follows. 
As can be found from Eq.~(\ref{eq:P22approx}) and 
Eq.~(\ref{eq:P13approx}), the one-loop power spectra, 
$P^{(22)}_{\rm cb}$ and $P^{(13)}_{\rm cb}$, which give nonlinear
corrections to the total matter,  are roughly 
proportional to squares of the linear power spectrum, $P^{\rm L}_{\rm cb}$, 
and therefore the suppression effect on the growth rate is enhanced
in the weakly nonlinear regime, compared to the model without massive
neutrinos. \par

The left panel also shows the halofit results. Note that, for this case,
the numerator and denominator of $P_{f_\nu\neq 0}/P_{f_\nu=0}$ are both
computed by the halofit. Unexpectedly the halofit results fairly well
reproduce the suppression features given by SPT, although the power
spectra themselves show a moderate difference 
in these two models as implied in
Fig.~\ref{fig:SPT vs HALOFIT}.  
The right panel of Fig.~\ref{fig:matter suppression} shows the results
for redshifts $z=0$, 1 and $3$.  
The apparent 
agreement between halofit and SPT can be seen only for redshifts $z=1$
and $3$, and the difference appears clear for 
for the $z=0$ results. 
Recent studies of $N$-body simulation in a MDM model also show a similar 
behaviour of the enhanced neutrino suppression 
\cite{Brandbyge:2008rv,Brandbyge:2008js}. 
A quantitative comparison among SPT, halofit and $N$-body results 
will be reported elsewhere. \par

Is the neutrino effect on total matter power spectrum measurable for a
future galaxy survey? To obtain an insight on this question, the shaded
boxes around the SPT curve with $f_\nu=0.01$ display 
expected $1$-$\sigma$ uncertainties in measuring band powers of the
power spectrum at each wave number bins, assuming survey parameters of
WFMOS-like low-$z$ survey (see Table~\ref{table:survey} for the details). 
To be more explicit, the fractional errors of measuring the power
spectrum, $P_{\rm m}(k)$, averaged over a spherical shell of each radial bin $k$
with bin width $\Delta k$
are, in an ideal case, given as
\be
 \left[\frac{\sigma_{P}}{P_{\rm m}(k)}\right]^{2}=\frac{4\pi^{2}}
{V_{\rm s}k^{2}\Delta k}
 \left[\frac{1+\bar{n}_{g}P_{\rm m}(k)}{\bar{n}_{g}P_{\rm m}(k)}\right]^{2}, 
\ee
where  $V_{\rm s}$ and $\bar{n}_{\rm g}$ are the comoving survey volume 
and number density of target galaxies.
Note that, for the measurement errors above, we assumed the
Gaussian errors for simplicity,  
and ignored the non-Gaussian contributions (see
\cite{Takahashi:2009bq} for the detailed study). 
The neutrino suppression appears to be greater than the errors at 
$k\gtsim 0.06h$Mpc$^{-1}$. Another intriguing consequence of 
the nonlinear clustering is that the amplified power of $P^{\rm NL}_{\rm m}(k)$ 
reduces the relative importance of the shot noise contamination.
Note that in reality matter power spectrum should be replaced with galaxy one 
and we address this issue when forecasting constraints on neutrino masses. 
Thus, extending available range of wavenumber, 
the constraint on neutrino masses can be improved.\par

Finally, it would be worth noting that wiggles in the curves reflect shifts 
in the BAO peak locations caused by the scale-dependent suppression 
effect of neutrinos. 
The amount of the modulations, however,  is smaller than the measurement 
errors. Hence the uncertainty in neutrino mass is unlikely to largely 
degrade the power of BAO experiments, at least for an expected small 
$f_{\nu}$. 

\section{Nonlinear Galaxy Bias}
\label{subsec:Nonlinear galaxy biasing}

To model galaxy clustering relevant for actual galaxy surveys we need to
further include a galaxy bias effect. According to
\cite{McDonald:2006mx} (also see Appendix~\ref{sec:nonlinear-bias} for
the details), we below describe the modeling of nonlinear galaxy bias
in a MDM model, which is done in
a self-consistent manner with the modeling of nonlinear matter
clustering presented up to the preceding section.\par

We assume the local bias: the galaxy distribution at a given spatial
position is locally related to the underlying matter distribution at the
same position, which would be a good approximation at least on large
length scales of interest. 
In this modeling the galaxy density fluctuation field, 
$\delta_{\rm g}(\bfx)$, is given in terms of the matter field 
$\delta_{\rm m}(\bfx)$ in a Taylor expansion as \cite{Fry:1993pt} 
\be
 \deltag(\bfx)=\epsilon+c_{1}\deltam(\bfx)+
 \frac{1}{2}c_{2}\deltam^{2}(\bfx)+
 \frac{1}{6}c_{3}\deltam^{3}(\bfx)+\dots, 
\label{eqn:galaxy_bias}
\ee
where $c_{n}$ are the $n$-th order bias parameters, 
and $\epsilon$ represents 
the stochasticity of galaxy bias which is a statistical noise
originating from the fact that the relation between $\deltag$ and $\deltam$ 
is not perfectly deterministic. 
Here the stochasticity is assumed to be 
white noise and be uncorrelated with the density fluctuations, 
$\langle\epsilon\deltam\rangle=0$. Note that the galaxy bias parameters 
and the stochasticity depend on galaxy type, and vary with time 
\cite{Fry:1996fg,Tegmark:1998wm,Hui:2007zh}.\par

Eq.~(\ref{eqn:galaxy_bias}) of galaxy bias relation has an analogous form
to the perturbative expansion. Therefore we can include the bias
contribution up to the one-loop corrections in terms of the matter
density fluctuations.
According to the methods developed
in \cite{McDonald:2006mx,Jeong:2008rj}, the galaxy power spectrum
including the one-loop corrections can be computed as 
\ba
 P_{\rm g}(k) &= &b_{1}^{2}\left[
 P^{\rm NL}_{\rm m}(k)+b_{2}P_{\rm b2,\delta}(k)
 +b_{2}^{2}P_{\rm b22}(k)
 \right]+N, 
\label{eq:Nonlinear_galaxy_Pk}
\ea
where the functions $P_{\rm bs,\delta}$ and $P_{\rm b22}$ are defined as
\ba
 P_{\rm b2,\delta}(k)&\equiv&2\int\frac{d^{3}\bfq}{(2\pi)^{3}}~
   P^{\rm L}_{\rm m}(q)P^{\rm L}_{\rm m}(|\bfk-\bfq|)
   \calF^{(2)}_{\delta}(\bfq,\bfk-\bfq), \nonumber \\
 P_{\rm b22}(k)&\equiv&\frac{1}{2}\int\frac{d^{3}\bfq}{(2\pi)^{3}}~
   P^{\rm L}_{\rm m}(q)
   [P^{\rm L}_{\rm m}(|\bfk-\bfq|)-P^{\rm L}_{\rm m}(q)].
\label{eqn:Pb2}
 \ea
The detailed derivation is shown in Appendix~\ref{sec:nonlinear-bias}.
In Eq.~(\ref{eq:Nonlinear_galaxy_Pk}) $P^{\rm NL}_{\rm m}(k)$ is the
nonlinear matter power spectrum given by
Eq.~(\ref{eq:Nonlinear_matter_Pk}), and the kernel $F_\delta$ used in
Eq.~(\ref{eqn:Pb2}) is given by Eq.~(\ref{eq:C3}).
Note that $P_{\rm b2,\delta}>0$ and $P_{\rm b22}<0$ at scales of interest, 
$0.01~h{\rm Mpc}^{-1}\ltsim k \ltsim 0.2~h{\rm Mpc}^{-1}$. 
Eq.~(\ref{eq:Nonlinear_galaxy_Pk}) shows that the nonlinear
galaxy power spectrum is modeled by the three parameters $b_1$, $b_2$
and $N$ once the matter power spectra are specified for a given
cosmological model, where the parameters $b_1, b_2$ and $N$ are
redefined from the original parameters in Eq.~(\ref{eqn:galaxy_bias})
and the linear mass power spectrum as shown in
Appendix~\ref{sec:nonlinear-bias} (also see \cite{McDonald:2006mx} for
the detailed derivation).

For the limit of very small $k$, Eq.~(\ref{eq:Nonlinear_galaxy_Pk})
recovers the linear regime result, but with correction term: 
\be
 P_{\rm g}(k) \to b_{1}^{2}
 P^{\rm L}_{\rm m}(k)+N. \label{eq:linearlimit_galaxy_Pk}
\ee
Thus $b_1$ acts as an effective {\em linear} 
bias parameter for the power spectrum and
$N$ adds a shot noise contamination arising from stochastic bias and
nonlinear clustering (also see
\cite{Heavens:1998es,Seljak:2000gq,Smith:2006ne}). 
The terms that depend on $b_2$ are proportional to the one-loop
corrected mass power spectrum give an effect of scale-dependent
bias due to the nonlinear clustering. These parameters $b_1, b_2$ and
$N$ change with galaxy type we are working on, so need to be treated as
free parameters for each galaxy type. In fact, as carefully studied in
\cite{Jeong:2008rj}, the galaxy power spectrum
(\ref{eq:Nonlinear_galaxy_Pk}) can fairly well reproduce the
semi-analytic simulation results in the weakly nonlinear regime, if the
parameters are properly chosen so as to match the simulation results. 

\begin{figure}[t]
\begin{center}
\includegraphics[width=0.45\textwidth]{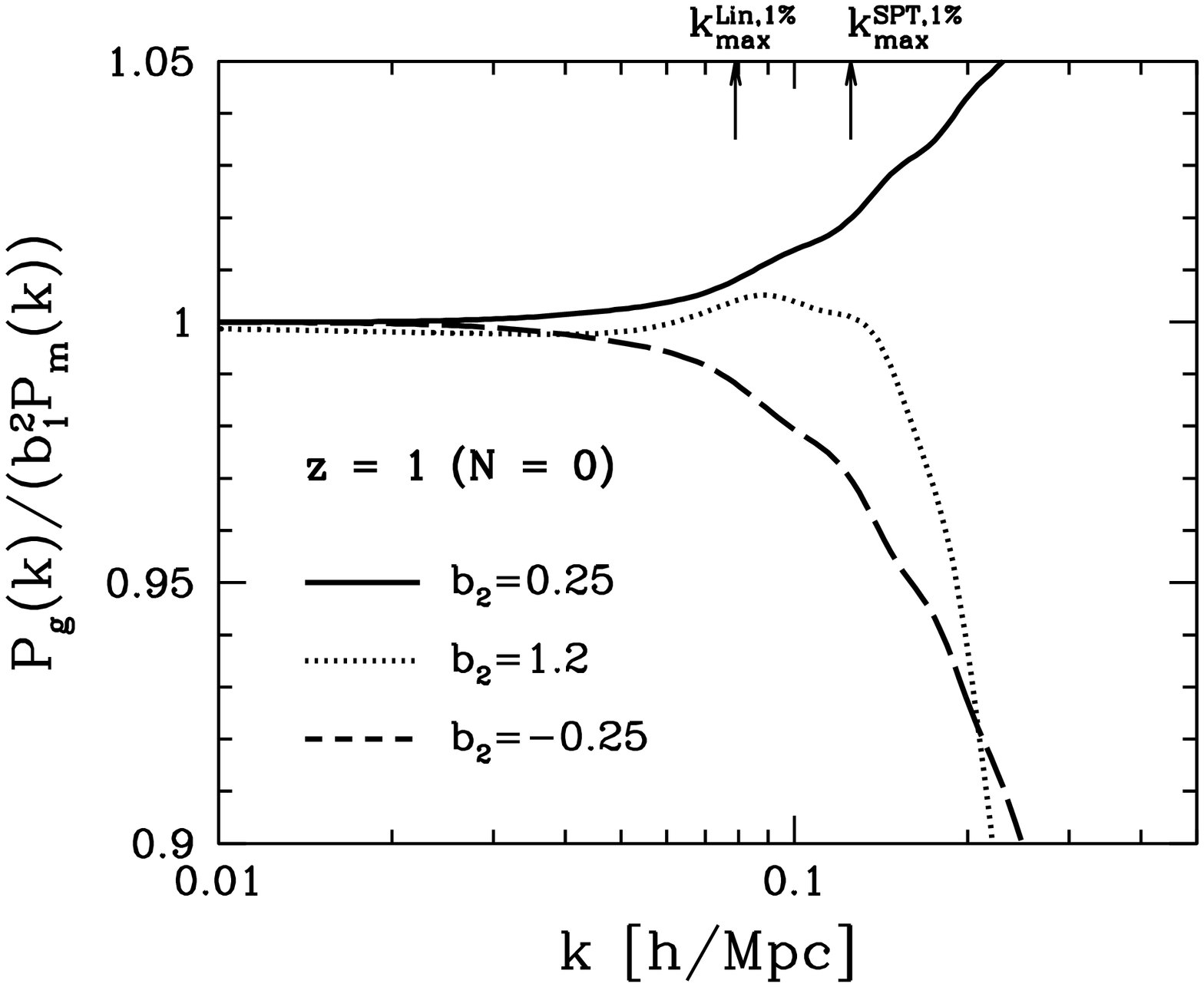}\\
\includegraphics[width=0.45\textwidth]{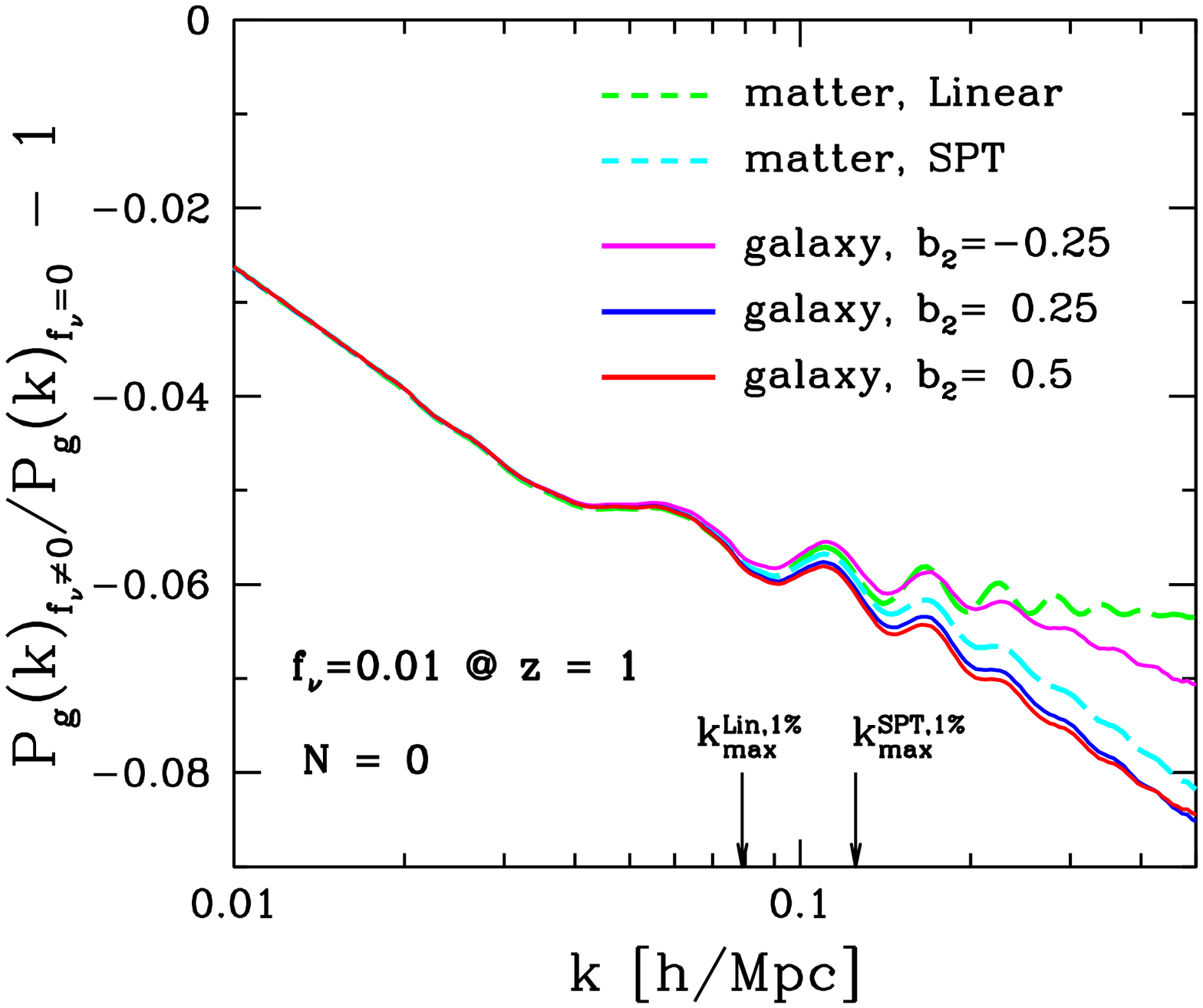}\\
\includegraphics[width=0.45\textwidth]{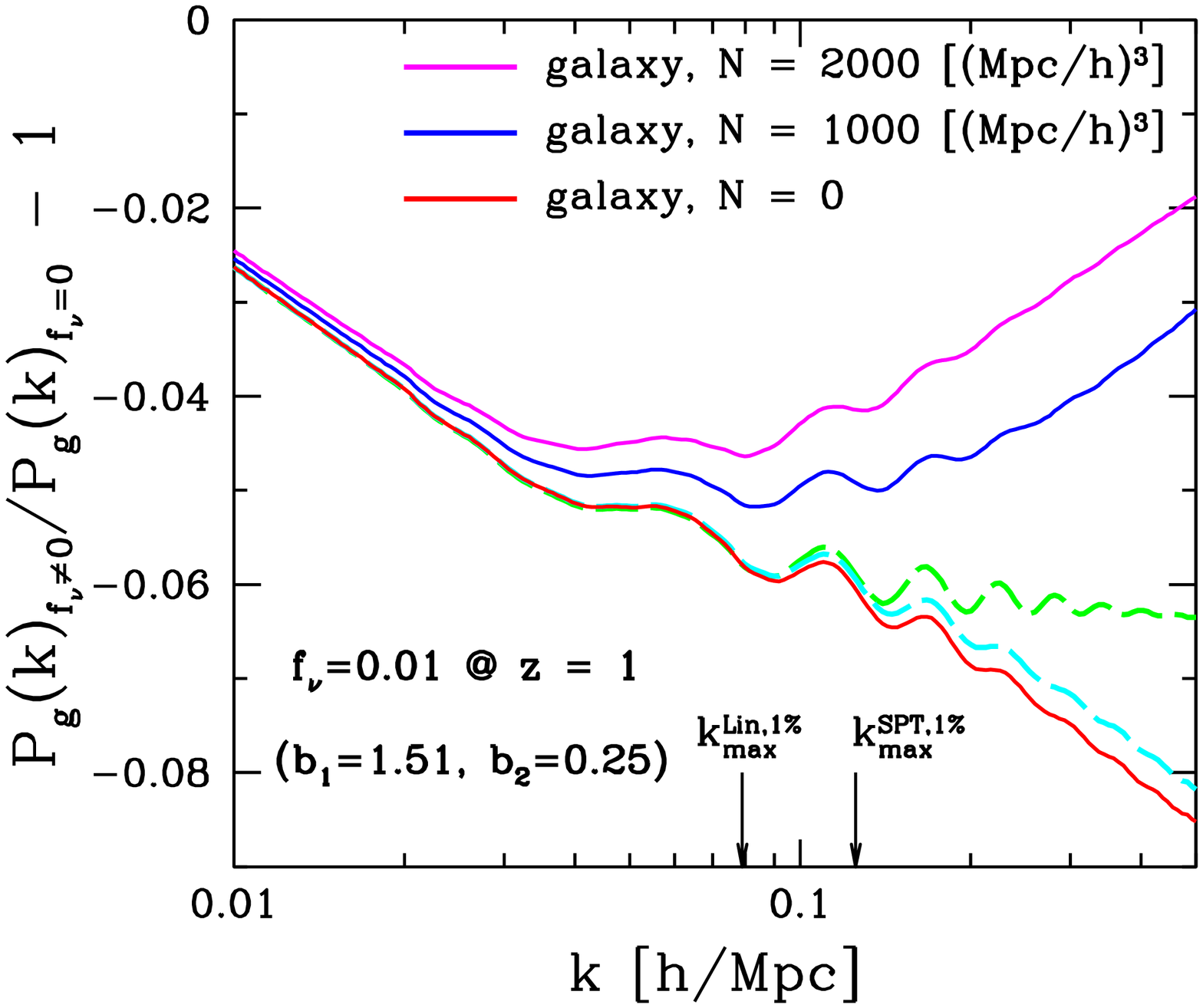}
\end{center}
\vspace*{-2em}
\caption{ {\it Top panel}: The perturbation theory predictions for
 nonlinear galaxy power spectrum at redshift $z=1$, which are computed from
 Eq.~(\ref{eq:Nonlinear_galaxy_Pk}) assuming the three fiducial values
 of nonlinear bias parameter, $b_2=-0.25, 0.25$ and $1.2$,
 respectively. The results are divided by the nonlinear mass power
 spectrum multiplied by the same linear bias parameter $b_1^2$ such that the
 deviation from unity represents the nonlinear, scale-dependent bias
 effect. The positive and negative $b_2$ values, with $|b_2|<1$,
 cause enhanced and suppressed power spectrum amplitudes on smaller
 scales compared to
 the linearly biased  power spectrum. The model with $b_2>1$ causes
 a complex scale-dependent bias (also see text for the details). 
 The valid range of linear theory and SPT are indicated by the two arrows
 in the upper horizontal axis (see text for the definition). 
{\it Middle panel}: The neutrino suppression features for the
 nonlinear galaxy power spectra for different fiducial values of $b_2$. 
For comparison the two dashed curves are the results for mass power
 spectrum computed from the SPT and linear theory as in
 Fig.~\ref{fig:matter suppression}.
{\it Bottom panel}: The effect of residual shot noise contamination
that arises from nonlinear clustering, which is modeled as
 $P_{\rm g}\rightarrow P_{\rm g}+N$. The three solid curves show the
 results for  $N=0,1000$ and 2000, respectively, where 
 $b_{1}=1.51$ and
 $b_{2}=0.25$ are kept fixed.  } 
\label{fig:galaxy suppression}
\end{figure}

The top panel of Fig.~\ref{fig:galaxy suppression} explicitly shows 
how a scale-dependent bias in the galaxy power spectrum is modeled 
by $b_2$. Note that we consider $z=1$ and $N=0$, 
and the spectra plotted are divided by
$b_1^2P_{\rm m}(k)$ such that deviation from unity represents 
the effect of scale-dependent bias.
Here we consider $b_2=\pm 0.25$ and 1.2 as a working example. First of
all, it is worth noting that 
the nonlinear galaxy bias of $b_2\sim 0.1-1$ causes a modification in the galaxy
power spectrum shape at BAO scales, and the effect may need to be 
taken into account for BAO surveys. For a case of $b_2<1$,
as can be seen from the results of $b_2=\pm 0.25$, a positive (negative)
$b_2$ enhances (suppresses) the power spectrum amplitudes increasingly
at larger $k$, relative to the linear bias case.  These features are
from the second term in the bracket on the r.h.s of
Eq.~(\ref{eq:Nonlinear_galaxy_Pk}) because $P_{\rm b2,\delta}>0$. 
On the other hand, when $b_2>1$, the nonlinear bias causes a complicated
modification in the spectrum shape, because the third term in
Eq.~(\ref{eq:Nonlinear_galaxy_Pk}) becomes dominant over the second term
at larger $k$. Note that the third term is always negative, so always
suppresses the power spectrum amplitudes.  

The results in the top panel imply that, even if a linear bias parameter
is well determined, the scale-dependent bias may cause a degeneracy with
the effect of finite neutrino masses, thereby degrading the ability of
future surveys for constraining neutrino masses. In particular a negative
$b_2$ causes a suppression in the power spectrum amplitudes, similarly
to the neutrino effect, so this case may cause a stronger degeneracy.
To obtain an insight on this, the middle panel of 
Fig.~\ref{fig:galaxy suppression} studies the neutrino suppression 
effect on galaxy power spectrum in the
presence of nonlinear bias. Shown here is the fractional difference of
galaxy spectra with and without neutrinos of $f_\nu=0.01$, for three
cases of $b_2$.  For comparison, the dashed curves show the results for
``{matter}'' power spectra employing linear theory and SPT.  While the
neutrino suppression effect is preserved, the nonlinear bias alters the
features in the weakly nonlinear regime. 
This figure shows that a negative (positive) $b_2$ weakens (strengthens)
the suppression effect. 

The bottom panel shows the dependence on the residual shot noise
contamination, given by the term including the parameter $N$ in
Eq.~(\ref{eq:Nonlinear_galaxy_Pk}). The shot noise term has no
wavenumber dependence for the power spectrum measurement, but the figure
implies that the shot noise residual with $N=O(10^3)~({{\rm Mpc}/h})^3$
may significantly alter the power spectrum shape over a wide range of
wavelengths where the neutrino suppression effect appears. This residual
shot noise effect arising from nonlinear clustering can be studied by
using semi-analytic $N$-body simulations where 
 galaxies are populated with
halos, so may be not so a serious source of systematics in the end
(e.g. \cite{Smith:2006ne}).

Thus nonlinear bias effects cause additional modification on the galaxy
power spectrum shape.
Therefore uncertainties in
the nonlinear bias parameters need to be properly taken into account in
extracting cosmological parameters from the measured power
spectrum. These will be carefully studied below.

\clearpage
\section{Parameter Forecasts} 
\label{sec:Forecast}

We now estimate the ability of future surveys for constraining neutrino
masses when using the SPT model predictions to be compared with 
the measurements. 

\subsection{Fisher Matrix Formalism}
For an actual galaxy redshift survey, there is another nonlinear effect
to be taken into account: redshift distortion effect due to the peculiar
velocities of galaxies.  The redshift distortion causes the
redshift-space power spectrum to be two-dimensional: the galaxy
clustering strength is varying as a function of two wavenumbers
perpendicular and parallel to the line-of-sight direction.  
The redshift-space power spectrum would be 
more prominent than the real-space one to carry useful cosmological 
information including dark energy parameters 
because it contains the geometrical distortions in directions both  
along and perpendicular to the line-of-sight, the so-called 
Alcock-Paczynski test \cite{AP} (also see 
\cite{Seo:2003pu,Matsubara:2002rf}). 
However the distortion effect in
the nonlinear regime is not yet fully understood, and a more careful
study based on high-resolution $N$-body simulations is needed to develop
the accurate modelling \cite{Taruya:2009ir}.  
Hence, in this paper for simplicity we focus on the one-dimensional  
real-space power spectrum. This roughly corresponds 
to the monopole power spectrum 
obtained by averaging the redshift-space power 
spectrum over the spherical shell of
a given wavenumber in radius, in combination with the proper weighting
as well as with the
Finger-of-God compression algorithm \cite{Tegmark:2004}, 
as developed in
\cite{Tegmark:2006az,Eisenstein:2005su}. 
Note that, after the spherical shell average, the residual 
Kaiser's effect of redshift distortion \cite{Kaiser:1984aa} 
behaves like the linear bias parameter. 
That is, we include only the
nonlinear galaxy bias effect.

We can not measure directly the length scale in real space from the
observed galaxy distribution; rather we measure the angular positions of
galaxies on the sky, and the radial position in redshift space. To
convert the observed position to the real-space position, one needs to
assume a reference cosmological model which generally differs from the
underlying true cosmology. An incorrect mapping causes an apparent
distortion in the measured power spectrum, known as the geometrical
distortion \cite{Seo:2003pu,Matsubara:2002rf}. Since in this paper we
focus on the one-dimensional, real-space power spectrum that is given as
a function of wavenumber, the wavenumber estimated from the reference
cosmology, $k_{\rm ref}$, is related to the true wavenumber $k$ as
\begin{equation}
k=\frac{D_V(z)_{\rm ref}}{D_V(z)}k_{\rm ref}
\label{eq:k_relation}
\end{equation}
where $D_V(z)$ is the effective distance factor accounting for the
spherical shell average in redshift space, and is given in terms of the
angular diameter distance and the Hubble expansion rate as
$D_V(z)\propto [D_A^2(z)/H(z)]^{1/3}$ \cite{Eisenstein:2005su}. The
quantities with subscript ``ref'' denote the quantities for the
reference cosmology. Further taking into account the amplitude
shift caused by assuming the reference cosmology, the galaxy power
spectrum estimated from a galaxy redshift survey, $P^{\rm est}_g$, is
related to the true spectrum as
\begin{equation}
P^{\rm est}_{\rm g}(k_{\rm ref})=\frac{D_V^3(z)_{\rm ref}}
{D_V^3(z)}P_{\rm g}(k,z).
\end{equation}

In order to estimate the accuracies of neutrino mass determination, we
adopt the Fisher matrix formalism (e.g. see \cite{Takada:2005si}).  
The Fisher formalism gives minimal attainable errors on the parameters by
means of a set of observables considered. However, this method becomes
inaccurate in a case that only an upper bound on neutrino masses rather
than the detection can be obtained for a given survey. In this case we
need to take into account the non-Gaussian effect of the likelihood,
i.e. a sharp cutoff at $f_\nu=0$ in parameter space. 
A more
accurate parameter estimation can be obtained, e.g. by using
a Markov-Chain Monte-Carlo based method
\cite{Perotto:2006rj}. 
The Fisher matrix formalism is sufficient for 
our purpose, which is to estimate ability of future surveys and to 
examine the impact of the refined model predictions on parameter 
estimation compared to the linear theory based method. 

The Fisher matrix for the galaxy power spectrum measurement 
for a given survey is expressed in \cite{Tegmark:1997rp} as
\be
 F^{\rm galaxy}_{\alpha\beta} = \sum_{i}\frac{V_{\rm s}(z_{i})}{4\pi^2}
 \int^{k_{\rm max}(z_{i})}_{k_{\rm min}}
 k^{2}dk~\frac{\partial\ln P^{\rm est}_{\rm g}(k;z_{i})}{\partial p_{\alpha}}
 \frac{\partial \ln P^{\rm est}_{\rm g}(k;z_{i})}{\partial p_{\beta}}
 \left[\frac{\bar{n}_{\rm g}(z_{i})P^{\rm est}_{\rm g}(k;z_{i})}
 {\bar{n}_{\rm g}(z_{i})P^{\rm est}_{\rm g}(k;z_{i})+1}
 \right]^2, 
 \label{eq:FisherGalaxy}
\ee
where $p_{\alpha}$ represents a set of free parameters, $V_{\rm
s}(z_{i})$ and $\bar{n}_{\rm g}(z_{i})$ are the comoving survey volume
and number density of galaxies, respectively, at $i$-th redshift bin
defined as $[z_{i}-\Delta z/2:z_{i}+\Delta z/2]$, and the summation runs
over redshift slices.  Note that $P_{\rm g}$ is given by
Eq.~(\ref{eq:Nonlinear_galaxy_Pk}) and the argument $k$ in 
$P_{\rm g}^{\rm est}$ is the reference wavenumber $k_{\rm ref}$ in
Eq.~(\ref{eq:k_relation}), but we omitted the subscript for notational
simplicity. The partial derivative of the power spectrum with respect to
parameter $p_\alpha$ is computed by infinitesimally varying the
parameter $p_\alpha$ around the fiducial model assumed, with other
parameters $p_\beta (\beta\ne\alpha)$ being kept to the fiducial values,
such that the Fisher matrix estimates the parameter accuracies around
the fiducial model.

To compute the Fisher matrix for a given survey we need to specify lower
and upper wavenumber bounds in the $k$-integration of
Eq.~(\ref{eq:FisherGalaxy}).  We set $k_{\rm min}=10^{-4}~h{\rm
Mpc^{-1}}$, and have checked that choosing the smaller $k_{\rm min}$
little changes the results.  On the other hand, one should be careful in
choosing the maximum wavenumber for each redshift slice, $k_{\rm
max}(z_{i})$, which needs to be chosen from the range of wavenumbers
where the model predictions, linear theory or perturbation theory, 
are reliable and accurate. One way to determine $k_{\rm max}(z_i)$ is
using $N$-body simulations in comparison with the model predictions. 
However, high-precision simulations for a MDM model are not yet fully 
explored (see \cite{Brandbyge:2008rv,Brandbyge:2008js} for the recent 
attempts based on the initial pioneer work \cite{Klypin:1992sf}). 
Here we simply  employ the following method for a CDM model in
\cite{Nishimichi:2008ry} to specify $k_{\rm max}(z_i)$ 
for each redshift slice: 
\be
 \frac{k_{\rm max}(z_{i})^{2}}{6\pi}\int^{k_{\rm max}(z_{i})}_{0}
 P^{\rm L}_{\rm m}(q;z_{i})dq = C_{\rm max},
 \label{eq:Cmax}
\ee
where $P^{\rm L}_{\rm m}$ is the input linear mass power spectrum at
redshift $z_i$. 
Since $C_{\rm max}$ is a monotonically increasing function with
$k_{\rm max}$, we will study how a choice of $k_{\rm max}$ (or
equivalently a choice of $C_{\rm max}$) affects our results. 
\cite{Nishimichi:2008ry} carefully showed that, for a CDM model,  
the SPT results fairly well agree with $N$-body 
simulations up to a maximum wavenumber corresponding to 
$C_{\rm max}=0.18~(0.3) $ to within up to $\sim 1\%$ (3\%) accuracy, 
while the corresponding valid ranges for linear theory are given by the
smaller values
$C_{\rm max}=0.06~(0.13)$. 
They also showed that another criterion derived by \cite{Jeong:2006xd} 
seems optimistic, where it was proposed that SPT 
predictions agree well with simulations up to $k_{\rm max}$ given by
$\Delta^{2}(k_{\rm max})=\left.
k^{3}P(k)/2\pi^{2}\right|_{k=k_{\rm max}}\ltsim 0.4$. 

We also comment that the Gaussian error covariance for galaxy power
spectrum is assumed in Eq.~(\ref{eq:FisherGalaxy}), where the power
spectra of different wavenumbers are assumed to be independent. 
Non-linearities of structure formation cause correlated errors of
different band powers, i.e. non-Gaussian errors for power spectrum
measurement, due to the nonlinear mode-coupling. 
The non-Gaussian errors are not negligible even on BAO scales,
comparable with neutrino free-streaming scales, and are more significant
at higher $k$ due to stronger nonlinearities. 
However at scales of our interest the impact of non-Gaussian errors on
parameter estimation is expected to be insignificant, so we employ the
Gaussian error assumption for simplicity 
(see \cite{Takada:2008fn,Takahashi:2009bq} for more detailed studies on
the non-Gaussian errors).\par

A galaxy survey alone cannot determine all the cosmological 
parameters simultaneously due to severe parameter degeneracies. A useful
way to break the parameter degeneracies is combining the galaxy survey
constraints with the constraints obtained
 from CMB temperature and polarization anisotropies.
In this paper we include information from the CMB temperature
anisotropy, $C_l^{\rm TT}$, $E$-mode polarization, $C_l^{\rm EE}$, 
and their cross correlation, $C_l^{\rm TE}$, where
we use the range of multipoles $10\le l\le 1500$ for 
$C_l^{\rm TT}$ and $C_l^{\rm TE}$ and use $2\le l\le 1500$ for $C_l^{EE}$, 
respectively.  
To compute the CMB fisher matrix, $\bmf{F}^{\rm CMB}_{\alpha\beta}$,
we adopt the noise per pixel and the angular resolution for the Planck
experiment that were assumed in \cite{Planck:2006uk}.\par 

To model the galaxy power spectra and CMB spectra we include all the key
parameters that affect the observables within the CDM and dark energy
cosmological framework. Our fiducial model is based on the WMAP 5-year
results \cite{Komatsu:2008aa}: the density parameters for total matter
and baryon are $\Omega_{\rm m0}(=0.24)$, $\Omega_{\rm m0}h^2(=0.1277)$, and
$\Omega_{\rm b0}h^2(=0.0223)$ (note that we assume a flat universe); the
primordial power spectrum parameters are the spectral tilt, $n_s(=1.0)$,
the running index, $\alpha_s(=0)$, and the normalization parameter of
primordial curvature perturbations, 
$\Delta^2_{\cal R}(k_{0})(=2.35\times 10^{-9})$ 
(the values in the parentheses denote the fiducial model). We employ the
transfer function computed from the CAMB code, and note that the primordial
spectrum amplitude is normalized at $k_0=0.002~{\rm Mpc}^{-1}$ following
the convention in \cite{Komatsu:2008aa}. The redshift evolution of dark
energy density is given by $\Omega_{\rm de}(=1-\Omega_{\rm m0})$ and the
equation of state parameter $w_0(=-1)$. When computing the CMB spectra,
we further include the Thomson scattering optical depth to the last
scattering surface, $\tau(=0.089)$. For neutrino parameters we assume the
standard three neutrino species and vary the fiducial value of total
neutrino mass, $f_\nu$. In summary, for a galaxy surveys with $N_{z}$
redshift slices in combination with the hypothetical Planck constraints, the
model parameters we consider are given as
\be
p_{\alpha}=\{
\Omega_{\rm m0}, \Omega_{\rm m0}h^{2}, \Omega_{\rm b0}h^{2}, 
w_{0}, f_{\nu}, n_{\rm S}, \alpha_{\rm S}, \Delta^{2}_{\cal R}, \tau, 
b_{1}(z_{i}), b_{2}(z_{i}), N(z_{i})
\},
\label{eq:palpha}
\ee
where $z_i=z_1, z_2, ...., z_{N_z}$.  In total, we include $(9+3N_z)$
free parameters for our Fisher matrix analysis. Note that, for the
linear theory analysis for the parameter forecasts, we consider
$(9+2N_z)$ free parameters (the parameters above minus the nonlinear
bias parameters $b_2(z_i)$).  The fiducial values of galaxy bias and
shot noise parameters change with a galaxy survey specification 
and are described in the next subsection.

The full Fisher matrix for the joint experiment of galaxy survey 
and CMB can be obtained simply by adding the Fisher matrices: 
$\bmf{F}_{\alpha\beta}=\bmf{F}^{\rm galaxy}_{\alpha\beta}
+\bmf{F}^{\rm CMB}_{\alpha\beta}$. 
The unmarginalized error on a given parameter $p_\alpha$ is
given as $\sigma(p_\alpha)=(F_{\alpha\alpha})^{-1/2}$, which corresponds
to the accuracy of determining $p_\alpha$ when other parameters are
perfectly known. On the other hand, the
marginalize error including uncertainties of other parameters is given
as $\sigma(p_\alpha)=[(\bmf{F}^{-1})_{\alpha\alpha}]^{1/2}$, where
$\bmf{F}^{-1}$ denotes the inverse of the Fisher matrix.  
The correlation coefficient $r$ between two parameters, $p_{\alpha}$ and 
$p_{\beta}$, is also useful to study how the parameters are degenerate
with each other: 
\be
r(p_{\alpha},p_{\beta}) \equiv 
\frac{(\bmf{F}^{-1})_{\alpha\beta}}
{\sqrt{(\bmf{F}^{-1})_{\alpha\alpha}(\bmf{F}^{-1})_{\beta\beta}}}. 
\label{eq:coeff}
\ee
If $r=+1~(-1)$, the parameters are totally correlated (anti-correlated), 
while $r=0$ means no correlation between the two parameters.

\subsection{Survey Parameters}
\label{subsec:survey parameters}

\begin{table}[t]
\begin{tabular}{c|cccccccccc}
\hline\hline
Survey & $z_{\rm c}$ & $\Delta z$ 
& \shortstack{$\bar{n}_{\rm g}$ \\ $10^{-4}(h^3{\rm Mpc}^{-3})$} 
& \shortstack{Survey Area \\ $({\rm deg}^{2})$} 
& \shortstack{$V_{\rm s}$ \\ $(h^{-3}{\rm Gpc}^3)$} & $b_{1}$ & $b_{2}$ 
& \shortstack{$N$ \\ $10^4(h^{-3}{\rm Mpc}^{3})$}
& \shortstack{$k^{\rm SPT3\%}_{\rm max}$ \\ $(h{\rm Mpc}^{-1})$}
& \shortstack{$\bar{n}_{\rm g}P_{\rm g}$ \\ $(k^{\rm SPT3\%}_{\rm max})$} \\
\hline
 ~SDSS LRG~ & ~0.3~ & 0.2 & 
 1.0 & 10000 & 1.17 & ~2.10~ 
 & ~0.336~ & ~0.0778 ~ & ~0.120~ & ~1.67~
\\ 
($0.2<z<0.4$)&\\ \cline{1-11}
 \multirow{2}{*}{~BOSS~} & ~0.45~ & 0.1 & 
 3.0 & 10000 & 1.13 & ~2.13~ 
 & ~0.140~ & ~0.0062 ~ & ~0.127~ & ~3.94~ 
\\ 
$(0.4<z<0.7)$ & ~0.55~ & 0.1 & 3.0 & 10000 & 1.53 & 2.21 & 0.211 & 0.0125 
 & 0.133 & 3.57
\\ 
 & ~0.65~ & 0.1 & 3.0 & 10000 & 1.94 & 2.29 & 0.263 & 0.0194 
 & 0.138 & 3.27 
\\ \cline{1-11}
 \multirow{3}{*}{~Stage-III low-$z$~} & ~0.8~ & 0.2 &
 4.0 & 3200 & 1.61 &
  ~1.41~ & ~0.295~ & 0.0177 
 & 0.146 & 1.31
\\ 
& ~1.0~ &0.2 & 4.0& 3200& 2.06 & ~1.51~ & ~0.443~ & 0.0332
 & 0.158 & 1.15
\\ 
$(0.7<z<1.6)$ & ~1.2~ & 0.2& 4.0& 3200& 2.42 & ~1.63~ & ~0.572~ & 0.0524
 & 0.170 & 1.07
\\ 
 & ~1.45~ & 0.3 & 4.0& 3200& 4.15 & ~1.77~ & ~0.760~ & 0.0851
 & 0.184 & 0.97
\\ \cline{1-11}
 ~Stage-III high-$z$~ 
& ~2.9~ & ~0.8~ & $2.5$ & 300 & 1.23 & ~3.30~ & ~2.215~ & 0.2719
 & 0.275 & 0.43
\\ 
($2.5<z<3.3$)&\\ \cline{1-11}
 \multirow{8}{*}{~Stage IV~} & ~0.6~ & 0.2 & 
200 & 20000 & 6.94 & 1.31 & -0.409 & 0.0124
 & 0.134 & 69.4
\\ 
& ~0.8~ & 0.2  & 200  & 20000 & 10.07 & 1.41 & -0.384 & 0.00933 
 & 0.146 & 57.2 
 \\ 
 & ~1.0~ & 0.2  & 200  & 20000 & 12.85 & 1.51 & -0.345 & 0.00594
 & 0.158 & 49.3
\\ 
 & ~1.2~ & 0.2  & 200  & 20000 & 15.14 & 1.63 & -0.299 & 0.00383 
 & 0.170 & 45.5
 \\ 
($0.5<z<2.1$) 
 & ~1.4~ & 0.2  & 200  & 20000 & 16.94 & 1.74 & -0.242 & 0.00217
 & 0.182 & 40.4
 \\ 
 & ~1.6~ & 0.2  & 200  & 20000 & 18.29 & 1.86 & -0.177 & $9.96\times 10^{-4}$
 & 0.195 & 34.8
\\ 
 & ~1.8~ & 0.2  & 200  & 20000 & 19.27 & 1.99 & -0.096 & $2.38\times 10^{-4}$
 & 0.206 & 31.2
\\ 
 & ~2.0~ & 0.2  & 200  & 20000 & 19.94 & 2.11 & -0.016 & $0.06\times 10^{-4}$
 & 0.219 & 28.4
\\
\hline\hline
\end{tabular}
\caption{ Survey parameters that we assume in this paper to make
parameter forecasts. The survey parameters are chosen such that the
surveys fairly well represent the existing survey (SDSS LRG), the
near-future planned survey (BOSS), and the 5-10 year time-scale future
surveys which we call Stage-III and -IV surveys, respectively, according
to Dark Energy Task Force Report \cite{Albrecht:2006um}.  We employ the
method described in Appendix~\ref{sec:halo-bias} (also see text) in
order to determine the fiducial values of the linear and nonlinear bias
parameters $b_1$ and $b_2$ for each redshift slice of the respective
survey. We also include the residual shot noise contamination arising
from nonlinear clustering, which is parametrized by $N$, and we
determine the fiducial value of each redshift slice according to the
method in Appendix~\ref{sec:halo-bias}. The values in the column
labelled by $k_{\rm max}^{\rm SPT 3\%}$ denote the maximum wavenumber up
to which the standard perturbation theory (SPT) is expected to be
reliable to within a few $\%$ accuracy compared to $N$-body simulation
results at each redshift (we determined the $k_{\rm max}$ values
following using Eq.~\ref{eq:Cmax}). We also
show the quantity $\bar{n}_{\rm g}P_{\rm g}(k_{\rm max})$ at the maximum
wavenumber for each redshift slice: if 
$\bar{n}_{\rm g}P_{\rm g}(k_{\rm max})\ge 1$, the power spectrum measurement 
is in the sample variance limited regime. 
\label{table:survey}
}
\end{table}

To make meaningful parameter forecasts, we consider survey parameters
that fairly well represent future surveys being planned or under
serious consideration.  The hypothetical surveys considered in this
paper are intended to resemble BOSS, WFMOS-like survey, and the ideal
space-based BAO experiment such as those proposed by JDEM and Euclid
missions, which are roughly categorized as the Stage-III and -IV
surveys, respectively, in the DETF report \cite{Albrecht:2006um}.

The survey parameters are summarized in Table~\ref{table:survey}. The
survey area, redshift range and number densities of target galaxies were
taken from the proposed survey design of each survey. Just briefly, the
BOSS-like survey samples luminous red galaxies (LRGs) over a range of
redshifts $0.4<z<0.7$ extending the SDSS-I and -II surveys. 
A ground-based Stage-III survey with optical spectroscopy
may be designed to survey galaxies for two different slices: one is for
galaxies over $0.7<z<1.6$ with survey area $3200 $ deg$^2$, and the
other is for high-redshift Lyman-$\alpha$ emission or Lyman break
galaxies over $2.5<z<3.3$. The survey parameters for the Stage-IV type
survey are taken from \cite{bennet}.  Having multiple redshift slices is
useful to improve the accuracies of parameter estimation by breaking the
parameter degeneracies because the sensitivity of each redshift slice to
cosmological parameters is slightly different as will be shown below
(also see \cite{Matsubara:2004} for the related discussion).  
These surveys are complementary to each other in redshift ranges
covered. 
It is also worth commenting
that a high-redshift survey with $z>1$ has a potential to explore an
early dark energy model where dark energy may be more rapidly evolving at higher
redshifts than naively expected.

We further need to specify galaxy bias parameters. However, because we
have a limited knowledge on galaxy formation, it is difficult to predict
galaxy bias parameters with certainty. Here we rather employ a crude
method used in \cite{Seo:2003pu,Takada:2005si} to estimate the linear
galaxy bias parameter $b_1$ for each redshift slice, where $b_1$ is
estimated by imposing the rms number density fluctuations of galaxies
within a sphere of $8h^{-1}$Mpc radius to be unity:
$\sigma_{g8}^2=1$. However the LRG bias is relatively well understood
based on the existing SDSS sample such as $b_1\approx 2.1$ in
 \cite{Tegmark:2006az}.
We assume $b_1=2.1$ for the fiducial value of SDSS LRG bias, 
from
which we compute a correction factor that needs to be multiplied by
$\sigma_{g8 }^2=1$ to obtain $b_1=2.1$ 
for our fiducial cosmological model. Similarly, for
BOSS LRGs, we multiply $\sigma_{g8}^2=1$ by the same correction factor
to estimate the linear bias $b_1$
(see Appendix~\ref{sec:halo-bias} 
for more details).  The nonlinear bias parameter $b_2$
and the residual shot noise parameter $N$ are more uncertain. We define
their fiducial values based on the prescription described in
Appendix~\ref{sec:halo-bias}, but will study how our results change with
different fiducial values of $b_2$ and $N$.  Note that the parameter $N$
is estimated based on the perturbation theory, but we will employ the
same fiducial value for the linear theory based forecasts.
When $N\ge 1/\bar{n}_{\rm g}$, the residual shot
noise contamination is dominant. 
Our survey parameters imply that, for BOSS and
Stage-III surveys, $N<1/\bar{n}_{\rm g}$. On the other hand, the
residual shot noise contamination is significant for some redshift
slices having higher number densities of galaxies for the Stage-IV survey.

\subsection{Parameter Forecasts}
\label{subsec:forecast results}

\subsubsection{Summary of constraints on neutrino mass}

\begin{table}[t]
Expected marginalized error on total neutrino mass: $\sigma(m_{\nu,{\rm
 tot}})$ (eV)
\begin{tabular}{l|ccccc}
\hline\hline
Survey
& \hspace{1em}\shortstack{Linear 1\% \\ $C_{\rm max}=0.06$} 
& \hspace{2em}\shortstack{Linear 3\% \\ $C_{\rm max}=0.13$} 
& \hspace{2em}\shortstack{SPT 1\% \\ $C_{\rm max}=0.18$}
& \hspace{2em}\shortstack{SPT 3\% \\ $C_{\rm max}=0.3$} 
& \hspace{2em}\shortstack{+ $\sigma(\Omega_{\rm m0})=0.01$\\ 
$C_{\rm max}=0.3$} 
\\
\hline
 BOSS        
 &
0.161
& 
0.111
& 
0.095
&  
0.088
&
0.082
\\ 
 Stage-III (low-z slices alone) 
 & 
0.173
&
0.123
&
0.110
&
 0.096
 &
 0.082
\\
 Stage-III (low- + high-z) 
 &
0.161
&
 0.122
&
 0.107
&
  0.091
 &
 0.081
\\
 Stage IV
 &
 0.067
&
0.059
&
 0.053
&
 0.046
 &
 0.046
\\
\hline
\hline
\end{tabular}
\caption{
\label{table:sigmanu}
 Marginalized 1$\sigma$ error on total neutrino masses, 
 $\sigma(m_{\nu,{\rm tot}})~[{\rm eV}]$,
expected from
each hypothetical survey when combined with the Planck 
 and $z\sim 0.3$ SDSS LRG information. The errors are derived 
 including the galaxy power
 spectrum information over $10^{-4}\le k\le k_{\rm max}~h{\rm
 Mpc}^{-1}$, where $k_{\rm max}$ is determined by Eq.~(\ref{eq:Cmax})
 from the input linear mass power spectrum. 
 As implied, the linear theory and perturbation theory are expected to be
 accurate 
up to the given
 $k_{\rm max}$ 
to within a given $\%$ accuracy compared to $N$-body simulations 
 \cite{Nishimichi:2008ry}. For these results we assume 
 $m_{\nu,{\rm tot}}=0.12{\rm eV}~(f_{\nu}=0.01)$ for the fiducial value
 of neutrino mass, therefore the errors shown roughly correspond to the
 expected $1\sigma$ upper limit on neutrino mass if 
 $\sigma(m_{\nu,{\rm tot}})\gtsim 0.12~{\rm eV}$. The last column
 labelled by ``$+\sigma(\Omega_{\rm m0})=0.01$'' shows an improvement in
 the neutrino mass constraint for the case $C_{\rm max}=0.3$ if 
 the prior $\sigma(\Omega_{\rm m0})=0.01$ is added. 
}
\end{table}

Table~\ref{table:sigmanu} summarizes forecasts for the marginalized
errors on total neutrino mass for each of hypothetical galaxy surveys
listed in Table~\ref{table:survey}, combined with the Planck and SDSS
LRG information.  To derive these errors we determined $k_{\rm max}$ for
each redshift slice based on the criteria (\ref{eq:Cmax}) and then
included the power spectrum information over $10^{-4}\le k\le k_{\rm
max}(z_i)~h\rm{Mpc}^{-1}$.  We compare the expected constraints obtained
when using the linear theory and SPT models, over a range of wavenumbers
where the respective models seem reliable as indicated from the assumed
value of $C_{\rm max}$ (the corresponding $k_{\rm max}$ values when
$C_{\rm max}=0.3$ for each redshift slice
are listed in Table~\ref{table:survey}). 
Note that we assume $f_\nu=0.01$ ($m_{\nu, {\rm
tot}}\simeq 0.12~{\rm eV}$) for the fiducial value, and the number of
free parameters is different in between the linear theory and SPT as
described around Eq.~(\ref{eq:palpha}) (SPT additionally includes the
nonlinear bias parameter $b_2$ for each redshift slice). 

It is clear that the use of SPT allows for an improvement 
in the neutrino mass constraint compared to the linear theory results: 
roughly a factor of 1.3 improvement
if SPT can be used up to the maximum wavenumbers 
where SPT seems reliable to within a few \% accuracy corresponding 
to $C_{\rm max}=0.3$. We have checked that the accuracy of neutrino mass
determination, $\sigma(m_{\nu,{\rm tot}})$, little changes 
for each survey even if the fiducial value of $f_\nu$ is varied
within the current limit $f_\nu\ltsim 0.05$.
Hence Table~\ref{table:sigmanu} implies 
the BOSS and Stage-III type survey may allow for
the accuracy of $\sigma(m_{\nu,{\rm tot}})\simeq 0.1~$eV, while the
Stage-IV survey $\sigma(m_{\nu,{\rm tot}})\simeq 0.05~$eV.
In particular the expected accuracy for a Stage-IV type 
survey is compatible with
the lower limit implied from the normal mass hierarchy. That is, 
Stage-IV may allow for a detection of total neutrino mass at more than 
 $1$-$\sigma$ significance; if neutrinos obey the inverted mass
 hierarchy implying the lower limit $m_{\nu,{\rm tot}}\gtsim 0.1~$eV, a
 2-$\sigma$ level detection may be achieved. 

Note that the forecasted constraints here are much weaker than those obtained
in our previous work \cite{Saito:2008bp}. The differences are (1) we
here consider the one-dimensional power spectrum as the observable
rather than the full two-dimensional power spectrum in redshift space
and (2) we include the nonlinear bias parameters. 
The full analysis including the
two-dimensional redshift power spectrum information will be presented
elsewhere (Saito et al. in preparation). 

\subsubsection{Degeneracy between neutrino mass and other parameters}

\begin{figure}[t]
\begin{center}
\includegraphics[width=0.45\textwidth]{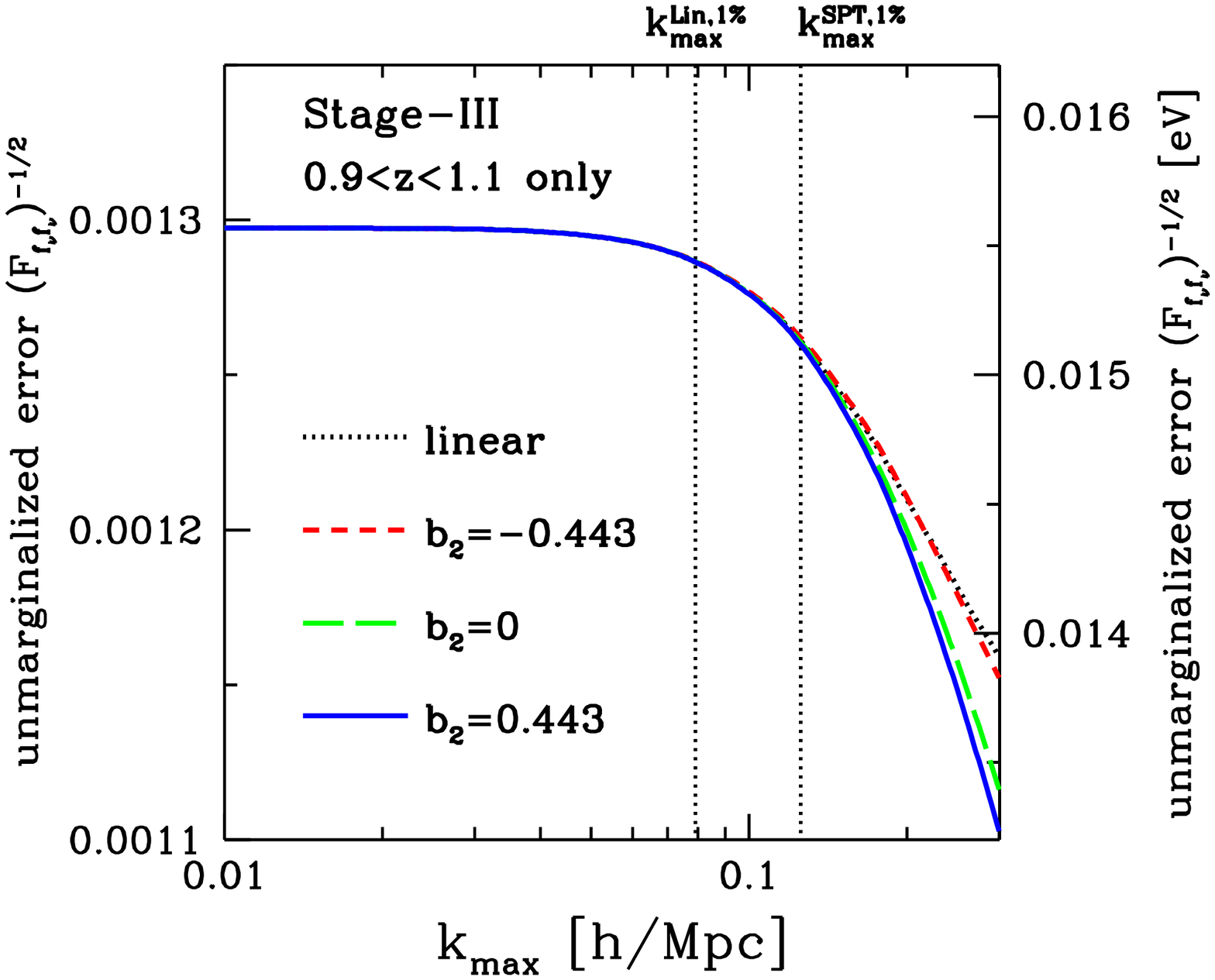}
\includegraphics[width=0.45\textwidth]{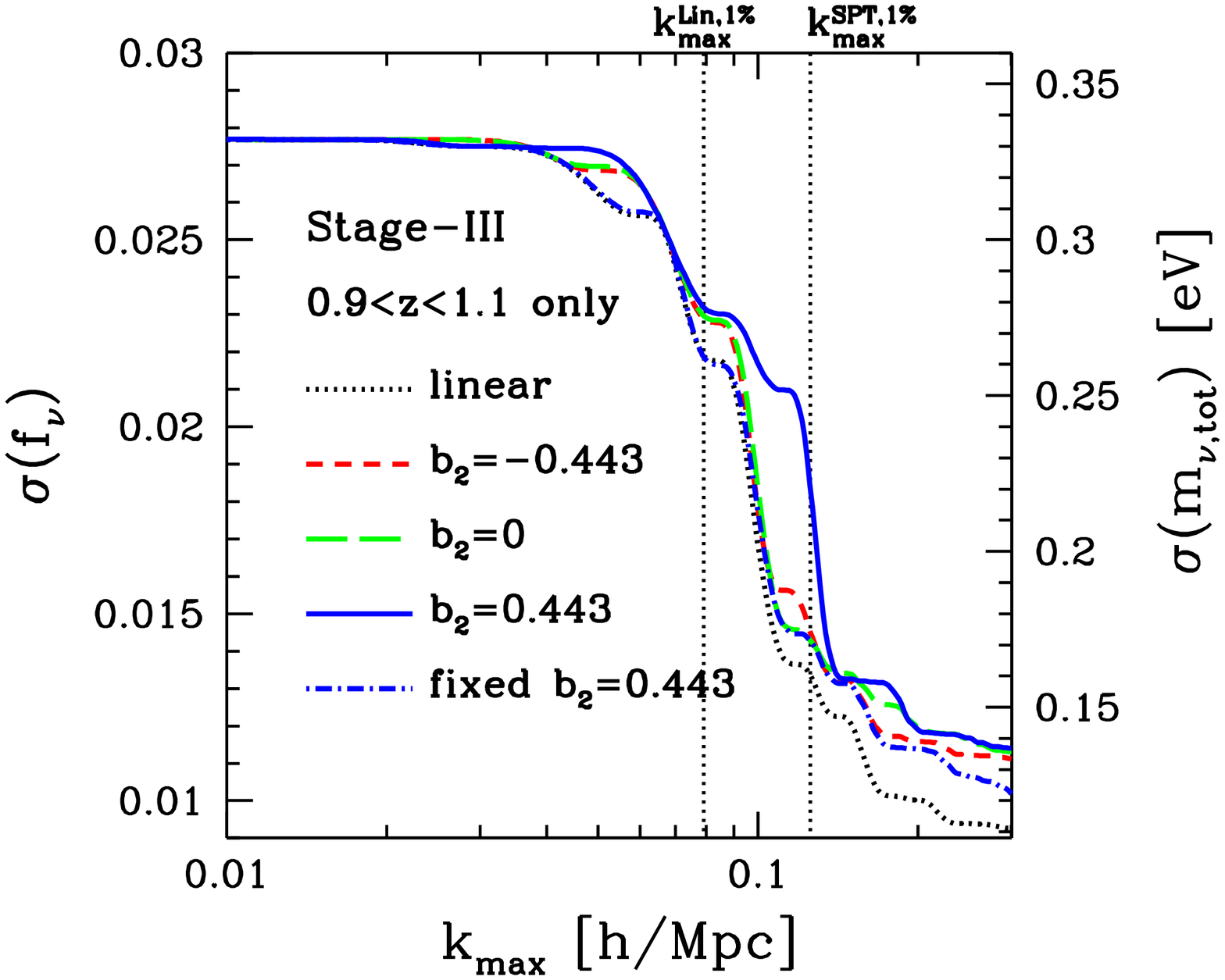}\\
\includegraphics[width=0.45\textwidth]{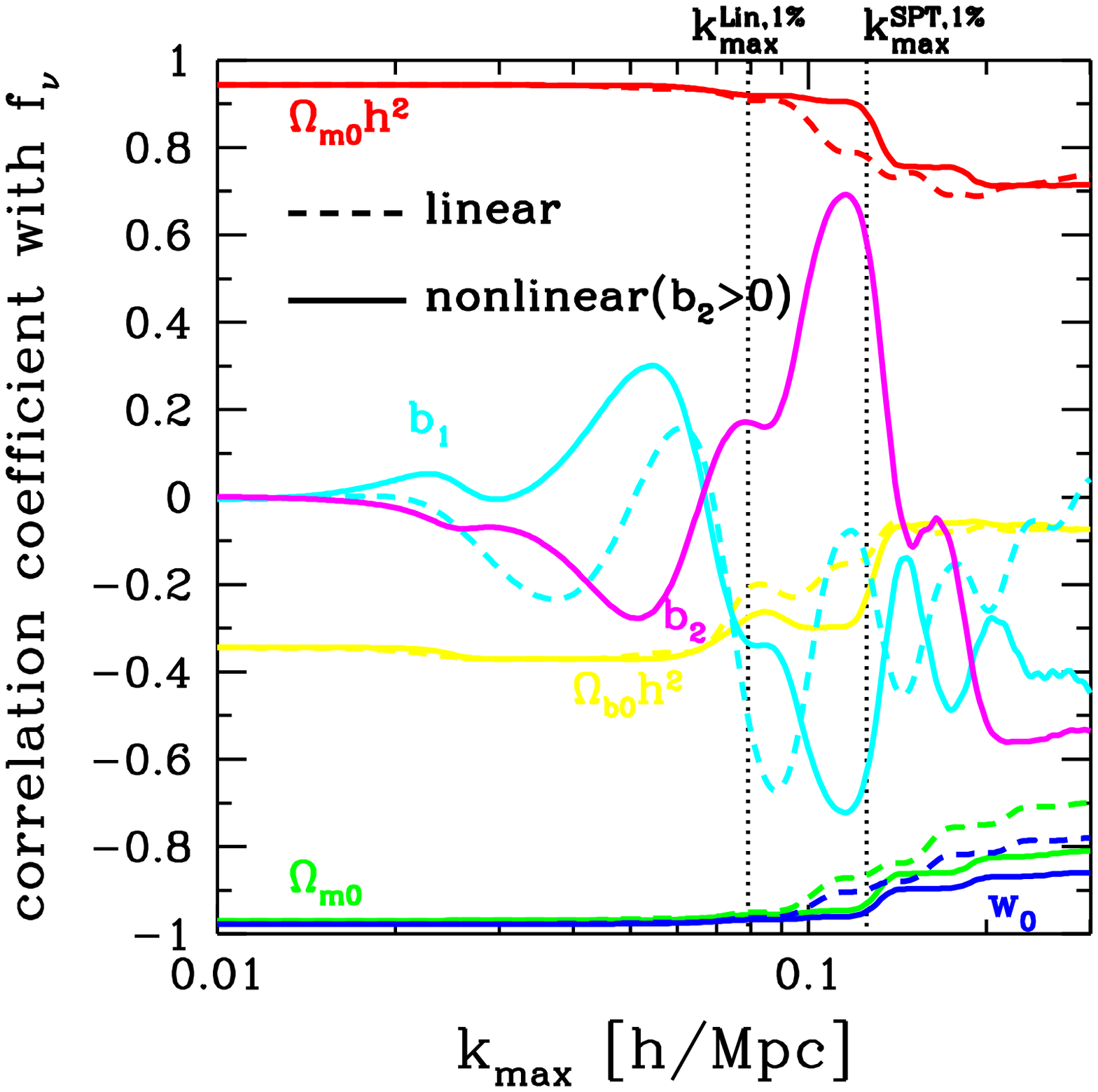}
\end{center}
\vspace*{-2em}
\caption{
 {\it Top-left panel}: The unmarginalized $1\sigma$ error 
 on $f_{\nu}$ as a function of the maximum wavenumber $k_{\rm max}$, 
 expected from a single redshift slice around the centering
 redshift $z_{\rm c}=1$ of Stage-III low-$z$ survey in
 Table~\ref{table:survey}. 
 The short-dashed, long-dashed and solid curves show how the neutrino
 mass error changes if the nonlinear bias
 parameter is changed from the fiducial choice $b_2=0.443$ to the other
 choices  $b_2=-0.443$ and 0, where other survey parameters are fixed as
 shown in Table~\ref{table:survey}. 
 The neutrino mass constraint is sensitive
 to the underlying $b_2$, i.e. galaxy types. For comparison, the dotted
 curve shows the linear theory result. Note that the two vertical dotted
 lines show the maximum wavenumbers up to which the perturbation theory
 is expected to be accurate to within the given accuracy for this
 redshift slice. 
 {\it Top-right panel}: The similar plot, but shows the marginalized
errors on neutrino mass. The step-like features are apparent: the
plateau shape is due to strong parameter degeneracies and a sudden drop
of the error at some particular wavenumbers imply an improvement in the
parameter errors because the parameter degeneracies can be to some
extent broken by including the BAO features with increasing 
$k_{\rm max}$. Compared to the top-left panel, the marginalized error is 
not necessarily more stringent in the weakly nonlinear regime than the
linear theory extrapolated error due to the stronger parameter
degeneracies, although the linear theory ceases to be reliable in the
nonlinear regime. 
 {\it Bottom panel}: The correlation coefficients of neutrino mass
 with other parameters, $r(f_\nu,p_\alpha)$, defined by
 Eq.~(\ref{eq:coeff}), displaying complex degeneracy behaviors as a
 function of $k_{\rm max}$. 
}
\label{fig:z1.0 only}
\end{figure}

To develop a better understanding of the forecasted neutrino mass
errors, we study how parameters are degenerate with each other and how
the degeneracies can be broken when the galaxy power spectrum
information ranging from the linear to non-linear regime are combined
with the CMB information. 
First, the top panel of Fig.~\ref{fig:z1.0 only} 
shows the unmarginalized errors on neutrino mass as a
function of the maximum wavenumber, for the single $z=1$ slice of 
the Stage-III low-$z$ survey in Table~\ref{table:survey}, 
where the linear theory result 
is compared with the SPT results obtained assuming various fiducial values 
of nonlinear bias parameter $b_2$. 
For $k_{\rm max}\ltsim0.07~ h{\rm Mpc}^{-1}$, the neutrino mass
constraint does not depend on $k_{\rm max}$, implying that the
constraint is mostly from the CMB information. For the larger 
$k_{\rm max}$ the galaxy power spectrum is becoming to be more powerful 
to constrain the neutrino mass due to the increased independent Fourier
modes. From comparison between the linear theory and SPT results,
one can find that 
 the unmarginalized error on neutrino mass
is improved in the weakly nonlinear regime due to the improved
signal-to-noise ratio of power spectrum measurement, 
except for the case of $b_2<0$. The case of $b_2<0$ causes
a suppression in the power spectrum amplitudes, as implied in the top
panel of Fig.~\ref{fig:galaxy suppression}. 
As a result the information content of the
power spectrum does not increase so much in the weakly nonlinear regime
compared to the linear theory, although the linear theory breaks 
down in the regime. 
Thus the neutrino mass constraints are sensitive to
galaxy bias parameters or equivalently galaxy types.

The upper-right panel shows the neutrino mass errors marginalized over
other parameter uncertainties. Again notice that the results are only
for one $z=1$ slice of the Stage-III low-$z$ survey corresponding to the
survey volume $2.1~ h^{-3}{\rm Gpc}^3$ (see below for the full forecast
for all the redshift slices combined). Compared to the unmarginalized
errors, the neutrino mass error is significantly degraded due to strong
parameter degeneracies. The plot also shows a clear plateau feature in
the error for $k_{\rm max}\ltsim 0.1~h{\rm Mpc}^{-1}$, and then shows a
step-like improvement in the error at some particular $k_{\rm max}$
values, which are found to correspond to the BAO peaks. Namely, when
the BAO peaks are included by increasing $k_{\rm max}$, the accuracies
of constraining cosmological parameters are dramatically improved by
breaking the parameter degeneracies via the Alcock-Paczynski test.
Comparing the linear theory and SPT results manifests that, in contrast
to the results for the unmarginalized errors, the neutrino mass error
does not improve by using SPT, due to the significant parameter
degeneracies and the addition of nonlinear bias parameter $b_2$. 
The effect of $b_2$ can be explicitly studied 
by the dot-dashed curve, where $b_2$
is kept fixed. Fixing $b_2$ does improve the
neutrino mass constraints, implying a strong degeneracy between neutrino
mass and $b_2$ in the nonlinear power spectrum. However, the SPT result
with $b_2$ being fixed is still apparently 
worse than the linear theory extrapolated result
in the weakly nonlinear regime (although the linear theory
breaks down in the regime). This may be understood as follows. 
As discussed, the neutrino mass constraints are sensitive to an
inclusion of BAO features which helps break parameter degeneracies. 
However, the nonlinear mode coupling somewhat smooths out BAO features in the
weakly nonlinear regime, which degrades 
the constraining power of galaxy
surveys in the weakly nonlinear regime. 

The bottom panel of Fig.~\ref{fig:z1.0 only} explicitly studies 
the Fisher correlation coefficients of neutrino mass
with other parameters, $r(f_\nu,p_\alpha)$,  
as a function of $k_{\rm max}$ for the $z=1$ slice. 
The neutrino mass appears to be significantly 
degenerate with some parameters such as
$\Omega_{\rm m0} h^2$, $w_{0}$, $\Omega_{\rm m0}$ and $b_2$  
showing almost perfect degeneracy of $|r|\sim 1$. 
The degeneracies show complex behaviors as a function of $k_{\rm max}$,
where the 
oscillatory features of $r$ correspond to the BAO features.

We comment on the parameter $N$ which models the residual shot noise
contamination to the power spectrum measurement
 due to nonlinear
clustering of galaxies.
For the assumed Stage-III survey, the residual shot noise contamination
arising from the nonlinear galaxy clustering is smaller than the standard
shot noise $1/\bar{n}_{\rm g}$. 
In addition the sample variance gives a dominant contribution
to the power spectrum covariance over all the scales we consider, 
$k\ltsim 0.3~h{\rm Mpc}^{-1}$. Therefore the effect of $N$ is
insignificant for the results shown here. However, the genuine effect
needs to be studied using $N$-body simulations, since this shot noise
contamination is not yet fully explored. 

\begin{figure}[t]
\begin{center}
\includegraphics[width=0.45\textwidth]{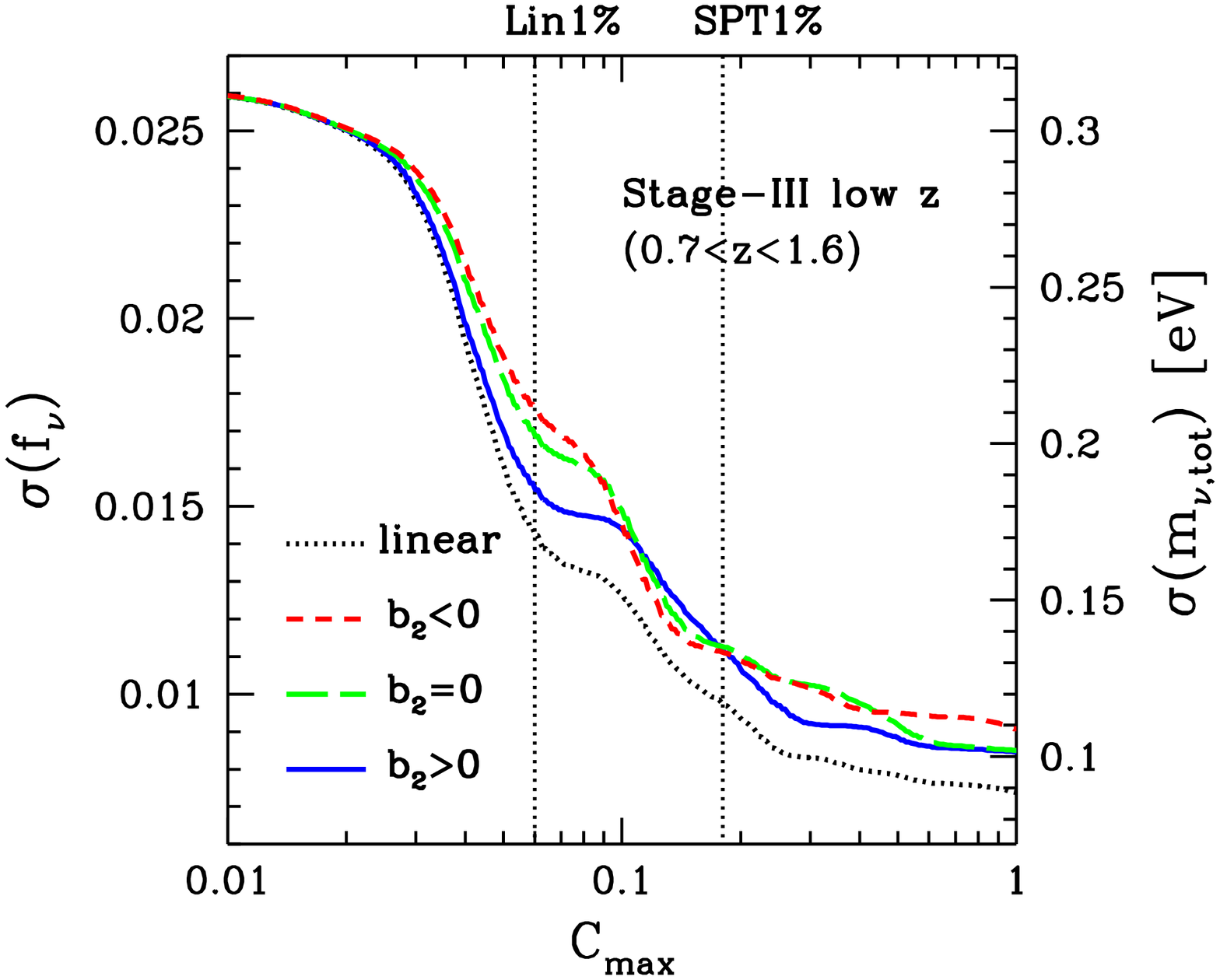}
\includegraphics[width=0.45\textwidth]{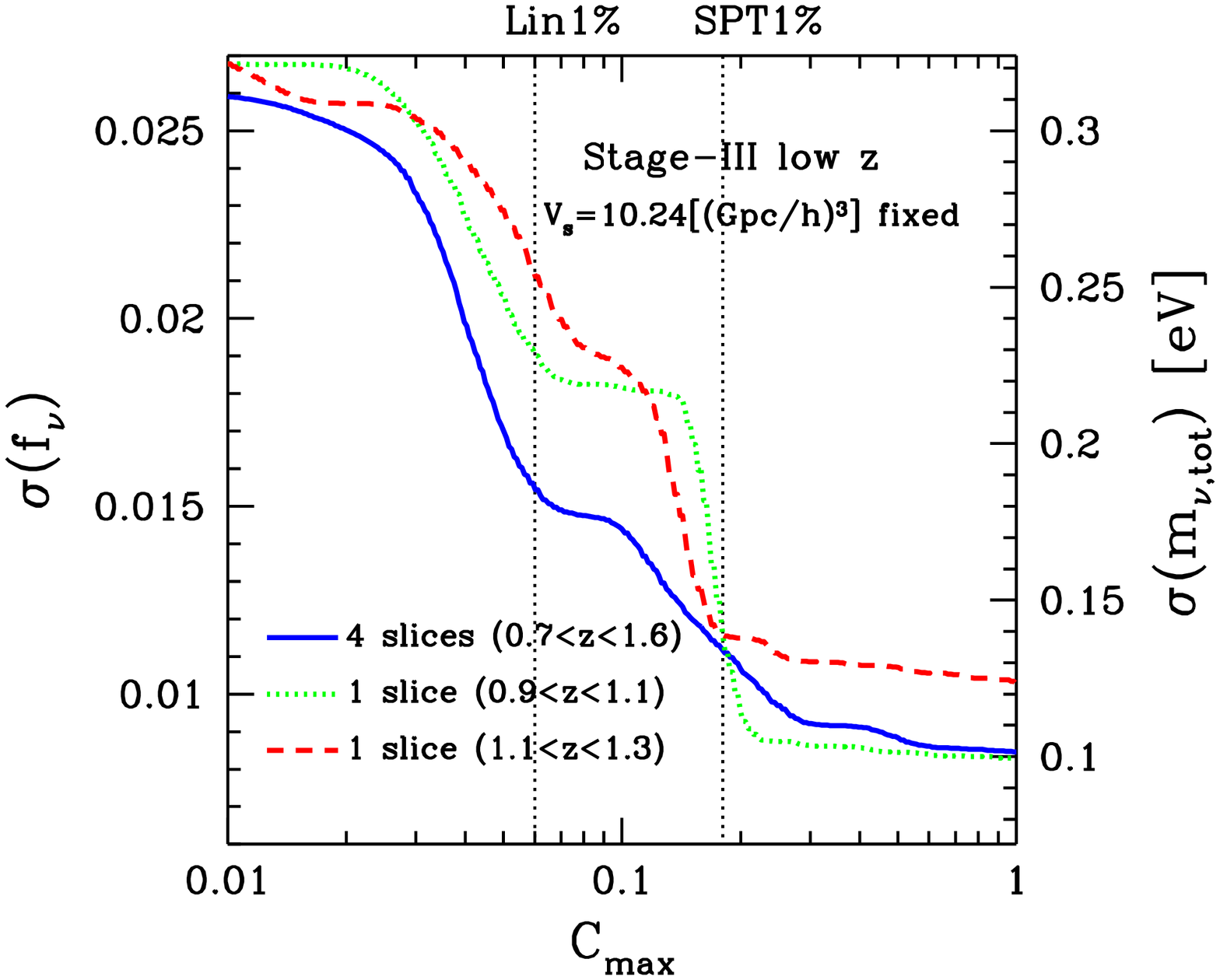}
\end{center}
\vspace*{-2em}
\caption{
 {\it Left panel}: As in the top-right panel of the previous figure,
 this plot shows the marginalized error on $f_{\nu}$ obtained by 
combining the four different  
 redshift slices of Stage-III low-$z$ survey in Table~\ref{table:survey}
 as a function of $C_{\rm max}$, where the maximum wavenumber of each
 redshift slice, $k_{\rm max}(z_i)$, is computed using
 Eq.~(\ref{eq:Cmax}) for an input $C_{\rm max}$ given 
in the horizontal axis. 
 Note that for reference $C_{\rm max}=0.1~(1.0)$ corresponds to 
 $k_{\rm max}=0.097~(0.266)~h{\rm Mpc}^{-1}$ at $z=1$. 
The fiducial values of nonlinear bias parameters $b_2$ for each redshift
slice, which are all positive,  
are given in Table~\ref{table:survey}. For comparison, the
 short- and long-dashed curves show the results obtained when the sign
 of $b_2$ is flipped or assuming $b_2=0$ for all the slices,
 respectively. 
 {\it Right panel}: The complementarity of different redshift slices is
 more explicitly studied. The solid curve is same as the solid curve in
 the left panel. The dotted and dashed curves show the results for only
 one redshift slice with the centering redshifts $z_c=1$ and $1.2$,
 respectively, keeping the survey comoving volume fixed. 
}
\label{fig:combining slices}
\end{figure}

Fig.~\ref{fig:combining slices} shows the results combining 
all the four redshift slices for Stage-III low-$z$ survey 
(the total survey volume 
$V_s=10.24~h^{-3}{\rm Gpc}^3$).  
The left panel demonstrates the marginalized
errors on neutrino mass as in Fig.~\ref{fig:combining slices}, 
but as a function of $C_{\rm max}$, 
where $k_{\rm max}$ for each redshift slice is specified
by using Eq.~(\ref{eq:Cmax}). 
Compared to Fig.~\ref{fig:galaxy suppression}, 
the accuracies of neutrino mass determination continue 
to improve with increasing $k_{\rm max}$ or adding more galaxy 
power spectrum information; there is 
no regime dominated by CMB information 
over the scales we have considered. 
This is because the galaxy power spectra at different redshifts 
depend on cosmological parameters in different ways, so combining 
the different redshift information helps break the parameter 
degeneracies. The solid, short- and long-dashed curves compare 
the results for different fiducial values of $b_2$: 
In the first case we adopt the fiducial values of $b_2$ 
given in Table~\ref{table:survey}, while in the second case 
we multiply the minus sign in the fiducial values of $b_2$ 
in Table~\ref{table:survey} 
(all the $b_2$ values are negative for this case). 
In the third one, we set $b_2=0$ for all redshift slices. 
As also discussed in Fig.~\ref{fig:galaxy suppression}, 
the neutrino mass determination accuracies are found to be 
sensitive to the fiducial values of $b_2$, or equivalently 
to galaxy types targeted for future surveys.
However the differences due to different values of $b_{2}$ become milder
by combining different slices.  

The SPT results can be compared with the linear theory result 
(the dotted curve). As in Fig.~\ref{fig:galaxy suppression}, 
the nonlinear regime suffers from severe parameter degeneracies, 
yielding less stringent parameter constraints than naively expected 
by linear theory. For this reason the parameter forecasts in
the previous studies 
may be somewhat too optimistic, if the forecasts 
are derived based on the linear theory and the linear bias parameter 
(e.g., \cite{Takada:2005si,Hannestad:2007cp,Saito:2008bp}). 
However, we again note that the full information on galaxy clustering 
is inherent in the two-dimensional redshift space, while we 
consider the one-dimensional power spectrum in this paper.  

The usefulness of combining different redshift slices is explicitly
shown in the right panel of Fig.~\ref{fig:combining slices}.  The plot
compares between the results of different redshift slicing, where the survey
volume is kept fixed to $V_s\simeq 10.24~h^{-3}{\rm Gpc}^3$. However,
note that the maximum wavenumber $k_{\rm max}$ is different for
different redshift slices, therefore the effective survey volume is 
different.  The solid curve is the result of our fiducial Stage-III
low-$z$ survey, while the dotted and dashed curves are the results
assuming single redshift slice which have different centering redshifts
$z_c=1.0$ and $1.2$ with width $\Delta z=0.2$, respectively.  The single
redshift slice cases correspond to survey areas $\Omega_{\rm s}=15900$
and $13500$ deg$^2$, respectively, compared to the fiducial area
$\Omega_{\rm s}=3200$ deg$^2$ over redshift range $0.7\le z\le 1.6$.  It
is clear that the neutrino mass constraint is improved by 
combining the different redshift slices.
  Also, comparing the dotted and dashed curves
clarifies that a choice of redshift slices affects the constraining
power in the weakly nonlinear regime.

\subsubsection{Impact of massive neutrinos on dark energy constraints}

\begin{figure}[t]
\begin{center}
\includegraphics[width=0.45\textwidth]{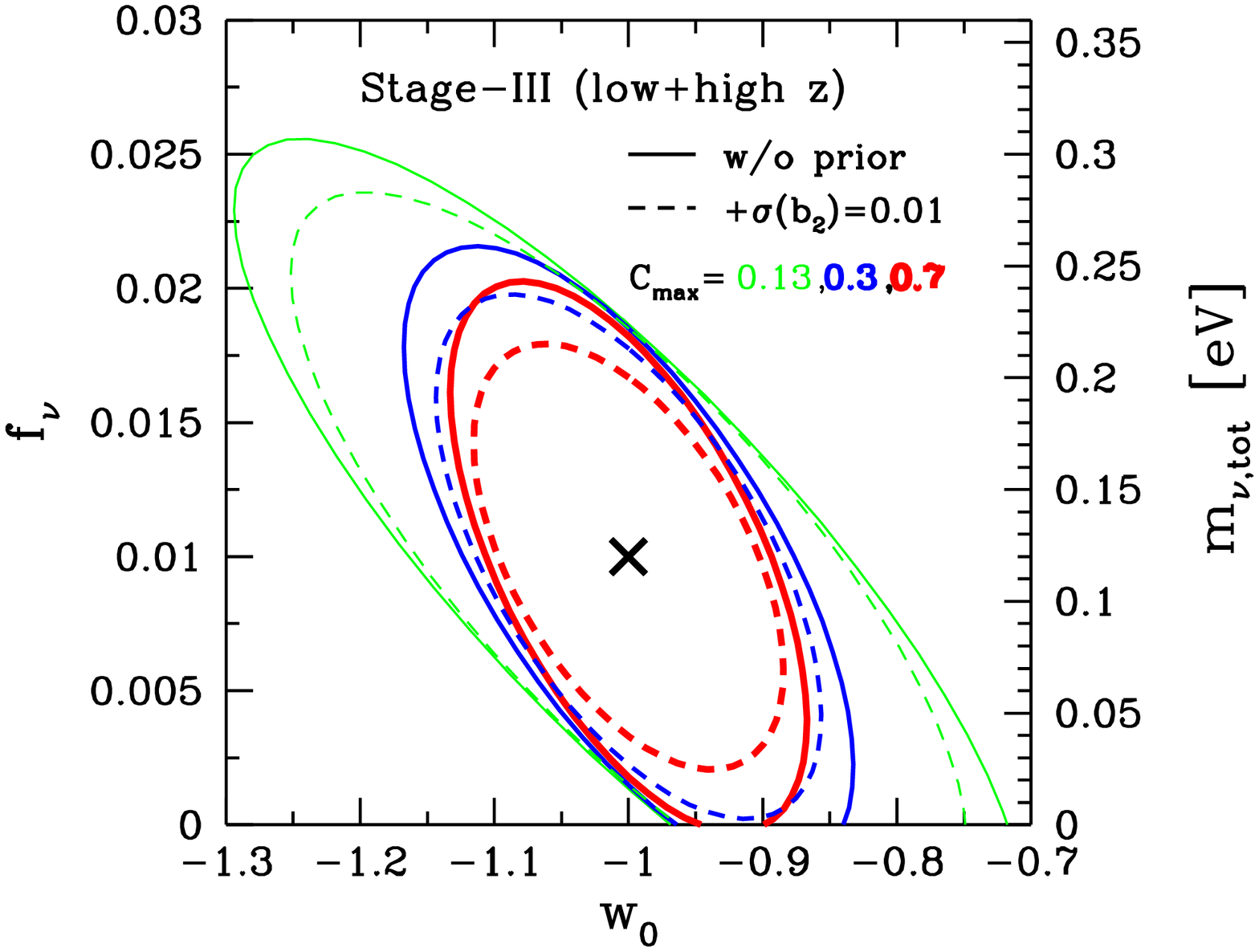}
\includegraphics[width=0.45\textwidth]{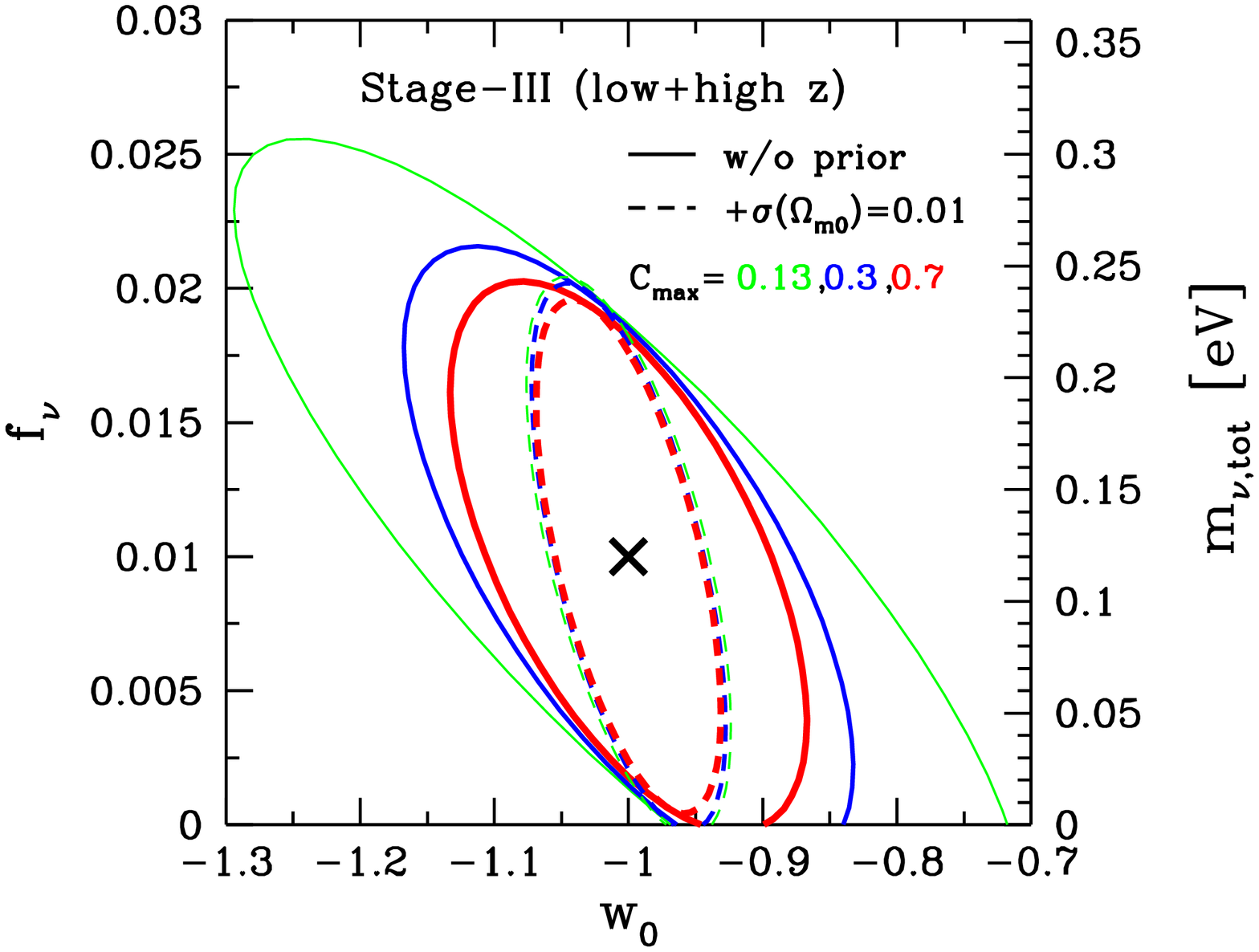}\\
\includegraphics[width=0.45\textwidth]{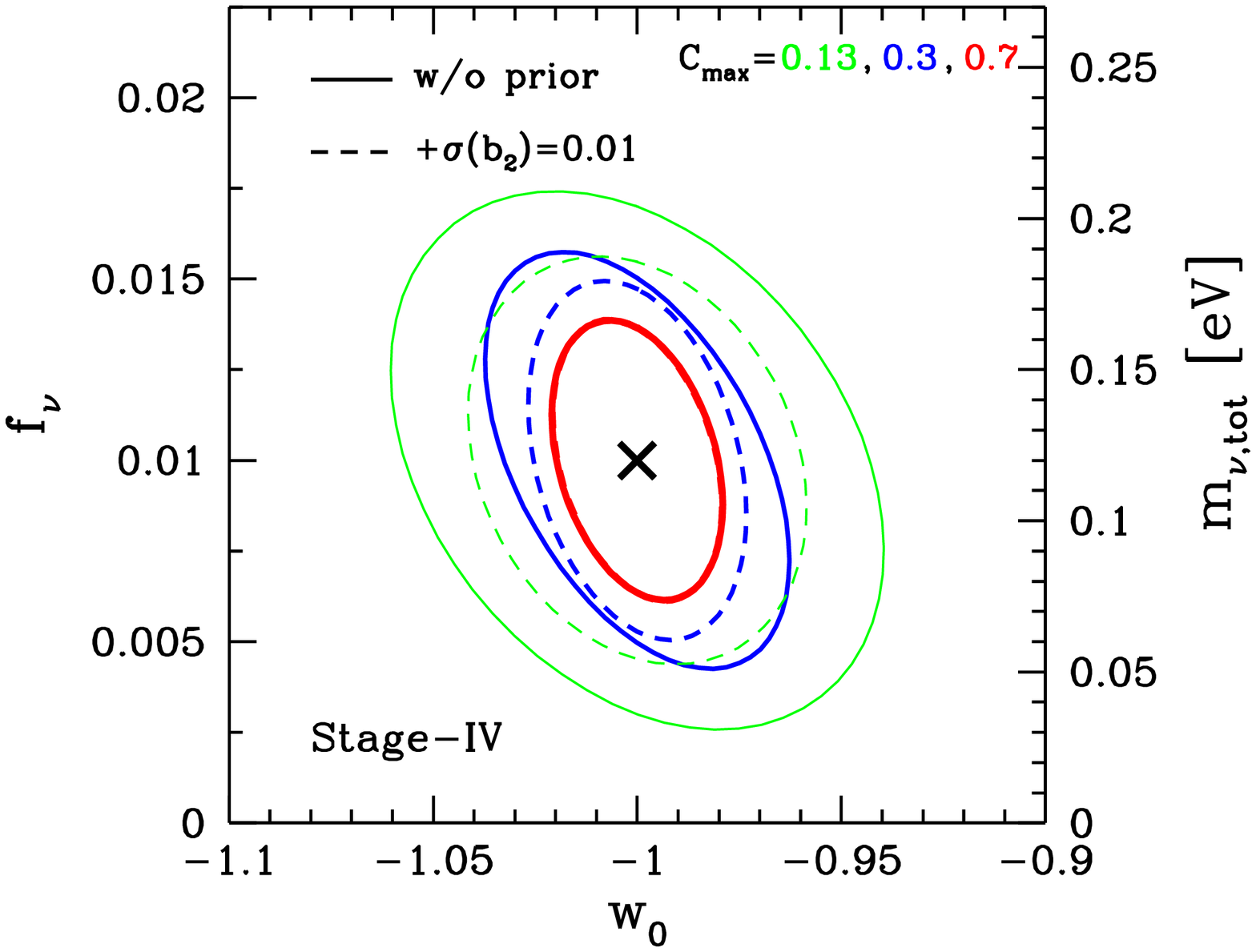}
\includegraphics[width=0.45\textwidth]{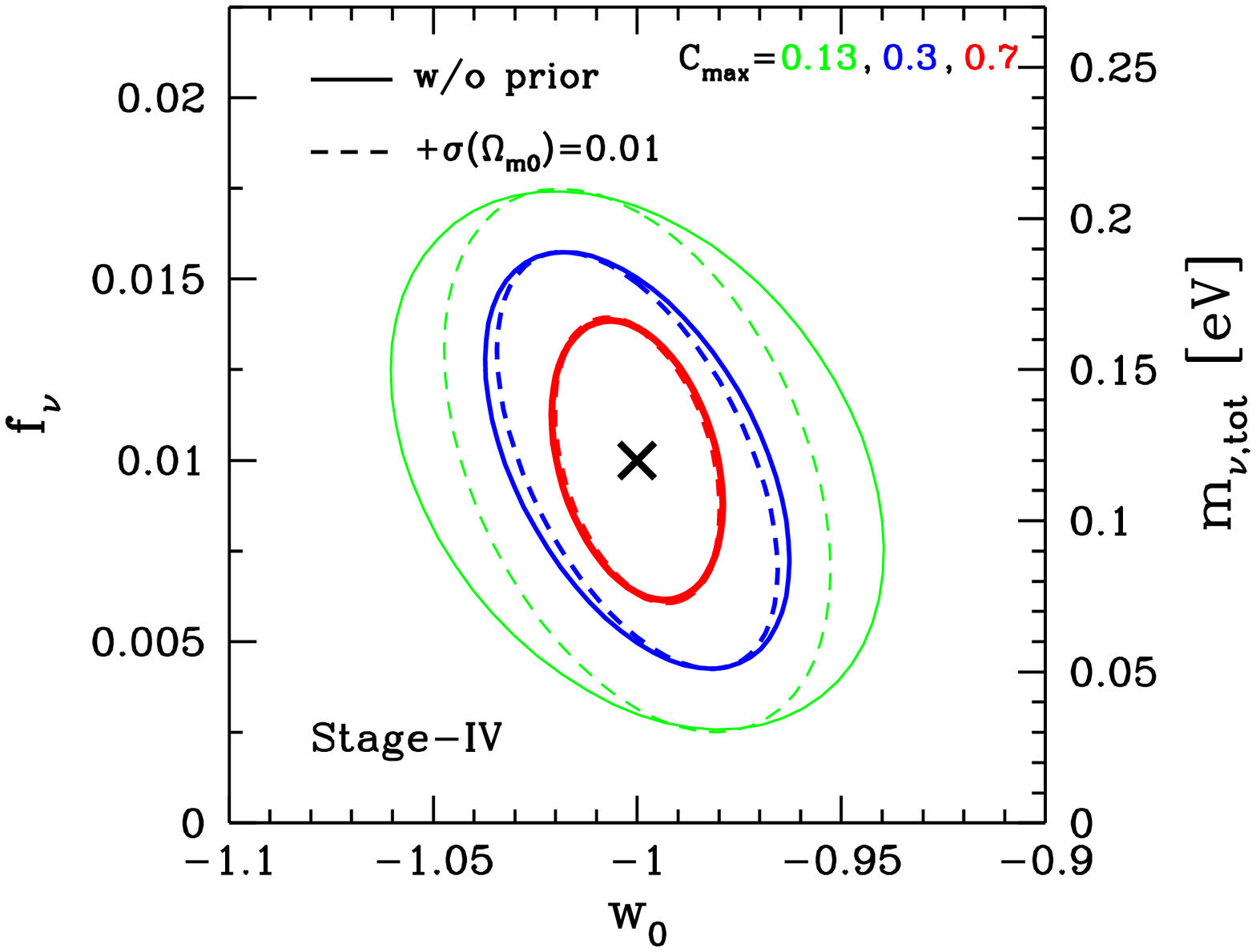}
\end{center}
\vspace*{-2em}
\caption{
Forecasted $1\sigma$ error ellipses in  
$(w_{0},f_{\nu})$-subspace
 for Stage-III ({\it top two panels}) and Stage IV ({\it bottom}),
 respectively. The outermost, intermediate and innermost contours show
 the results assuming $C_{\rm max}=0.13$, $0.3$ and $0.7$, respectively,
 that correspond to the higher $k_{\rm max}$ in each redshift slice.  
 Again note that 
$C_{\rm max}=0.13$ and 0.3 roughly correspond to the maximum wavenumbers
 that the linear theory and SPT are reliable with a few 
\% accuracy. 
The case of 
 $C_{\rm max}=0.7$ may be feasible if the refined 
 model of nonlinear power spectrum can be used (see text for the details). 
The dashed contours show the errors when the prior of $\sigma(b_2)=0.01$
 (left panels) or $\sigma(\Omega_{\rm m0})=0.01$ (right panels) is added.
}
\label{fig:w0-fnu}
\end{figure}

The primary science goal of future surveys is constraining the
nature of dark energy via the BAO experiment.  However, the dark energy
constraints may be biased if the model fitting ignores neutrino mass
contribution. Fig.~\ref{fig:w0-fnu} presents the marginalized error ellipses
in a sub-space of the neutrino mass $f_\nu$ and the dark energy equation
state parameter $w_0$ for the Stage-III and -IV
surveys, respectively. Note that the dark energy constraints shown here are
from both the BAO peak locations and the power spectrum
amplitude information.  There appears to be a significant correlation
between  $w_0$ and neutrino mass as expected. For example, 
a model with $w_0>-1$ or greater $\Omega_{\rm de0}$
yields smaller amplitudes in the galaxy power spectrum,
because such a model causes dark energy to be more significant from
earlier epochs and therefore 
the greater cosmic acceleration suppresses the
clustering growth rate for the CMB normalization of linear
power spectrum amplitude. This dark energy effect can be compensated
by lowering the neutrino mass (i.e. the smaller $f_\nu$) that leads to a
less suppression in the power spectrum amplitudes at the larger $k$. 
One can also find that having larger $k_{\rm max}$ 
(equivalently larger $C_{\rm max}$) yields
more stringent constraints on these parameters. 
In particular, it should be noted that 
a Stage-IV type survey may allow for a stringent test of 
neutrino mass,
even from the
1D power spectrum information over a range of wavenumbers where SPT
seems reliable.

In Fig.~\ref{fig:w0-fnu} we also study  how the parameter constraints are
improved  by adding an external prior of $b_{2}$ or 
$\Omega_{\rm m0}$. These priors may be delivered from the galaxy
bispectrum analysis \cite{Sefusatti:2007}, the SN survey and/or weak
lensing surveys \cite{Takada:2008fn}. Adding the priors shrinks areas of
the error ellipses,
because $\Omega_{\rm m0}$ and $b_2$ are degenerate 
 with  neutrino mass and dark energy parameters in
the galaxy power spectrum as implied in Fig.~\ref{fig:z1.0 only}. 
In
particular, for a Stage-III type survey, the prior of precision
$\sigma(\Omega_{\rm m0})\sim 0.01$ can efficiently break the
$m_{\nu,{\rm tot}}$-$w_0$ degeneracy, 
 thereby yielding the accuracies of $\sigma(m_{\nu, {\rm tot}})\simeq
 0.1~$eV and $\sigma(w_0)\simeq 0.05$, respectively. 
 For a Stage-IV type survey, the constraining power
is already sufficient, so such a prior does not much help
improve the parameter constraints. 

A more important question is how the
uncertainty of neutrino mass affects
dark energy constraints from future galaxy surveys.  
Table~\ref{table:sigmaw0} addresses this issue. 
First, comparing between the third and fourth columns clarifies that the
accuracy of $w_0$ determination is affected by including neutrino mass
parameter in the model fitting.  If neutrino mass is ignored, the
error of $w_0$ is apparently tightened by a factor of 1.2--1.4 for 
the galaxy surveys we consider here. 
It should be noted that the tighter constraints correspond to a 
case that the neutrino mass is sufficiently well 
determined by a laboratory experiment
(i.e. in this case $f_\nu$ is no longer a free parameter 
in the galaxy power spectrum). 

More importantly, ignoring neutrino mass in the model galaxy power
spectra likely results in a biased best-fit value of $w_0$. 
According to the method described in
Appendix~\ref{sec:systematic bias}, the column labelled as 
``$\delta w_0$ ignoring $f_\nu$'' estimates an amount of the possible 
bias, that is, the difference between
the input $w_0$ and the best-fit value obtained from the model fitting
without neutrino mass parameter: 
 $w_{0}^{\rm best-fit}=-1+\delta w_0$. To be more explicit, we here 
estimate the bias caused 
when the template of galaxy power spectrum assuming 
$f_{\nu}=0$ is fitted to
the observed spectrum having the true neutrino 
contribution of $f_\nu=0.01$ 
($m_{\nu, {\rm tot}}=0.12$~eV). The table shows a positive bias $\delta w_0$:
$w_0^{\rm best-fit}>w_0^{\rm input}=-1$, because a model
with $w_0>-1$ predicts galaxy spectra with smaller amplitudes due to the
suppressed growth rate, which mimics the 
neutrino suppression effect inherent in the (presumably here) measured 
spectrum. 
For BOSS and Stage-III type surveys, the bias is not significant because
$|\delta w_{0}/\sigma(w_{0})|< 1$, while a Stage-IV type survey may
suffer from a significant bias as $|\delta w_0|\sim \sigma(w_0)$. 
Table~\ref{table:sigmaw0} also shows that a $1\%$-level
prior of $\Omega_{\rm m0}$ helps reduce the statistical error 
$\sigma(w_{0})$, but also
make the systematic bias more significant at the same time. 

\begin{figure}[t]
\begin{center}
\includegraphics[width=0.45\textwidth]{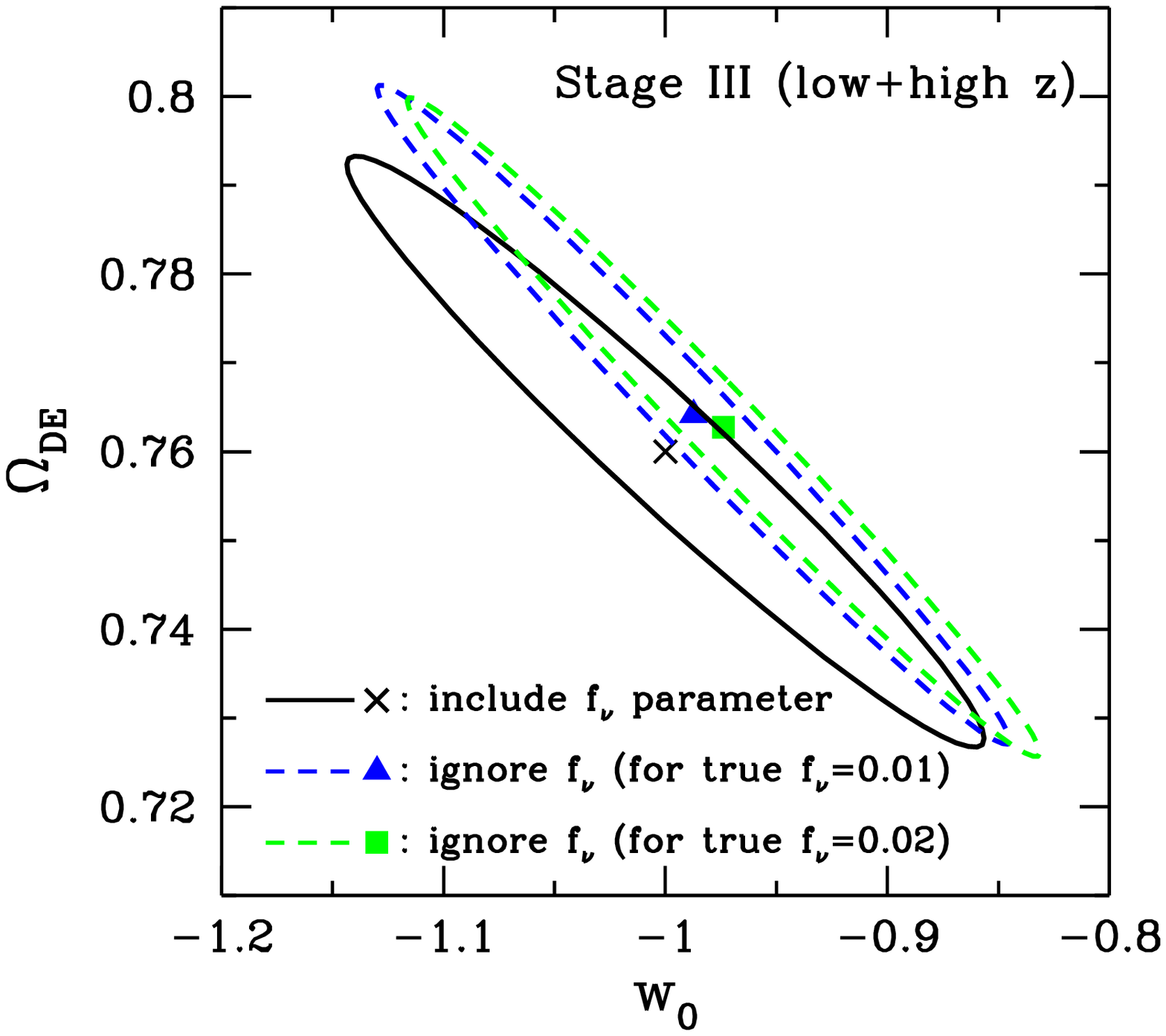}
\includegraphics[width=0.45\textwidth]{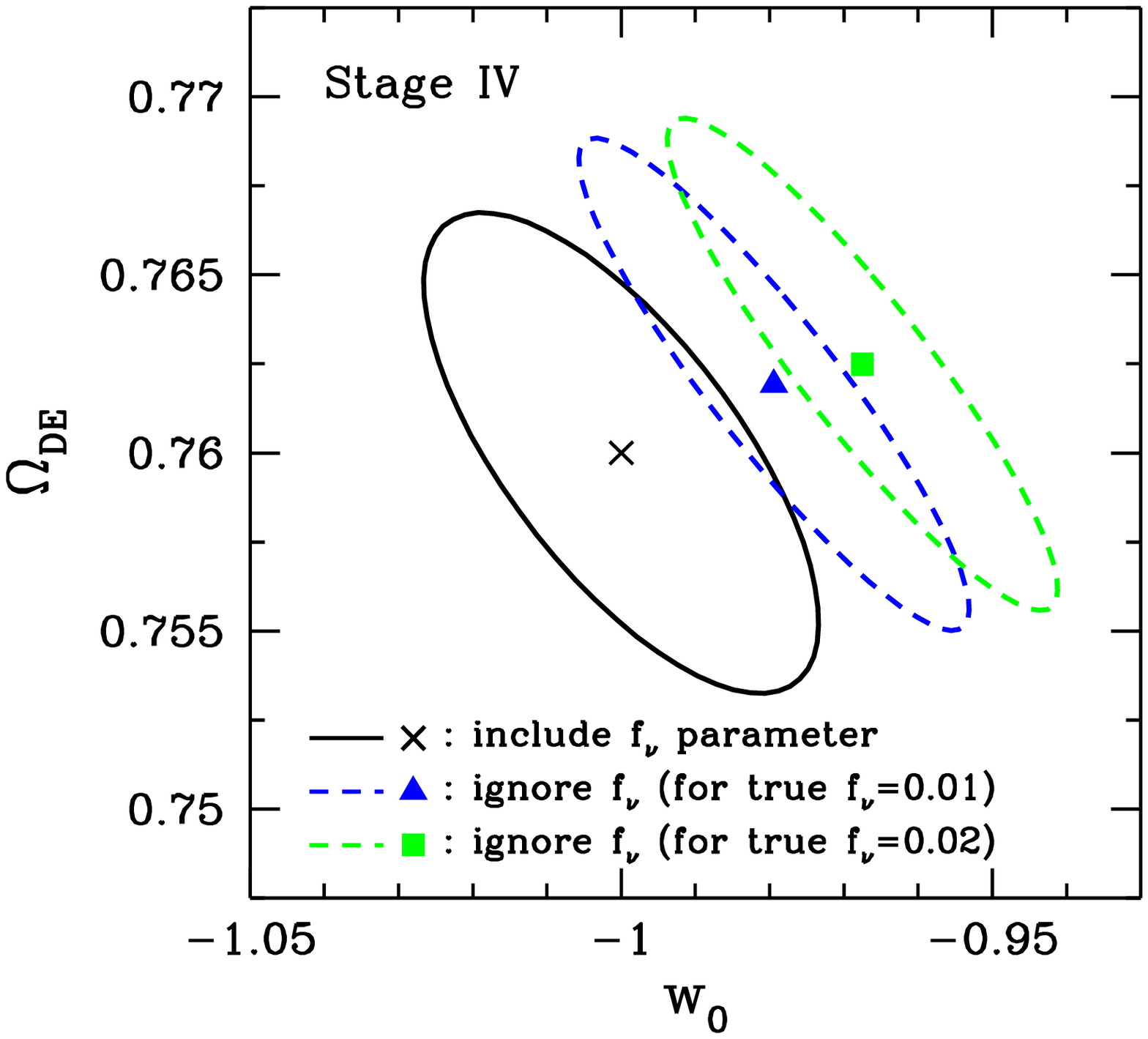}
\end{center}
\vspace*{-2em}
\caption{
The projected $1\sigma$ error ellipses in 
$(w_{0},\Omega_{\rm DE})$-plane for Stage-III 
(left panel) and Stage-IV surveys (right panel), respectively. 
Note that $C_{\rm max}=0.3$ is assumed.  The solid contours in each
panel shows the result for our fiducial method where neutrino mass
contribution to the galaxy power spectrum is properly taken into account
and the errors on dark energy parameters are derived by marginalizing 
over other parameter uncertainties.  The dashed contours and the
triangle or square symbols show the worst-case results: apparently tighter
constraints (smaller error ellipses) and biased best-fit values may be
caused if neutrino mass contribution is ignored in the model galaxy
power spectra. The triangle and square symbols show the biased values
when the underlying true cosmology has $f_\nu=0.01$ 
($m_{\nu,{\rm tot}}=0.12$~eV) and $0.02$ (0.24~eV), respectively. 
For a Stage-IV type survey, ignoring neutrino mass may cause a false 
best-fit model that is away from the true model $w_0=-1$ by more than 
the 1-$\sigma$ statistical errors. 
}
\label{fig:w0-Omegam0}
\end{figure}

Fig.~\ref{fig:w0-Omegam0} more nicely illustrates the impact of neutrino
mass uncertainty on dark energy parameter estimation from future galaxy
surveys, showing the projected error ellipses in
 ($w_0,\Omega_{\rm de0}$)-plane. It is clear that
ignoring $f_\nu$ 
leads to model fitting apparently with smaller error
ellipses and biased best-fit values for these parameters. In particular,
for a Stage-IV type survey, the biased best-fit dark energy model
confined by the 1-$\sigma$ statistical error bounds may 
happen to be outside from the underlying true model (the input value 
$w_0=-1$ in our case). The amount of bias would become greater for the
greater values of true neutrino mass. Thus Table~\ref{table:sigmaw0} and
Fig.~\ref{fig:w0-Omegam0} imply that neutrino mass contribution is not
negligible and needs to be included in the model interpretation for
future galaxy surveys in order not to have too optimistic and biased
dark energy constraints.

Note that the parameter biases studied here are mainly from the power
spectrum amplitude information. If the dark energy parameters
are estimated from BAO peak locations being marginalized over a
sufficient number of
nuisance parameters that include power spectrum amplitude parameters
\cite{Seo:2008yx}, 
the dark energy parameter biases can be minimized, 
although the constraining power is significantly weakened.
This is beyond the scope of this paper, but would be worth carefully 
studying.


\begin{table}[t]
\begin{tabular}{l|cccccc}
\hline\hline
\hspace{0.5em}Survey\hspace{0.5em}
& \hspace{0.5em}\shortstack{range of $k$ \\ $(C_{\rm max})$}\hspace{0.5em}
& \hspace{0.5em}\shortstack{$\sigma(w_{0})$ \\ ~}\hspace{0.5em}
& \hspace{0.5em}\shortstack{$\sigma(w_{0})$ \\ 
ignoring $f_{\nu}$
}\hspace{0.5em}
& \hspace{0.5em}\shortstack{$\delta w_{0}$ \\ 
ignoring $f_{\nu}$}\hspace{0.5em}
& \hspace{0.5em}\shortstack{$\sigma(w_{0})$ \\ 
+$\sigma(\Omega_{\rm m0})=0.01$}\hspace{0.5em}
& \hspace{0.5em}\shortstack{$\delta w_{0}$ \\ 
+$\sigma(\Omega_{\rm m0})=0.01$
}\hspace{0.5em}
\\
\hline
 \hspace{0.5em}BOSS\hspace{0.5em} & Linear 3\% (0.13)
& 0.1522 & 0.0978 & 0.0090 & 0.0507 & 0.0262 
\\ 
     & SPT 3\% (0.30) 
& 0.0768 & 0.0603 & 0.0141 & 0.0435 & 0.0243
\\
Stage-III & Linear 3\% (0.13)
& 0.1935 & 0.1067 & 0.0060 & 0.0503 & 0.0255
\\
    & SPT 3\% (0.30)
& 0.1103 & 0.0801 & 0.0125 & 0.0476 & 0.0254
\\
Stage-IV  & Linear 3\% (0.13)
& 0.0398 & 0.0375 & 0.0113 & 0.0311 & 0.0176 
\\
    & SPT 3\% (0.30)
& 0.0245 & 0.0223 & 0.0206 & 0.0226 & 0.0223
\\
\hline
\hline
\end{tabular}
\caption{
\label{table:sigmaw0}
The impact of massive neutrinos on determination of 
dark energy equation of state parameter, $w_{0}$. 
Note that $f_\nu=0.01$ is assumed for the fiducial model.  The fourth and fifth
columns, labelled as ``ignoring $f_\nu$'', show apparently tighter
constraints and biased best-fit values of $w_0$ caused when galaxy power
spectrum models without neutrino mass parameter is fitted to the true
spectrum with $f_\nu=0.01$, respectively, for each galaxy surveys.  The
sixth and seventh columns, labeled as 
``+$\sigma(\Omega_{\rm m0})=0.01$'', show the similar results when adding 
the prior $\sigma(\Omega_{\rm m0})=0.01$. 
}
\end{table}

\section{Summary and discussion}
\label{sec:Conclusion}
Following our earlier work \cite{Saito:2008bp}, in this paper we have
developed a method for computing nonlinear power spectra of total matter
and galaxies in a mixed dark matter (MDM) model (a model with CDM plus
finite-mass neutrinos) based on standard perturbation theory (SPT) approach. 
In particular we have carefully examined the validity of approximations
employed in our approach.

For our fiducial approach, we include only the linear-order neutrino
perturbations to compute the nonlinear power spectrum, where nonlinear
clustering is driven by the nonlinear growth of CDM plus baryon
perturbations. Our approach is motivated by the fact that the neutrino
free-streaming scale is sufficiently large for small neutrino mass
scales consistent with the current limit 
($m_{\nu,{\rm tot}}\ltsim 0.6$~eV) and the neutrinos are expected to 
more stay in the linear regime than CDM plus baryon. 
We carefully studied the validity of this assumption 
as briefly summarized in the following
(see Appendix~\ref{sec:validity_neutrino_linear}).  By solving the
hierarchical Boltzmann equations of neutrino perturbations including the
{\em nonlinear} gravitational potential contribution (due to the
nonlinear CDM and baryon density perturbations), we indeed found that
the amplitudes of neutrino density perturbations are enhanced by up to
$10\%$ at the weakly nonlinear scales up to $k\sim 0.5$Mpc$^{-1}$. Thus,
although the neutrino perturbation is indeed affected by nonlinear
clustering, the contribution to total matter power spectrum can be
safely ignored to less than a sub-percent level, for neutrino mass
scales of interest, because the neutrino perturbation contribution to
total matter clustering is suppressed by another small factor
$f_\nu=\Omega_{\nu0}/\Omega_{\rm m0}$ whose current limit 
$f_\nu\ltsim 0.05$.

Further we carefully studied the higher-order growth functions of CDM
plus baryon perturbations. Compared to the CDM case, the
finite-mass neutrinos cause a scale-dependent suppression in the
clustering growth rate, and therefore the higher-order growth rates
generally have complicated scale dependence. That is, the additional
nonlinear mode-coupling between perturbations of different wavenumbers
arise via the growth rates in a MDM model. We numerically solved the
differential equations of the higher-order growth rates, and found that
the higher-order growth rates are well approximated by the power of the
linear growth rate (see Fig.~\ref{fig:higher-order growth funtion}).

As a result the nonlinear power spectra can be approximately given by
rather simple forms (see Eqs.[\ref{eq:P22approx}], [\ref{eq:P13approx}]
and [\ref{eq:Nonlinear_matter_Pk}]) similarly as in the SPT approach for
a CDM model. The equation (\ref{eq:Nonlinear_matter_Pk}) is very
useful in a sense that the nonlinear power spectrum at a given redshift
$z$ can be computed from the linear transfer functions of CDM, baryon
and neutrino perturbations at the redshift $z$, which are standard
outputs of the publicly available codes, CMBFAST or CAMB. As in
\cite{Saito:2008bp}, we found that the neutrino suppression effect on
the total matter power spectrum amplitude is enhanced in the weakly
nonlinear regime than in the linear regime (see Fig.~\ref{fig:matter
suppression}). Note that the empirical halofit approach shows
$10\%$-level deviations from the SPT results in the weakly nonlinear
regime, although it qualitatively captures the neutrino effect in the
nonlinear regime (see Fig.\ref{fig:SPT vs HALOFIT}).

Thus we believe that our approach gives reliable, accurate model
predictions for the nonlinear matter power spectrum in a MDM model over
a wider range of 
scales, where the perturbation theory is valid, than the linear theory. 
Also important is the SPT
approach can explicitly tell the scales and redshifts where the 
linear theory ceases to be accurate or breaks down.  However, simulation
based studies are definitely needed to test and/or calibrate the SPT
predictions. An $N$-body simulations for a MDM model is still challenging, but
encouragingly the initial attempts are being explored in
\cite{Brandbyge:2008rv,Brandbyge:2008js}. Alternative approach to refine
the analytical modeling is to extend the SPT approach by including
higher-order loop corrections.  Recently there are several efforts made
in this direction for a CDM model: the Time-RG formalism
\cite{Pietroni:2008jx}, the renormalized perturbation
theory (RPT) \cite{Crocce:2007dt} and the closure theory approach
\cite{Taruya:2007xy} some of which show remarkable agreement with
$N$-body simulations over a wide range of BAO scales 
\cite{Carlson:2009it,Taruya:2009ir}. 
For example, an attempt to extend the PT approach for a MDM model 
has recently been made in the framework of the Time-RG formalism 
\cite{Lesgourgues:2009am}. 
Given the similarity between our approach and the PT of CDM model 
as described above, we hope that our method can be straightforwardly
extended to the improved nonlinear modeling.  This is our future
project, and will be presented elsewhere.

Another interesting result of this paper is we developed a method to
compute the nonlinear {\em galaxy} power spectrum in a MDM model by
taking into account the nonlinear biasing effect in a self-consistent
manner within the SPT framework. As given by
Eq.~(\ref{eq:Nonlinear_galaxy_Pk}), the nonlinear galaxy power spectrum
is modeled, in addition to cosmological parameters, by introducing the
linear and nonlinear bias parameters, $b_1$ and $b_2$, and one
additional parameter
to model the residual shot noise contamination $N$.
Once again,
although the validity of SPT approach needs to be tested by simulations,
our SPT approach is built on the physical foundation of large-scale
structure formation and therefore expected to be reasonably accurate in
the weakly nonlinear regime where the SPT approximately works out. 

After formulating the nonlinear galaxy power spectrum, we then estimated
the ability of future galaxy surveys for constraining neutrino masses
from the power spectrum information over scales ranging from the linear
regime to the weakly nonlinear regime.  In this paper for simplicity we
focused on the real-space power spectrum, i.e. ignored the redshift
distortion effect, because the nonlinear distortion effect, the
Finger-of-God effect, is not yet fully understood even in the weakly
nonlinear regime. We found that the accuracy of neutrino mass constraint
is indeed improved by including the power spectrum information up to the
weakly nonlinear regime compared to the linear regime, by a factor 1.3,
for all the planned BAO surveys (see
Table~\ref{table:sigmanu}).  However, the improvement is not so
significant because of severe parameter degeneracies in the nonlinear
regime (see Figs.~\ref{fig:z1.0 only} and \ref{fig:combining slices}). 
Thus the neutrino mass forecasts in the previous studies may
be too optimistic if the forecasts are derived assuming the linear bias
and the linear theory modeling.  Nevertheless it should be noticed that
Stage-III and -IV type surveys may allow for the neutrino mass
constraints to accuracies of $\sim 0.1$ and $0.05$~eV, respectively,
even from the 1D power spectrum information.  

We also studied how the finite-mass neutrinos affect the ability of
future surveys for constraining dark energy parameters. A change of dark
energy parameters such as $w_0>-1$ from cosmological constant model also
causes a suppression in the galaxy power spectrum amplitudes, because
the growth rate of mass clustering slows down due to the greater 
cosmic accelerating
expansion. Thus the dark energy constraints are likely correlated with
neutrino mass in the galaxy power spectrum (see Fig.~\ref{fig:w0-fnu}),
if the power spectrum amplitude information is included in parameter
estimation. In particular we pointed out that, if neutrino mass
parameter is ignored in the model fitting, the best-fit dark energy
parameters can be biased. For a Stage-IV type survey, the bias may be
greater than the statistical uncertainty: a false evidence of $w_0\ne
-1$ may be implied by the neutrino mass uncertainty, even if the
true model has $w_0=-1$. Thus our results
suggest that the neutrino mass contribution needs to be taken into
account for future BAO surveys and to be marginalized over in order to
obtain an unbiased constraint on dark energy parameters.

We believe that the SPT modeling of galaxy power spectrum can be a more
physically motivated model than other empirical approaches such as the
halo model approach or the method where 
 nuisance parameters such
as $Q_{\rm NL}$ in \cite{Tegmark:2006az} were empirically introduced to
model the nonlinear effects including the nonlinear bias effect.  The
method developed in this paper allows us to model the nonlinear galaxy
power spectrum self-consistently within SPT formulation without
introducing empirical nuisance parameters. Hence we hope that the use of
SPT model allows an unbiased extraction of cosmological parameters from
the measured galaxy power spectrum by marginalizing over the bias
parameters, as long as the analysis is restricted to scales where SPT is
valid.  We are planning to apply our method to the SDSS LRG power
spectrum. For the SDSS power spectrum measurement done in
\cite{Tegmark:2006az}, the redshift distortion effect is supposed to be
removed by using the Finger-of-Got compression algorithm
\cite{Tegmark:2004}. Note that the residual Kaiser's effect 
of redshift distortion is absorbed in the linear bias parameter 
after the spherical shell average of galaxy power spectrum 
 in redshift space. 
Therefore the
LRG power spectra are appropriate to compare with the SPT model
predictions studied in this paper. We will address how the use of our SPT
model changes the neutrino mass constraints as well as other
cosmological parameter estimation as a function of the maximum number
$k_{\rm max}$, compared to the previous results. This is now in progress
and will be presented elsewhere.

There are several other applications of our method. First is
gravitational lensing effects on CMB or distant galaxy images, which are
sensitive to total matter distribution and therefore known as a powerful
probe of neutrino mass being free of galaxy bias uncertainty
(e.g. \cite{Lewis:2006fu,Munshi:2006fn}). 
These lensing signals are affected by nonlinear
clustering, but the effect for a MDM model has not been fully explored.
Secondly, the formulation developed in this paper can be straightforwardly
extended to studying the higher-order correlations of total matter
and/or galaxy distribution, based on the SPT approach. The higher-order
correlations are expected to be very powerful to improve cosmological
constraints when combined with power spectrum information, and
especially to break degeneracies with galaxy bias parameters 
for a galaxy clustering case. 

\acknowledgments
We acknowledge D.~Eisenstein, O.~Lahav, E.~Komatsu, 
A.~Heavens, M.~Shoji, D.~Spergel and Y.~Suto
for useful discussion and valuable comments. 
S.S is supported by JSPS 
through research fellowships. 
This work is supported in part by Japan Society for Promotion
of Science (JSPS) Core-to-Core Program ``International Research 
Network for Dark Energy'', 
by Grant-in-Aid for Scientific Research on Priority Areas No. 467 
``Probing the Dark Energy through an Extremely Wide \& Deep Survey 
with Subaru Telescope'', 
and by World Premier International Research Center
Initiative (WPI Initiative), MEXT, Japan.
M.T. and A.T. are supported in part by a Grants-in-Aid for
Scientific Research from the JSPS (Nos. 17740129 and 18072001 for MT: 
No. 21740168 for AT).

\appendix 

\section{Nonlinear effect on neutrino perturbations}
\label{sec:validity_neutrino_linear}

Throughout the paper, we assumed that neutrino perturbations stay at 
linear level, and contribute to the higher-order CDM plus baryon 
perturbations only via the effect on the growth rate. 
We then simply used the result of neutrino perturbation in linear theory, 
$\delta^{\rm L}_{\nu}$. 
This assumption is essential for our formalism. 
In this Appendix, we discuss the validity of this assumption in some 
details. \par
 
Rigorously speaking, the higher-order Boltzmann equations for massive 
neutrinos must be solved for a quantitative estimate of 
the nonlinear effect on neutrino perturbations. 
However, there are at least two important facts 
that simplify the analysis. One is that the nonlinear 
gravitational instability is mainly driven by the CDM plus baryon 
perturbations, which have a dominant contribution to the total 
matter density, $f_{\rm cb}\gtsim 0.95$. Another important fact is 
the presence of the neutrinos' large free-streaming, 
which prevents the neutrinos from clustering 
together with CDM plus baryon on scales smaller than 
the neutrino free-streaming scale. Hence, to a good approximation, 
the impact of nonlinear clustering on the neutrino 
perturbations can be estimated from the nonlinear 
gravitational potential $\phi$ driven by the nonlinear 
CDM plus baryon perturbations, just ignoring the higher-order 
neutrino perturbations. Note that similar approach has been examined in 
\cite{Singh:2002de,Ringwald:2004np,Abazajian:2004zh}. 
Then, the Poisson equation would be modified as follows:
\ba
 &&-k^{2}\phi(k)  = 4\pi G a^{2}\rho_{\rm m}\left(
  f_{\rm cb}\delta^{\rm NL}_{\rm cb}(k)+f_{\nu}\delta_{\nu}(k)
  \right), \nonumber \\
 && \delta^{\rm NL}_{\rm cb}(k) \approx \sqrt{\frac{P^{\rm L}_{\rm cb}(k)
  +P^{(22)}_{\rm cb}(k)+P^{(13)}_{\rm cb}(k)}{P^{\rm L}_{\rm cb}(k)}}
 \delta^{\rm L}_{\rm cb}(k), 
 \label{eq:Poisson_NL}
\ea
where $P^{(22)}_{\rm cb}$ and $P^{(13)}_{\rm cb}$ describe the nonlinear 
CDM plus baryon density perturbations 
and are calculated from Eqs.~(\ref{eq:P22approx}) and (\ref{eq:P13approx}). 
Provided $P_{\rm cb}$ for a given cosmological model, 
we numerically solve the linearized Boltzmann hierarchies, 
Eqs.~(\ref{eq:Vlasov0})-(\ref{eq:Vlasov2}) coupled with
Eq.~(\ref{eq:Poisson_NL}), and obtain the solutions for $\Psi_{\ell}$ 
at a given redshift. We have used the CAMB code to implement this 
approach, modifying the corresponding parts in the code. 
Note that in the Poisson equation given above, the nonlinear 
corrections to the power spectrum $P_{\rm cb}$ are calculated 
assuming the linearity of neutrino perturbations. In this respect, 
our approach is not self-consistent, but is sufficient for our 
purpose to estimate the impact of the nonlinear clustering. 
In fact, the effect is found to be sufficiently small 
for scales of interest as shown below. Furthermore, if necessary, the 
correction to the CDM plus baryon perturbations due to the nonlinear 
neutrino perturbations can be computed iteratively 
in a perturbative manner.\par

\begin{figure}[t]
\begin{center}
\includegraphics[width=0.45\textwidth]{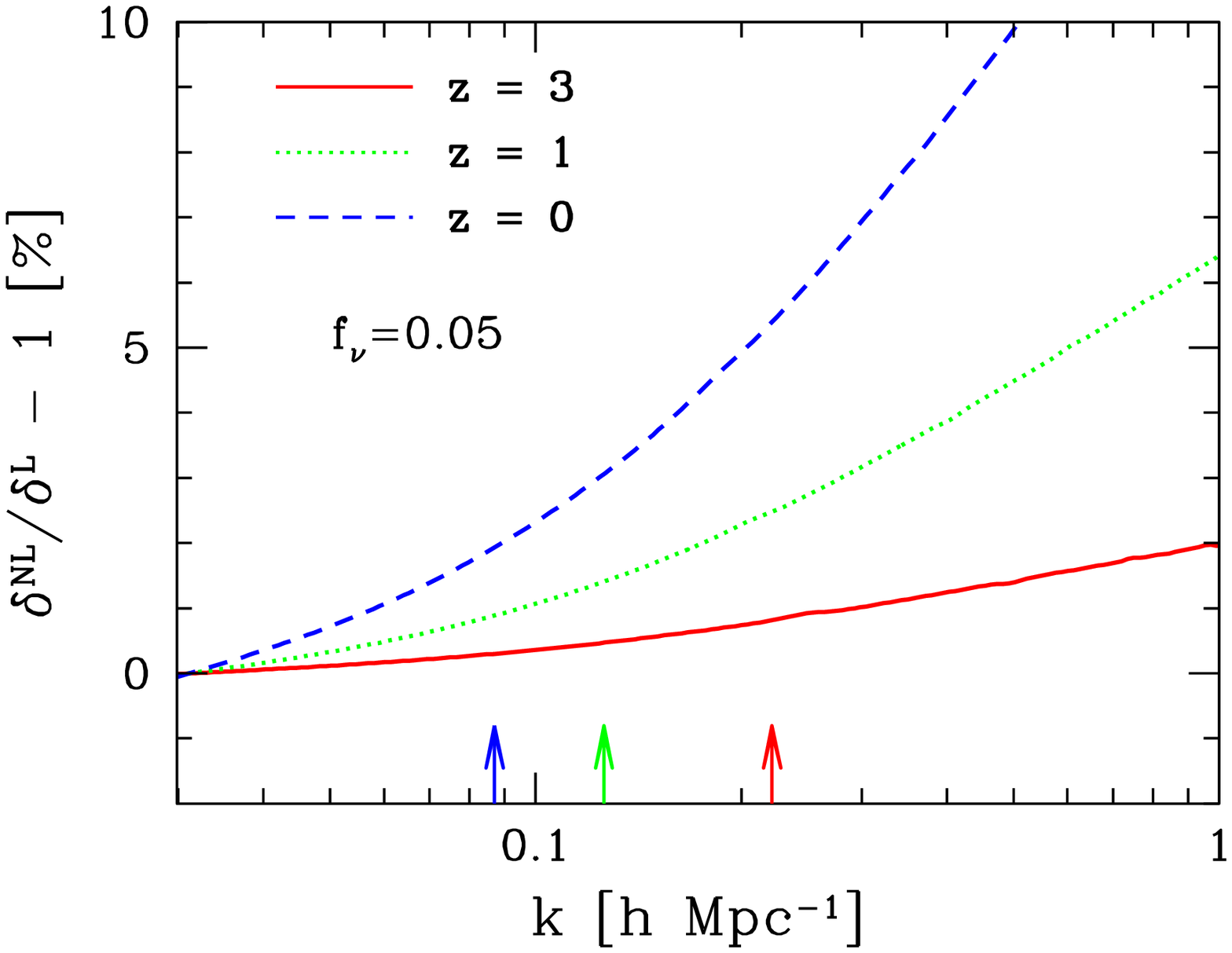}
\includegraphics[width=0.45\textwidth]{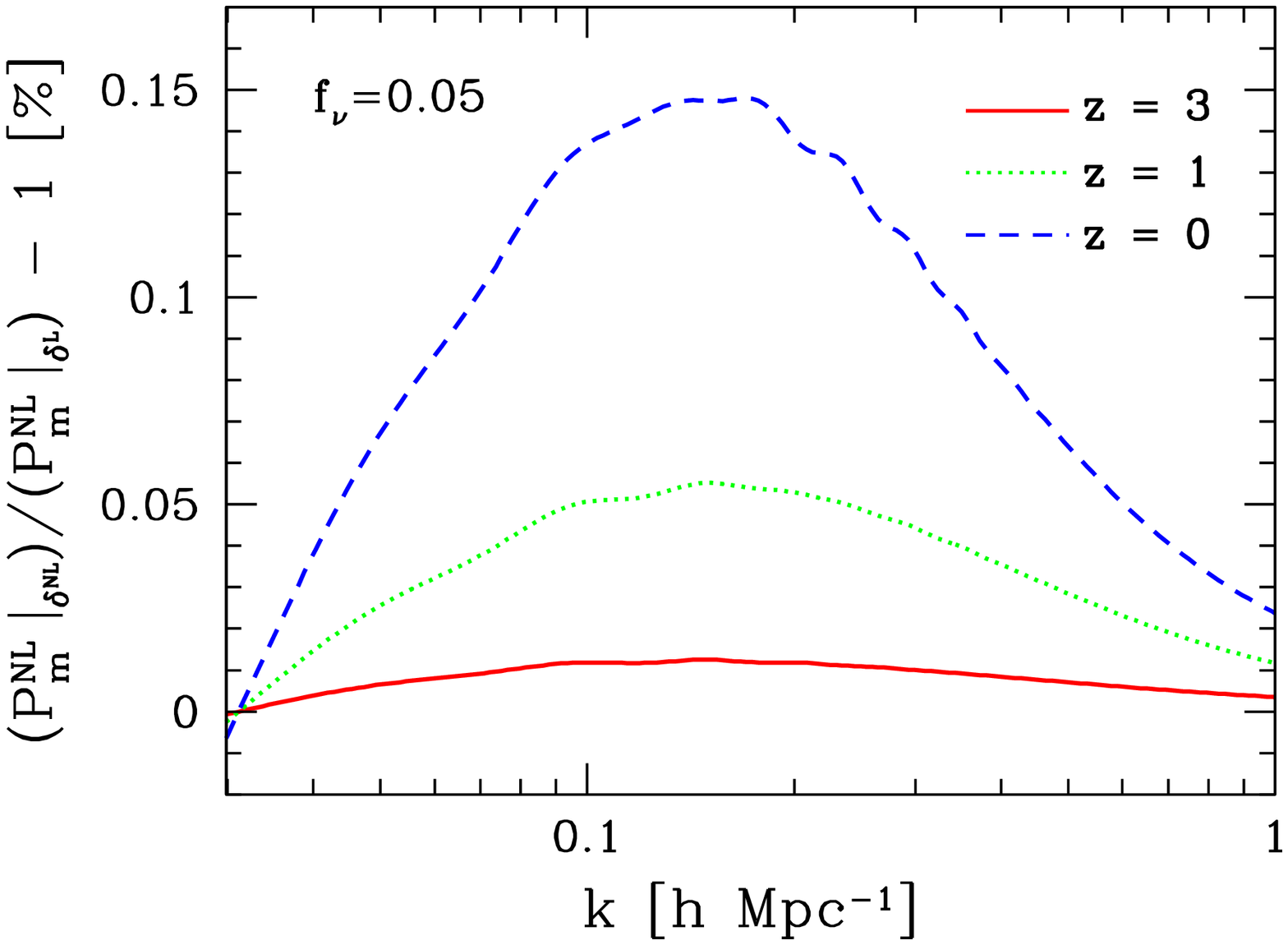}
\end{center}
\vspace*{-2em}
\caption{The fractional difference between the linear-order neutrino 
density perturbations $\delta^{\rm NL}_{\nu}$ computed using the approach 
described in text. We assume $f_{\nu}=0.05$ that corresponds roughly to the 
upper bound of current observational upper-bound. .
}
\label{fig.A}
\end{figure}

Fig.~\ref{fig.A} shows the fractional difference between the 
linear-order neutrino density perturbation $\delta^{\rm L}_{\nu}$ 
and the nonlinear perturbation $\delta^{\rm NL}_{\nu}$ obtained 
from the treatment mentioned above, where $\delta^{\rm NL}_{\nu}$ is 
calculated by inserting the solution for $\Psi_{0}$ into 
Eq.(\ref{eq:neutrino_delta}). We here chose a rather large neutrino 
mass, $f_{\nu}=0.05$,  close to the current upper bound.  
The plot clearly shows 
that nonlinear gravitational potential indeed enhances the 
neutrino perturbation by up to $\sim 10\%$ on scales where the PT 
is presumed to be valid. 
Since the contribution of 
neutrino perturbation to the total power spectrum 
always involves small additional factor $f_{\nu}$ 
(see Eq.~\ref{eq:total_matter_pk}),  
the result implies that influence of non-linearity on the total matter 
power spectrum is much more reduced. 
As a result, we found that the amplitude 
of $P_{\rm m}(k)$ increases only by $0.01\%$ compared to that obtained 
using the method of our paper. This effect gets even smaller as 
$\ltsim 0.01\%$ when $f_{\nu}\ltsim 0.05$. Hence, 
the error caused by the assumption that the 
neutrino perturbations stay at linear level is safely negligible,  
compared to the measurement errors at a percent level for a future survey. 
\par

Finally, we briefly comment on the recent work in \cite{Shoji:2009gg}. 
They discuss the effect of higher-order neutrino perturbations just 
treating the neutrinos as fluids with pressure. Strictly speaking, neutrinos 
cannot be treated as fluid, and the higher-order effect of 
moment hierarchy should be taken into account in a self-consistent way. 
Moreover, their formulation heavily relies on the assumption 
that neutrino perturbations stay at the same order as in the case of 
CDM plus baryon fluctuations, which is manifestly violated in the presence of 
the neutrino free-streaming. Even at the linear-order level, 
$\delta^{(1)}_{\rm cb}\gg \delta^{(1)}_{\nu}$ at the scales smaller than 
the neutrino free-streaming. Nevertheless, their results are qualitatively 
similar, and agree well with those examined here. 

\section{Higher-order growth functions in a MDM model}
\label{sec:Higher-order growth}

In this appendix, we summarize the basic equations for higher-order 
growth functions in a MDM model 
defined in Sec~\ref{subsec:Perturbative approach}, which were 
used for the analysis presented in Sec.~\ref{subsec:One-loop corrections}.  

\par 

Let us consider the second-order growth functions, $A^{(2)}_\delta$ 
and  $B^{(2)}_\delta$, defined in Eq.(\ref{eq:2nd_order_sol1}). From 
the perturbation equation for second-order quantity $\delta_{\rm cb}^{(2)}$ 
(see Eq.(\ref{eq:2nd_order_eq})), 
the governing equations for $A^{(2)}_\delta$ and  $B^{(2)}_\delta$ are 
obtained, and we have 
\ba
&&\ddot{A}^{(2)}_{\delta}(k_1,k_2;t) +2H\,\dot{A}^{(2)}_{\delta}(k_1,k_2;t)
-\frac{3}{2}H^2(1-\Omega_w)\,\fcb\,A^{(2)}_{\delta}(k_1,k_2;t) 
\nonumber\\
&&\quad\quad\quad\quad\quad\quad\quad\quad\quad= 
\left[H\,\frac{d\Dcb(k_1;t)}{d\ln a}\Dcb(k_2;t)\right]^{.}+
2H^2\,\frac{d\Dcb(k_1;t)}{d\ln a}\Dcb(k_2;t),
\label{eq:eq_for_A_delta}
\\
&&\ddot{B}^{(2)}_{\delta}(k_1,k_2;t) +2H\,\dot{B}^{(2)}_{\delta}(k_1,k_2;t)
-\frac{3}{2}H^2(1-\Omega_w)\,\fcb\,B^{(2)}_{\delta}(k_1,k_2;t) 
\nonumber\\
&&\quad\quad\quad\quad\quad\quad\quad\quad\quad
= \frac{1}{2}\,H^2\,\frac{d\Dcb(k_1;t)}{d\ln a}\frac{d\Dcb(k_2;t)}{d\ln a}.
\label{eq:eq_for_B_delta}
\ea
Note that in numerically solving the above equations, we retrieve only the 
inhomogeneous part of solutions so that the solution consistently 
approaches zero when going back to an initial time $t\to0$. This treatment 
just corresponds to picking up the growing-mode solution consistently, 
since the source terms of the evolution equations involve the 
growing-mode solution of linear perturbations.

Next write down the governing equations for third-order growth functions  
defined in Eq.(\ref{eq:3rd_order_sol1}),  
$\mathcal{I}^{(3)}_{\delta}(k_1,k_2,k_3)\,(\mathcal{I}=A,B,C,D,E,F)$,   
shortly abbreviated as $\mathcal{I}^{(3)}_{\delta}$.  
To do this, we first derive the 
perturbation equation for third-order quantity $\delta_{\rm cb}^{(3)}$.  
From Eq.(\ref{eq:2nd_df}), substitution of the linear and second-order 
solutions $\delta_{\rm cb}^{(1,2)}$ and $\theta_{\rm cb}^{(1,2)}$ leads to 
\ba
&&\ddot{\deltacb}^{(3)} + 2H\dot{\deltacb}^{(3)}
-\frac{3}{2}H^2(1-\Omega_w)\,f_{\rm cb}\deltacb^{(3)}
\nonumber\\
&&=\int\frac{d^3\bfk_1d^3\bfk_2d^3\bfk_3}{(2\pi)^6}\,\,
\delta_D(\bfk-\bfk_1-\bfk_2-\bfk_3) 
\hat{\Delta}(\bfk_1)\hat{\Delta}(\bfk_2)\hat{\Delta}(\bfk_3)
\nonumber\\
&&\quad\quad\times \,\,\Bigl[\,
\alpha_{1,23}\left\{\alpha_{2,3}\,\calS^{A}_{1,2,3}(t)
+\beta_{2,3}\,\calS^{B}_{1,2,3}(t)\right\}
-\alpha_{23,1}\left\{\alpha_{2,3}\,\calS^{C}_{1,2,3}(t)
+\beta_{2,3}\,\calS^{D}_{1,2,3}(t)\right\}\Bigr.
\nonumber\\
&&\quad\quad\quad\quad\quad\quad\quad\quad\quad\quad\quad\quad\quad
\quad\quad \Bigl.
-\beta_{1,23}\,
\left\{\alpha_{2,3}\,\calS^{E}_{1,2,3}(t)
+\beta_{2,3}\,\calS^{F}_{1,2,3}(t)\right\}
\Bigr],
\label{eq:3rd_order_eq}
\ea
where the quantities, $\alpha_{1,23}$, $\alpha_{23,1}$ and $\beta_{1,23}$ 
respectively indicate 
$\alpha(\bfk_1,\bfk_2+\bfk_3)$, $\alpha(\bfk_2+\bfk_3,\bfk_1)$ 
and $\beta(\bfk_1,\bfk_2+\bfk_3)$. 
Then, comparing 
the formal solution (\ref{eq:3rd_order_sol1}) with the above equation, 
we obtain the evolution equations for the third-order growth functions  
${\mathcal{I}}^{(3)}_{\delta}$: 
\be
\ddot{\mathcal{I}}^{(3)}_{\delta 1,2,3}
+2H\dot{\mathcal{I}}^{(3)}_{\delta 1,2,3}
-\frac{3}{2}H^{2}(1-\Omega_{w})f_{\rm cb}\mathcal{I}^{(3)}_{\delta 1,2,3}
=\calS^{\mathcal{I}}_{1,2,3}. 
\label{eq:eq_for_calI}
\ee
Again, the above equations must be solved just retrieving the 
inhomogeneous part of the solution. Here, the source functions, 
$\calS^{\mathcal{I}}_{1,2,3},\ (\mathcal{I}=A,B,C,D,E,F)$,  
are the scale- and time-dependent functions consisting of the linear and 
second-order growth functions. They are given by    
\ba
\calS^{A}_{1,2,3}(t)&=&
\left[H\frac{d\Dcb(k_1;t)}{d\ln a}A^{(2)}_{\delta}(k_2,k_3;t)\right]^{.}
  +2H^2\frac{d\Dcb(k_1;t)}{d\ln a}A^{(2)}_{\delta}(k_2,k_3;t),
\nonumber\\
\nonumber\\
\calS^{B}_{1,2,3}(t)&=&
\left[H\frac{d\Dcb(k_1;t)}{d\ln a}B^{(2)}_{\delta}(k_2,k_3;t)\right]^{.}
  +2H^2\frac{d\Dcb(k_1;t)}{d\ln a}B^{(2)}_{\delta}(k_2,k_3;t),
\nonumber\\
\nonumber\\
\calS^{C}_{1,2,3}(t)&=&
\left[H \Dcb(k_1;t) A^{(2)}_{\theta}(k_2,k_3;t)\right]^{.}
  +2H^2 \Dcb(k_1;t) A^{(2)}_{\theta}(k_2,k_3;t),
\nonumber\\
\nonumber\\
\calS^{D}_{1,2,3}(t)&=&
\left[H \Dcb(k_1;t) B^{(2)}_{\theta}(k_2,k_3;t)\right]^{.}
  +2H^2 \Dcb(k_1;t) B^{(2)}_{\theta}(k_2,k_3;t),
\nonumber\\
\nonumber\\
\calS^{E}_{1,2,3}(t)&=& H^2 \frac{d\Dcb(k_1;t)}{d\ln a} 
A^{(2)}_{\theta}(k_2,k_3;t),
\nonumber\\
\nonumber\\
\calS^{F}_{1,2,3}(t)&=& H^2 \frac{d\Dcb(k_1;t)}{d\ln a} 
B^{(2)}_{\theta}(k_2,k_3;t), 
\ea
In the above, the functions $A^{(2)}_{\theta}$ and $B^{(2)}_{\theta}$ are 
the second-order growth functions which appear in the solution of 
second-order velocity divergence $\theta^{(2)}_{\rm cb}$. These functions 
are related to the functions $A^{(2)}_{\delta}$ and $B^{(2)}_{\delta}$ 
through 
\be
A^{(2)}_{\theta}(k_1,k_2;t) = \frac{d\Dcb(k_1;t)}{d\ln a}\Dcb(k_2;t)-
H^{-1}\dot{A}^{(2)}_{\delta}(k_1,k_2;t), 
\quad
B^{(2)}_{\theta}(k_1,k_2;t) = -H^{-1}\dot{B}^{(2)}_{\delta}(k_1,k_2;t). 
\ee

Finally, we note that in the limit of $f_{\rm cb}\to1$ 
(i.e., case of massless neutrinos), there exist no free-streaming scales, 
and the linear growth function $D_{\rm cb}$ becomes independent of scales.   
From Eqs.(\ref{eq:eq_for_A_delta}), (\ref{eq:eq_for_B_delta}) and 
(\ref{eq:eq_for_calI}), this readily implies that all the second- and 
third-order growth functions become scale-independent. 
Then, employing the Einstein-de Sitter approximation, the 
analytical expressions for higher-order growth functions can be 
systematically obtained. In the Einstein-de Sitter approximation, 
all the calculations done in the Einstein-de Sitter universe are 
extended to apply to the other cosmological model by 
simply replacing the linear growth function in the Einstein-de 
Sitter universe with the one in the underlying cosmology.  
The detailed discussion on the validity of the Einstein-de Sitter 
approximation is given in 
~\cite{Bernardeau:1993qu,Takahashi:2008yk,Hiramatsu}.

\begin{figure}[t]
\begin{center}
\includegraphics[width=0.325\textwidth]{fig/A2.eps}
\includegraphics[width=0.325\textwidth]{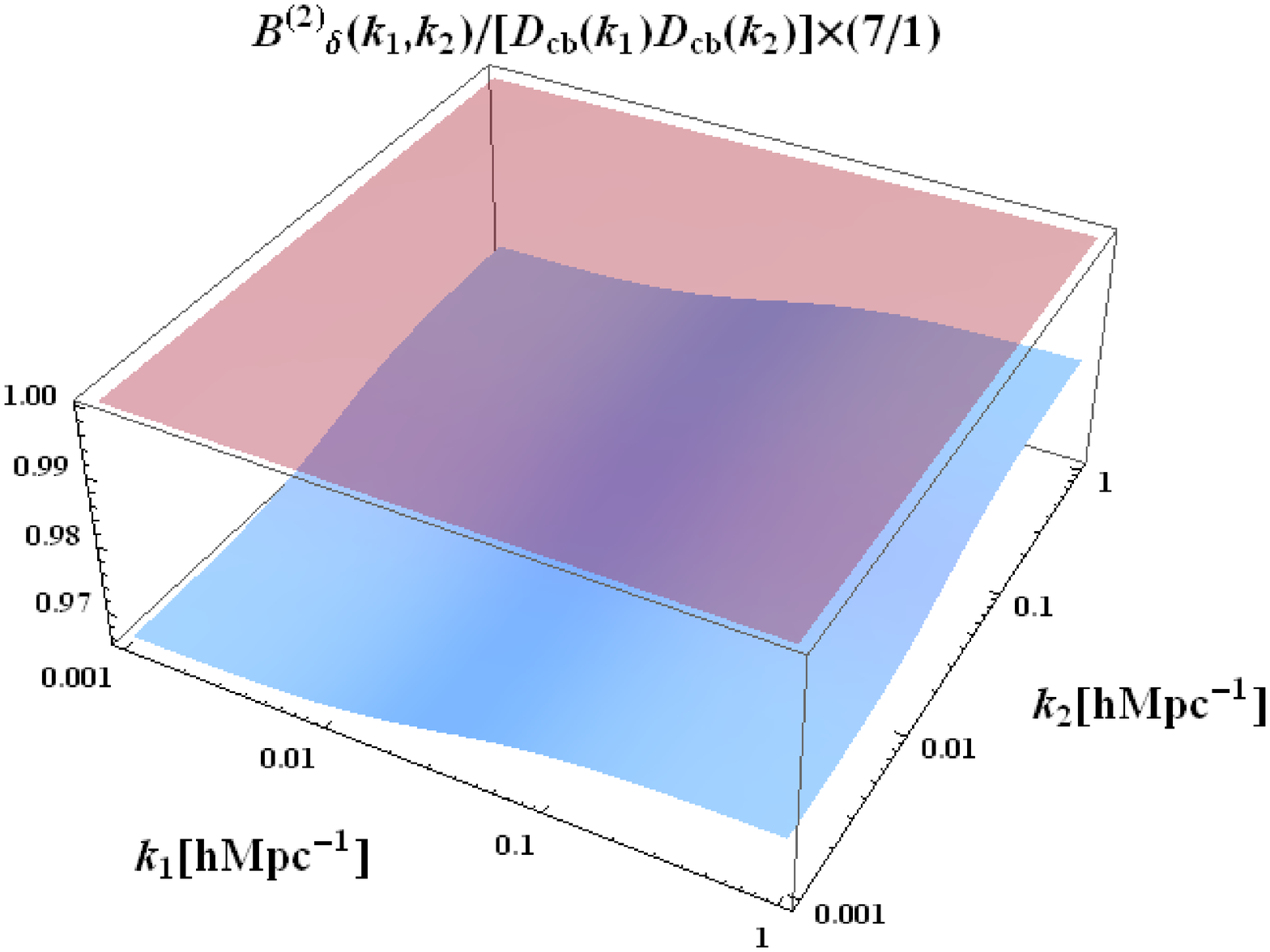}\\
\includegraphics[width=0.325\textwidth]{fig/A3.eps}
\includegraphics[width=0.325\textwidth]{fig/A3+.eps}
\includegraphics[width=0.325\textwidth]{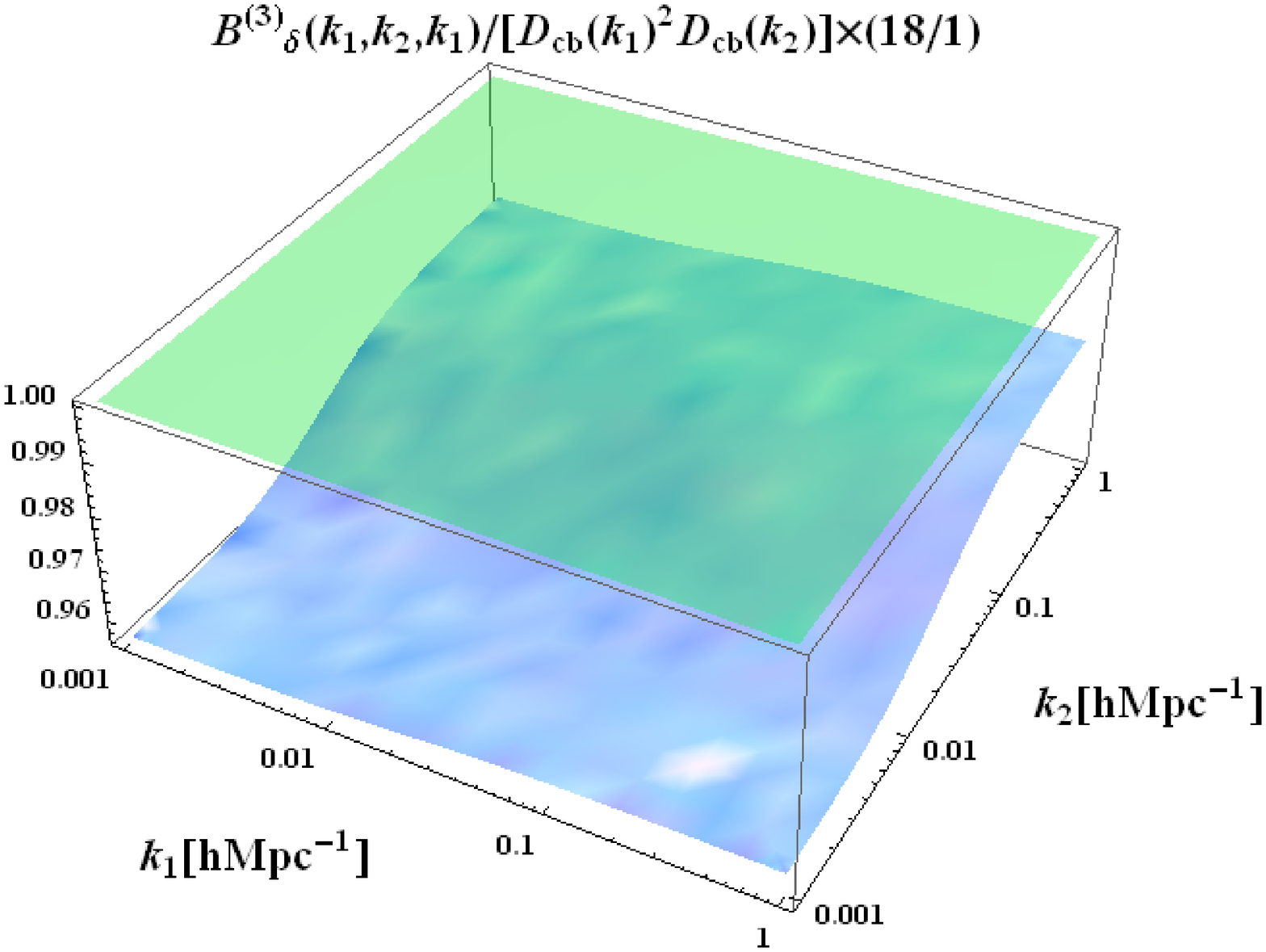}
\includegraphics[width=0.325\textwidth]{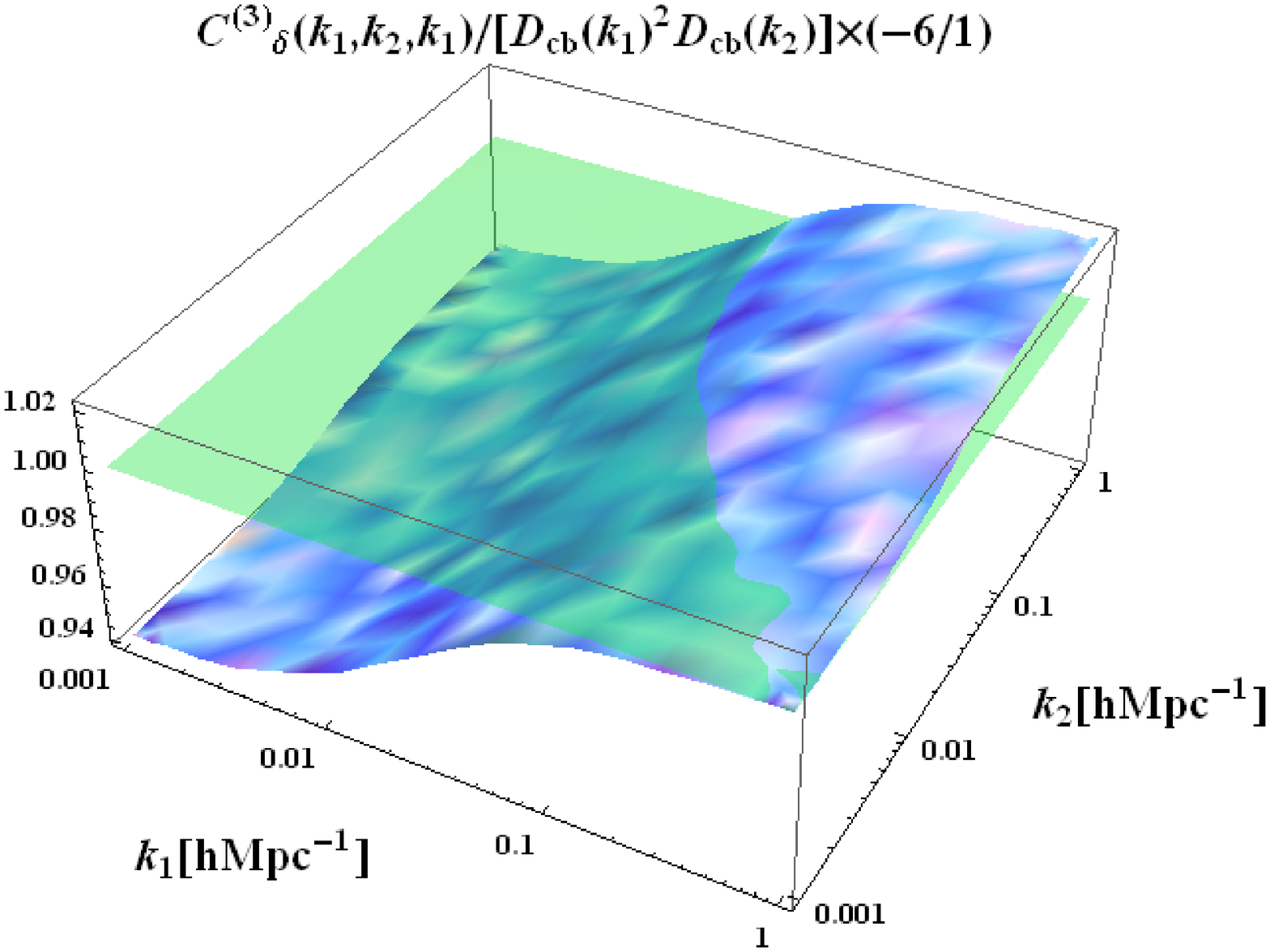}
\includegraphics[width=0.325\textwidth]{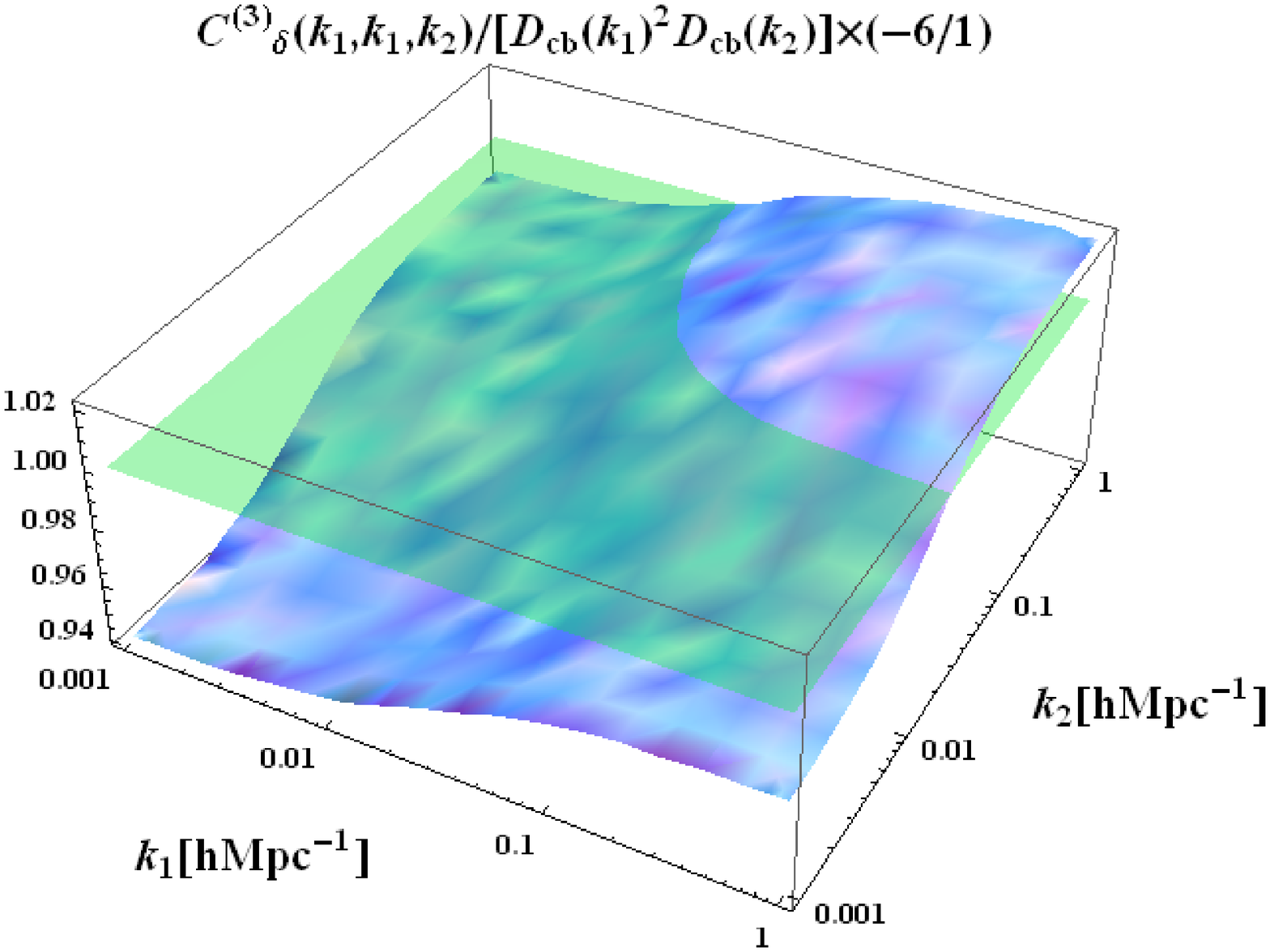}
\includegraphics[width=0.325\textwidth]{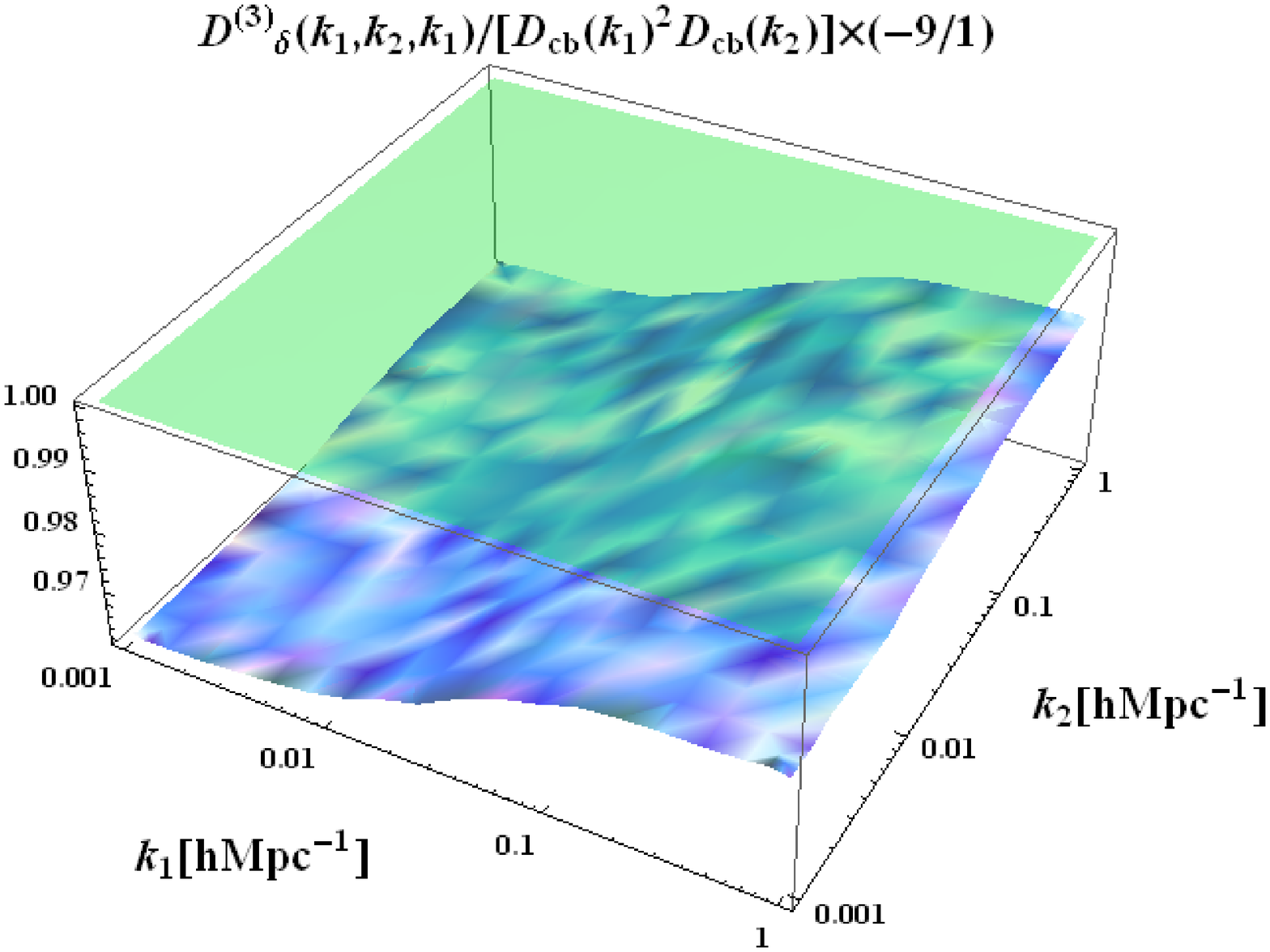}
\includegraphics[width=0.325\textwidth]{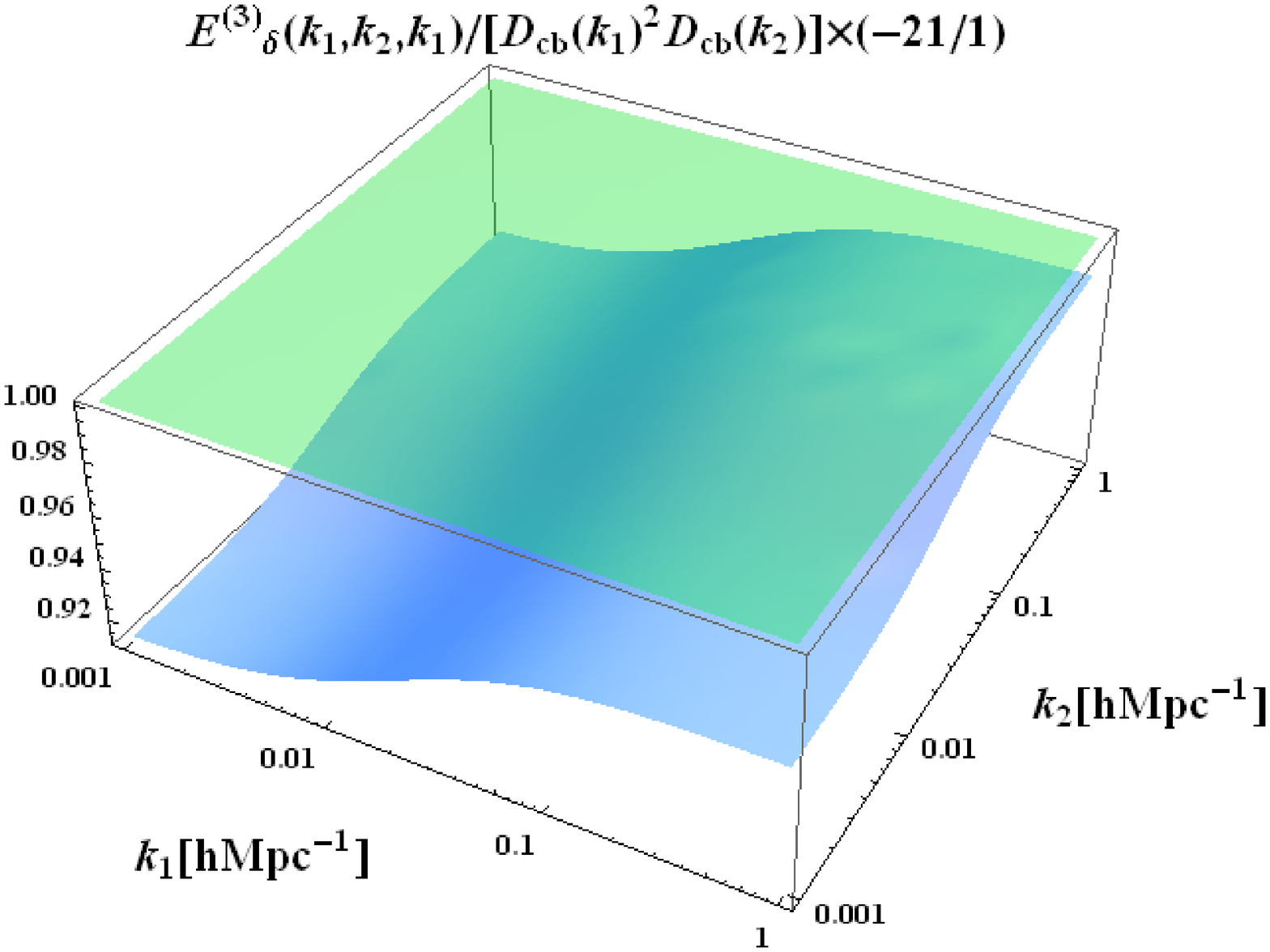}
\includegraphics[width=0.325\textwidth]{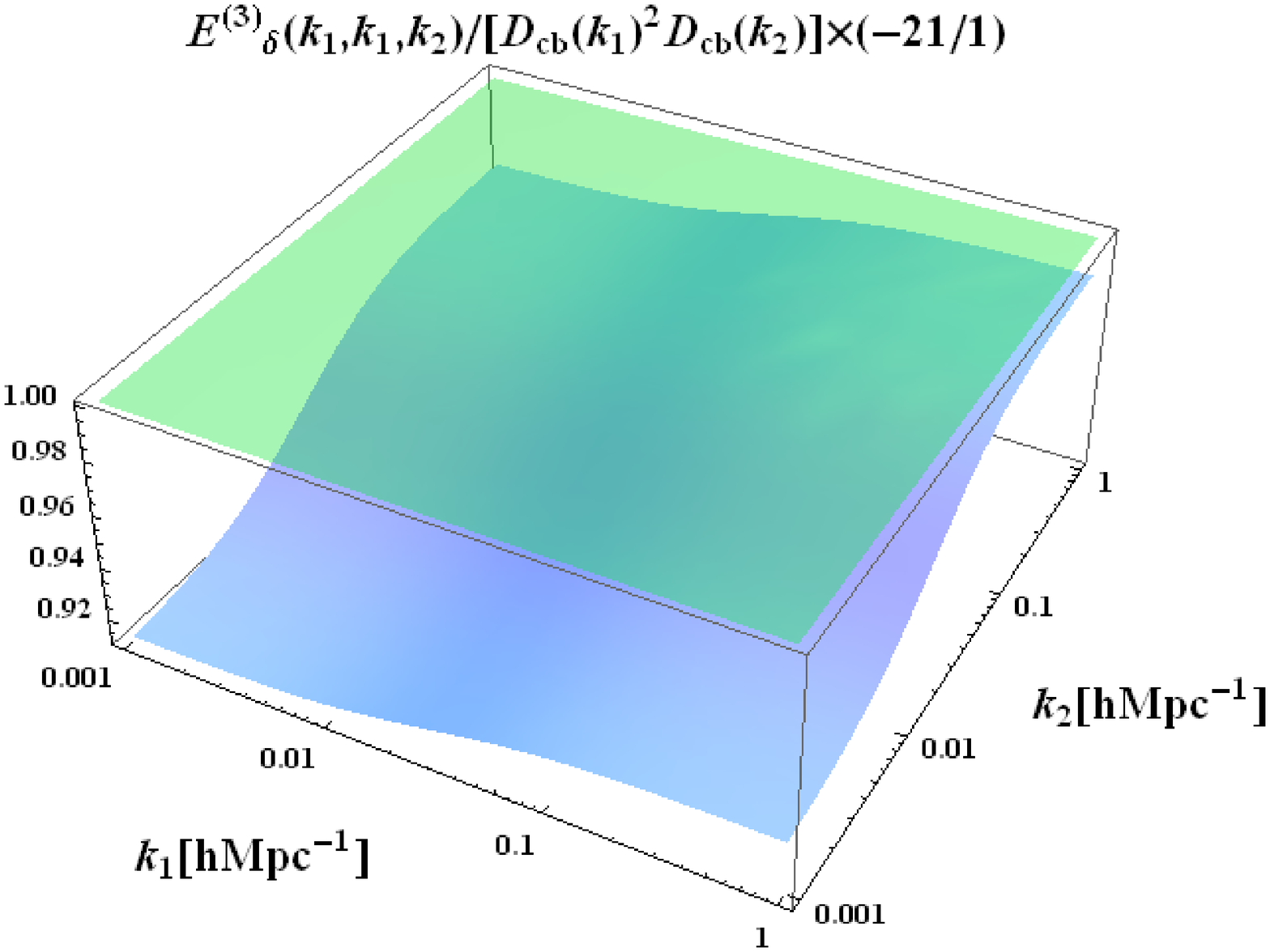}
\includegraphics[width=0.325\textwidth]{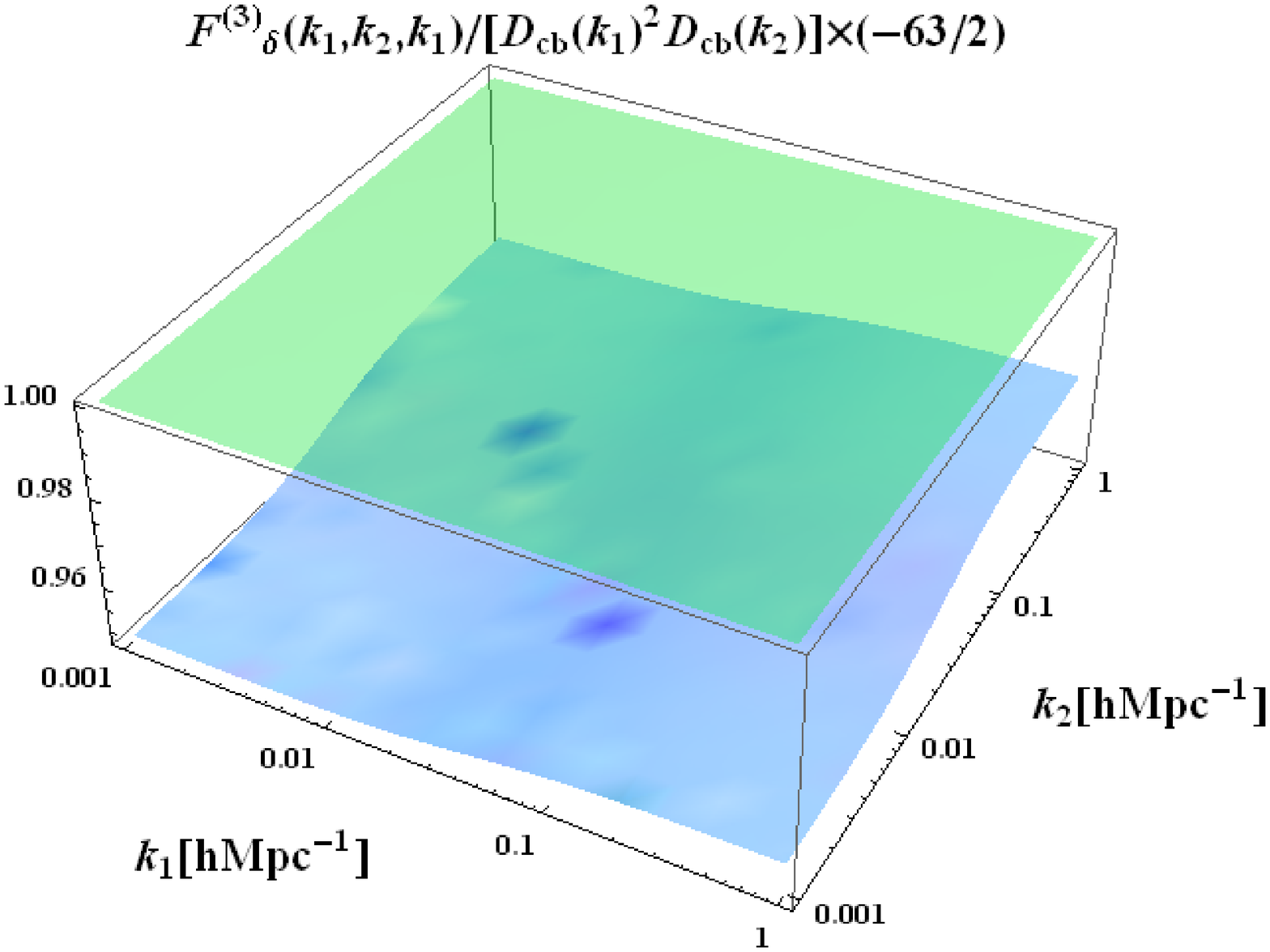}
\end{center}
\vspace*{-2em}
\caption{
 The 2nd-order and 3rd-order growth functions. 
 The ratios of 2nd-order growth functions to the square 
 of linear-oder growth functions, $A^{(2)}_{\delta}(k_{1},k_{2})
 /(D_{\rm cb}(k_{1})D_{\rm cb}(k_{2})$ ({\it top left}) and 
 $B^{(2)}_{\delta}(k_{1},k_{2})/(D_{\rm cb}(k_{1})D_{\rm cb}(k_{2})$ 
 ({\it top right}) are plotted ({\it blue} curved surfaces). 
 In the case of the 3rd-order growth functions, the ratios to the cubed 
 linear-oder growth functions, $\mathcal{I}^{(3)}_{\delta}
 /(D_{\rm cb}(k_{1})^{2}D_{\rm cb}(k_{2})),\ (\mathcal{I}=A-F)$ 
 are shown ({\it blue} curved surfaces). 
 Although the 3rd-order growth functions, 
 $\mathcal{I}^{(3)}_{\delta}$ depend on three specific wavenumbers, 
 it is sufficient to specify two wavenumbers to calculate 
 $P^{(13)}_{\rm cb}(k)$. That is why we show the ratios such as 
 the combination of $k_{1}$ and $k_{2}$, 
 $\mathcal{I}^{(3)}_{\delta}(k_{1},k_{2},k_{1})$ and 
 $\mathcal{I}^{(3)}_{\delta}(k_{1},k_{1},k_{2})$. 
 Note that $\mathcal{I}^{(3)}_{\delta}(k_{1},k_{2},k_{1})=
 \mathcal{I}^{(3)}_{\delta}(k_{1},k_{1},k_{2})$ for 
 $\mathcal{I}=B,D$ and $F$. 
 For the reference, the constant values ({\it red} plane for 2nd-order 
 and {\it green} plane for 3rd-order) are also shown. 
 These constant values corresponds to the SPT treatment in a CDM model.
}
\label{fig.B}
\end{figure}

As a result, higher-order growth functions in the $f_{\rm cb}\to1$ limit  
are analytically expressed as 
\be
A^{(2)}_{\delta}\to \frac{5}{7}D_{1}(t)^{2}, 
\quad\quad
B^{(2)}_{\delta}\to \frac{1}{7}D_{1}(t)^{2}. 
\ee
for the second-order growth functions, and 
\be
 \left\{A_{\delta}^{(3)},B_{\delta}^{(3)},C_{\delta}^{(3)},
 D_{\delta}^{(3)},E_{\delta}^{(3)},F_{\delta}^{(3)}\right\}
\to \left\{
 \frac{5}{18},\frac{1}{18},-\frac{1}{6},-\frac{1}{9},
 -\frac{1}{21},-\frac{2}{63}
\right\}\, D_{1}(t)^{3}. 
\label{eq:3rdNLgrowth_EdS}
\ee
for the third-order growth functions. For reference, in 
Fig.~\ref{fig.B} 
we show nonlinear growth functions 
as in Fig.~\ref{fig:higher-order growth funtion}. 

\section{Reparametrization of biasing parameters}
\label{sec:nonlinear-bias}

In this appendix, we review the reparametrized biasing parameters 
proposed by Ref.~\cite{McDonald:2006mx}. 
In this treatment, the galaxy power spectrum can be consistently 
related to the matter power spectrum calculated from SPT. \par

The starting point is that the fluctuation of galaxies is expanded in 
Taylor series assuming the local biasing prescription. In the local 
biasing scheme, 
the galaxy density field at a given position is described as the 
local function of matter fluctuation at the same position. 
In Fourier space, the galaxy density field is described as 
\ba
 \delta_{g}(\bfk)&=&c_{1}\delta_{\rm m}(\bfk)
  +\frac{c_{2}}{2}\int\frac{d^{3}\bfq}{(2\pi)^{3}}~
   \delta_{\rm m}(\bfq)\delta_{\rm m}(\bfk-\bfq)\nonumber\\
 && +\frac{c_{3}}{6}\int\frac{d^{3}\bfq_{1}d^{3}\bfq_{2}}{(2\pi)^{6}}~
   \delta_{\rm m}(\bfq_{1})\delta_{\rm m}(\bfq_{2})
   \delta_{\rm m}(\bfk-\bfq_{1}-\bfq_{2})+\epsilon(\bfk)+
   \mathcal{O}({\delta^{(1)}_{\rm m}}^{4}), 
\label{eq:local_bias}
\ea
where the $c_{i}$'s are the biasing parameters. The quantity 
$\epsilon$ represents the residual random field which cannot be represented 
by the matter fluctuations. We assume that randomness of $\epsilon$ is 
described by a white noise, and is uncorrelated with 
$\delta_{\rm m}$, i.e., $\langle\epsilon^2\rangle =N_{0}$ and 
$\langle\epsilon\delta_{\rm m}\rangle =0$. 
Then, the galaxy density power spectrum $P_{\rm g}$ up to the 
one-loop level is calculated as 
\ba
 P_{\rm g}&=&
 c_{1}^{2}P^{\rm NL}_{\rm m}(k)+
 \left(c_{1}c_{3}\sigma^{2}+\frac{68}{21}c_{1}c_{2}\sigma^{2}\right)
 P^{\rm L}_{\rm m}(k)\nonumber\\
 &&+2c_{1}c_{2}\int\frac{d^{3}\bfq}{(2\pi)^{3}}~
   P^{\rm L}_{\rm m}(q)P^{\rm L}_{\rm m}(|\bfk-\bfq|)
   {\cal F}^{(2)}_{\delta}(\bfq,\bfk-\bfq)\nonumber\\
 &&+\frac{c^{2}_{2}}{2}\int\frac{d^{3}\bfq}{(2\pi)^{3}}~
   P^{\rm L}_{\rm m}(q)P^{\rm L}_{\rm m}(|\bfk-\bfq|) + N_{0}, 
 \label{eq:exact-galaxy-density}
\ea
where the constant parameter $\sigma^{2}$ is defined as 
$\sigma^{2}\equiv\int d^{3}\bfq~P^{\rm L}_{\rm m}(q)/(2\pi)^{3}$, 
and the function $\calF^{(2)}_{\delta}(\bfk,\bfk')$ is the 
Fourier kernel of the second-order density perturbation:
\be
 \calF^{(2)}_{\delta}(\bfk,\bfk')\equiv 
 \frac{5}{7}+\frac{1}{2}\frac{\bfk\cdot\bfk'}{\bfk\bfk'}
 \left(\frac{k'}{k}+\frac{k}{k'}\right)+\frac{2}{7}
 \left(\frac{\bfk\cdot\bfk'}{kk'}\right)^{2}. 
 \label{eq:C3}
\ee
\par

While the calculation in the above is exact up to the fourth-order in density, 
due to the truncation at finite order,  
the expression (\ref{eq:exact-galaxy-density})  suffers from several 
unphysical behaviors such as an apparent divergence and anomalous 
low-$k$ power \cite{Heavens:1998es}. To remedy this, 
Ref.~\cite{McDonald:2006mx} proposed a way to regularize 
the expression (\ref{eq:exact-galaxy-density}) by 
reorganizing several terms and reparametrizing the biasing parameters. 
In this treatment, 
the first line of Eq.(\ref{eq:exact-galaxy-density}) is rewritten as 
\be
 c_{1}^{2}P^{\rm NL}_{\rm m}(k)+
 \left(c_{1}c_{3}\sigma^{2}+\frac{68}{21}c_{1}c_{2}\sigma^{2}\right)
 P^{\rm L}_{\rm m}(k)\to 
 b_{1}^{2}P^{\rm NL}_{\rm m}(k) \equiv
 \left[c_{1}^{2}+c_{1}c_{3}\sigma^{2}+\frac{68}{21}
 c_{1}c_{2}\sigma^{2}\right]P^{\rm NL}_{\rm m}(k). 
\ee
Further, the apparent divergence arising from the third line is 
absorbed by redefining the parameter $N_0$ as 
\be
 N \equiv N_{0}+\frac{c_{2}^{2}}{2}
 \int\frac{d^{3}\bfq}{(2\pi)^{3}}~P^{\rm L}_{\rm m}(q)^{2}. 
\label{eq:def_N}
\ee
Then, the galaxy power spectrum is re-expressed as follows:
\ba
 P_{\rm g}(k) &= &b_{1}^{2}\left[
 P^{\rm NL}_{\rm m}(k)+b_{2}P_{\rm b2,\delta}(k)
 +b_{2}^{2}P_{\rm b22}(k)
 \right]+N, \\
 P_{\rm b2,\delta}(k)&\equiv&2\int\frac{d^{3}\bfq}{(2\pi)^{3}}~
   P^{\rm L}_{\rm m}(q)P^{\rm L}_{\rm m}(|\bfk-\bfq|)
   \calF^{(2)}_{\delta}(\bfq,\bfk-\bfq), \\
 P_{\rm b22}(k)&\equiv&\frac{1}{2}\int\frac{d^{3}\bfq}{(2\pi)^{3}}~
   P^{\rm L}_{\rm m}(q)
   [P^{\rm L}_{\rm m}(|\bfk-\bfq|)-P^{\rm L}_{\rm m}(q)].  
\label{eq:reparameterized_pk_gal}
\ea
As a result, the galaxy power spectrum in the 
weakly nonlinear regime can be described with the only three 
parameters, $b_{1},b_{2}$ and $N$. Recently, 
the validity of the expression (\ref{eq:reparameterized_pk_gal}) 
has been examined in some details in Ref.~\cite{Jeong:2008rj}. 
They reported that this reparametrization scheme can fit well to the 
power spectrum of halos and galaxies in millennium simulations, 
and the cosmological parameters   
can be correctly estimated using Eq.(\ref{eq:reparameterized_pk_gal}) 
as a template in an unbiased fashion. \par

\section{Nonlinear bias parameters based on halo-model}
\label{sec:halo-bias}

We here summarize how to determine the fiducial values of the biasing 
parameters $b_1$, $b_2$ and $N$ listed in Table \ref{table:survey}, 
which are used in the Fisher matrix analysis 
in Sec.~\ref{subsec:forecast results}. \par

Following the treatment in Refs.~\cite{Seo:2003pu,Takada:2005si},  
we determine the linear biasing parameter $b_1$ 
at a given redshift so that the condition $\sigma_{8,\rm g}(z)=1$ 
is satisfied, where we define 
\be
 \sigma_{8,\rm g}(z)=b_{1}(z)\sigma_{8,\rm m}(z)
\sqrt{1+\frac{2F_{\rm m}(z)}{3}+\frac{F_{\rm m}(z)^{2}}{5}}, 
\label{eq:calc_b1}
\ee
with $F_{\rm m}(z)=-d\ln D_{1}(z)/d \ln (1+z)$. The function 
$D_1(z)$ is the linear growth rate for $\Lambda$CDM model, which we compute  
from the fiducial cosmological parameters just setting $f_\nu=0$. 
For SDSS LRG and BOSS surveys, their target samples are LRGs whose 
clustering properties are relatively known from the observations, and 
the linear biasing parameter is measured as $b_{1}\sim 2.10$ at $z=0.3$. 
Hence, when considering these surveys, we simply adopt this  
value, and the linear biasing parameters at different redshifts 
are determined from (\ref{eq:calc_b1}) just rescaling the 
condition $\sigma_{8,\rm g}(z)=1$ 
to $\sigma_{8,\rm g}(z)=\sigma_{8,\rm g}(0.3)$ with $b_{1}(0.3)=2.10$.  
\par

The non-linear biasing parameter $b_2$ in the expression 
(\ref{eq:reparameterized_pk_gal}) is related to 
the original parameters $c_i$ in Eq.(\ref{eq:local_bias}) as 
\be
 b_{2}=\frac{c_2}{c_1}. 
\ee
The biasing parameters $c_1$ and $c_2$ can be estimated from 
the halo-model approach (e.g., \cite{Scoccimarro:2000gm}). 
According to this prescription, we obtain
\be
 c_{i} = \frac{1}{\bar{n}_{\rm g}}\int_{M_{\rm min}}^\infty dM
 ~n_{\rm h}(M,z) 
 b^{\rm h}_{i}(M,z)\langle N\rangle_{M}, 
\label{eq:params_c_i}
\ee
where the function $n_{\rm h}(M,z)$ is the halo mass function 
for the given mass $M$ and redshift $z$, and the quantity 
$b^{h}_{i}(M,z)$ is the halo biasing parameter. The expectation value 
$\langle N\rangle_{M}$ is the so-called halo-occupation 
distribution, which describes the mean number of galaxies per halo 
with mass $M$. We here set  $\langle N\rangle_{M}=1$ for simplicity. 
We adopt the Sheth and Tormen formula for mass function $n_{\rm h}(M,z)$ 
\cite{Sheth:1999mn}:
\be
 n_{\rm h}(M,z)=-\frac{\bar{\rho}_{\rm m0}}{M^{2}}
 \frac{d\ln \sigma}{d\ln M}f(\nu)\,\,;\quad  
 f(\nu)=A\sqrt{\frac{2q}{\pi}}[1+(q\nu^{2})^{-p}]\nu e^{-q\nu^{2}/2},  
\label{eq:haloMF}
\ee
with $A=0.322$, $p=0.3$, and $q=0.707$. The density threshold $\nu$ is 
set to $\delta_{c}/\sigma(M,z)$ with $\delta_{c}=1.686$. Then,  
the halo biasing parameters $b^{\rm h}_{i}$ can be calculated from 
Eq.(\ref{eq:haloMF}) as 
\ba
  b^{\rm h}_{1}(M,z)&=&1+\epsilon_{1}+E_{1}, \\
  b^{\rm h}_{2}(M,z)&=&\frac{8}{21}(\epsilon_{1}+E_{1})
+\epsilon_{2}+E_{2},
\ea
where we define
\ba
 && \epsilon_{1}=\frac{q\nu^{2}-1}{\delta_{c}},\quad
 \epsilon_{2}=\frac{q\nu^{2}}{\delta_{c}}\frac{q\nu^{2}-3}{\delta_{c}},\\
 && E_{1}=\frac{2p}{\delta_{c}}\frac{1}{1+(q\nu^{2})^{p}},\quad
 \frac{E_{2}}{E_{1}}=\frac{1+2p}{\delta_{c}}+2\epsilon_{1}.  
\ea
In the expression (\ref{eq:params_c_i}),  there appears the 
minimum halo mass, $M_{\rm min}$, which can be determined from the 
condition, 
\be
 \overline{n}_{\rm g} =\int_{M_{\rm min}}dM~n_{h}(M,z)\langle N\rangle_{M}. 
\ee

Finally, it seems rather difficult to determine the fiducial 
value of the remaining parameter $N$, because physical meaning 
of the parameter $N$ is less clear. In this paper, we just adopt 
the relation (\ref{eq:def_N}), and compute $N$ assuming $N_{0}=0$: 
\be
 N=\frac{{c^{\rm h}_{2}}^{2}}{2}\int \frac{d^{3}{\bf q}}{(2\pi)^{3}}
 P^{\rm L}_{\rm m}(q)^{2}.  
\ee

\section{Systematic bias for the best-fit parameters}
\label{sec:systematic bias}

In this appendix, we briefly 
review how to estimate the biases in best-fit parameters 
arising from the systematic effects. We are especially 
concerned with the impact of neglecting massive neutrinos on the 
dark energy constraints. In this case, the biased parameter estimation 
is obtained by fitting the observational data to the 
power spectrum template incorrectly assuming $f_\nu=0$. 
Let us write down the observed power spectrum as 
\ba
 P^{\rm obs}_{\rm g}(k) 
& = & P^{f_{\nu}\neq 0}_{\rm g}(k) + P^{\rm noise}_{\rm g}(k)\nonumber\\
& = & P^{f_{\nu}=0}_{\rm g}(k)+P^{\rm sys}_{\rm g}(k)
+P^{\rm noise}_{\rm g}(k), 
\ea
where $P^{\rm noise}_{\rm g}(k)$ denotes the instrumental noise, while 
the systematics in power spectrum, $P^{\rm sys}_{\rm g}(k)$ is defined as 
$P^{\rm sys}_{\rm g}(k)\equiv P^{f_{\nu}\neq 0}_{\rm g}(k)-P^{f_{\nu}=0}_{\rm g}(k)$. 
Then, the systematic bias in a certain parameter $\delta p_{\alpha}$ 
is computed in (e.g., \cite{Joachimi:2009fr}) as 
\ba
 \delta p_{\alpha}=\sum_{\beta}({\bf F}^{-1})_{\alpha\beta}S_{\beta},
\label{eq:biased_params}
\ea
where ${\bf F}$ is the full Fisher matrix (namely, 
${\bf F}_{\alpha\beta}={\bf F}_{\alpha\beta}^{\rm galaxy}+
{\bf F}_{\alpha\beta}^{\rm CMB}$) in which the neutrino parameter $f_\nu$ 
is excluded from the matrix element. Note that in computing ${\bf F}$, 
the fiducial parameter for $f_{\nu}$ must be set to $f_\nu=0$, 
because we consider the situation that 
the observed power spectrum is incorrectly fitted to the 
template neglecting massive neutrinos.  Here, 
the vector quantity $S_{\alpha}$ is represented as 
$S_{\alpha}=S^{\rm CMB}_{\alpha}+S^{\rm galaxy}_{\alpha}$, which are 
respectively given by  
\ba
 S^{\rm CMB}_{\alpha}&=&\sum_{\ell}\sum_{X,Y}
 C^{X,{\rm sys}}_{\ell}
 \{\Xi(\hat{C}^{X}_{\ell},\hat{C}^{Y}_{\ell})\}^{-1}
 \frac{\partial C^{Y}_{\ell}}{\partial p_{\alpha}}, \\
 S^{\rm galaxy}_{\alpha}&=&
 \sum_{i}\frac{V_{\rm s}(z_{i})}{4\pi^2}
 \int^{k_{\rm max}(z_{i})}_{k_{\rm min}}k^{2}dk~
 \frac{P^{\rm est,sys}_{\rm g}(k;z_{i})}{P^{\rm est}_{\rm g}(k;z_{i})}
 \frac{\partial\ln P^{\rm est}_{\rm g}(k;z_{i})}{\partial p_{\alpha}}
 \left[\frac{\bar{n}_{\rm g}(z_{i})P^{\rm est}_{\rm g}(k;z_{i})}
 {\bar{n}_{\rm g}(z_{i})P^{\rm est}_{\rm g}(k;z_{i})+1}
 \right]^2, 
\ea
where the angular power spectrum for CMB, $C^{X,{\rm sys}}_{\ell}$, 
is defined similarly to the case of galaxy power spectrum, 
i.e., $C^{X,{\rm sys}}_{\ell}\equiv C^{X,f_{\nu}\neq 0}-C^{X,f_{\nu}=0}$. 
Note again that we set $f_{\nu}=0$ in computing 
$C^{X}_{\ell}$ and $P^{\rm est}_{\rm g}$.




\end{document}